\documentclass[reprint,
 amsmath,amssymb,
 aps,
 longbibliography,
]{revtex4-1}

\usepackage[dvipsnames]{xcolor}
\usepackage{xcolor}
\usepackage{graphicx}% Include figure files
\usepackage{dcolumn}% Align table columns on decimal point
\usepackage{bm}% bold math
\usepackage{hyperref}% add hypertext capabilities
\usepackage[normalem]{ulem} % \sout -> strike out text
\begin{document}

\preprint{}

\title{Entropy production in non-reciprocal polar active mixtures}

\author{Kim L. Kreienkamp}
\email{k.kreienkamp@tu-berlin.de}
\author{Sabine H. L. Klapp}
\email{sabine.klapp@tu-berlin.de}
\affiliation{%
 Institut f\"ur Physik und Astronomie, Technische Universit\"at Berlin
}%


\begin{abstract}
The out-of-equilibrium character of active systems is often twofold, arising from both the activity itself and from non-reciprocal couplings between constituents. A well-established measure to quantify the system's distance from equilibrium is the informatic entropy production rate. Here, we ask the question whether and how the informatic entropy production rate reflects collective behaviors and transitions in an active mixture with non-reciprocal polar couplings. 
In such systems, non-reciprocal orientational couplings can induce chiral motion of particles. At the field-theoretical level, transitions to these time-dependent chiral states are marked by so-called critical exceptional points. Here, we show that at a particle level, the entropy production rate within the chiral states increases with the degree of non-reciprocity, provided it is sufficiently strong. Moreover, even at small degrees of non-reciprocity, the transitions via exceptional points leave clear signatures in the entropy production rate, which exhibits pronounced peaks at coupling strengths corresponding to the field-theoretical exceptional points. Overall, the increase and peaks of the entropy production rate mirror the susceptibility of the polarization vector at the particle level. This correspondence is supported by a field-theoretical analysis, which reveals that, in the long-wavelength limit, the entropy production rate scales with the susceptibilities of the polarization fields.
\end{abstract}

\maketitle

% \tableofcontents	

\onecolumngrid

\section{\label{sec:introduction}Introduction}
The entropy production rate is a hallmark quantity for characterizing non-equilibrium systems, including nano- to microscale systems subject to strong fluctuations. By definition, it quantifies the breaking of time-reversal symmetry and, thereby, the system's distance from equilibrium \cite{seifert_2012_stochastic_thermodynamics,peliti_pigolotti_2021_stochastic_thermodynamics_introduction,fodor_cates_2022_irreversibility_active_matter,OByrne_vanWijland_2022_time_irreversibility_active_matter}.

Recently, two distinct sources of non-equilibrium have attracted particular interest, mostly in the level of single particles or a few interacting units: activity \cite{dabelow_eichhorn_2019_irreversibility_active_matter,pietzonka_seifert_2019_autonomous_engines_driven_active_matter,bebon_speck_2025_thermodynamics_active_matter_across_scales}, due to the self-propulsion of individual particles, and non-reciprocity \cite{loos_klapp_2020_irreversibility_heat_non-reciprocal_interactions,loos_klapp_2021_medium_entropy_recution_delay}, which breaks action-reaction symmetry. Non-reciprocity can arise, e.g., as an effective interaction mediated by a non-equilibrium environment \cite{saha_scalar_active_mixtures_2020,Ivlev_2015_statistical_mechanics_where_newtons_third_law_is_broken,scheibner_vitelli_2020_odd_elasticity,Bowick_2022_symmetry_thermodynamcs_topology_active_matter} or as a deliberate choice made by decision-making entities \cite{lotka_1920_lotka_volterra_model,volterra_1926_lotka_volterra_model,xiong_tsimring_2020_flower-like_patterns_multi-species_bacteria}.

Currently, the focus is shifting from single units to the entropy production associated with collective behavior and non-equilibrium phase transitions in interacting many-body systems. This includes, in particular, active matter systems \cite{pruessner_garcia-millan_2022_field_theories_active_matter_entropy_production,brossollet_biroli_2025_entropy_production_field_theories_interacting_particles} undergoing motility-induced phase separation \cite{nardini_cates_2017_entropy_production_field_theory} or flocking transitions \cite{borthne_cates_2020_time-reversal_symmetry_violations_entropy_production_field_theories_polar_active_matter,ferretti_walczak_2022_signatures_irreversibility_flocking,proesmans_fodor_2025_quantifying_dissipation_flocking_dynamics}, but also spin systems \cite{martynec_loos_2020_entropy_production_criticality_nonequilibrium_Potts} and phase transitions in interacting oscillator systems \cite{herpich_esposito_2018_collective_power,meibohm_esposito_2024_minimum-dissipation_principle_stochastic_oscillator,seara_murrell_2021_irreversibility_dynamical_phases_transitions}.
All of these studies address the overarching question of how collective behavior and phase transitions in non-equilibrium systems are reflected by the entropy production rate.

Very recently, particular attention has been devoted to the entropy production rate at non-reciprocal phase transitions \cite{dinelli_tailleur_2023_non-reciprocity_across_scales,alston_bertrand_2023_irreversibility_non-reciprocal_PT-breaking,suchanek_loos_2023_irreversible_fluctuations_emergence_dynamical_phases,suchanek_loos_2023_entropy_production_nonreciprocal_Cahn-Hilliard,loos_martynec_2023_long-range_order_non-reciprocal_XY_model}. Among the most intriguing phenomena in non-reciprocal systems are phase transitions to time-dependent states that break parity-time symmetry. At the field-theoretical level, these transitions are marked by so-called critical exceptional points (EPs) \cite{You_Baskaran_Marchetti_2020_pnas,fruchart_2021_non-reciprocal_phase_transitions,suchanek_loos_2023_irreversible_fluctuations_emergence_dynamical_phases}. A prototypical example is found in non-reciprocally coupled scalar Cahn-Hilliard systems, where increasing the degree of non-reciprocity transforms a static demixed state into a traveling wave state \cite{You_Baskaran_Marchetti_2020_pnas,suchanek_loos_2023_irreversible_fluctuations_emergence_dynamical_phases,suchanek_loos_2023_entropy_production_nonreciprocal_Cahn-Hilliard}. While the traveling state is clearly out of equilibrium, the static yet weakly non-reciprocal demixed state can superficially resemble an equilibrium state (even if it is not). Nevertheless, it turns out that the entropy production rate in Cahn-Hilliard systems is nonzero even for the slightest degree of non-reciprocity, revealing the underlying non-equilibrium character, and that it diverges at the EPs \cite{suchanek_loos_2023_irreversible_fluctuations_emergence_dynamical_phases,suchanek_loos_2023_entropy_production_nonreciprocal_Cahn-Hilliard,alston_bertrand_2023_irreversibility_non-reciprocal_PT-breaking}.

While most studies have so far considered the entropy production in either active or non-reciprocal systems, we here focus on systems exhibiting both, activity and non-reciprocity. 
These systems therefore possess a twofold non-equilibrium character. We consider, in particular, polar active matter, where sufficiently strong non-reciprocal orientational couplings can induce transitions to chiral states, in which particles tend to move on circular trajectories and synchronize \cite{fruchart_2021_non-reciprocal_phase_transitions}.
The transitions, which are signaled by EPs in the corresponding field theory, are accompanied by peaks in the angular velocities at the particle level \cite{kreienkamp_klapp_2024_synchronization_exceptional_points}.

These observations motivate the following questions: Is the exceptional transition in active matter systems also reflected in the entropy production rate at the particle level? How does the interplay between two distinct sources of non-equilibrium -- activity and non-reciprocity -- manifest in the system's entropy production rate?

To address these questions, we study a minimal model of an active mixture with non-reciprocal orientational couplings. The dynamical behavior of such systems has been studied in detail in Refs.~\cite{kreienkamp_klapp_2024_non-reciprocal_alignment_induces_asymmetric_clustering,kreienkamp_klapp_2024_dynamical_structures_phase_separating_nonreciprocal_polar_active_mixtures,kreienkamp_klapp_2024_synchronization_exceptional_points}. Here, we compute the informatic entropy production rate from particle-based simulations as a measure of the system's distance from equilibrium.

We find that, above a certain threshold of non-reciprocity, the entropy production rate increases with the strength of the non-reciprocal couplings. 
However, below this threshold, the entropy production rate remains largely unaffected by the strength of non-reciprocity, provided that couplings between different species are antisymmetric, i.e., equal in magnitude but opposite in sign.
Nevertheless, even below the threshold, at coupling strengths corresponding to exceptional points in the corresponding field theory, the particle-level entropy production rate exhibits pronounced peaks.

Comparing the particle-level entropy production rate with conventional observables that characterize the collective behavior, we find that, overall, it behaves similarly to the susceptibility of the polar vector as a function of the non-reciprocal coupling strength.

To rationalize these particle-level findings, we derive an analytical expression for the entropy production rate from corresponding field equations. In the long-wavelength limit, we show that the entropy production rate scales with the susceptibilities of the polarization perturbation field. This result provides a qualitative explanation for the observed correspondence between polarization vector susceptibility and entropy production rate on the particle level.

The close link between the susceptibility and the entropy production rate allows us to explain the observed peaks in entropy production near critical EPs. From a dynamical systems perspective, these critical EPs mark points where the Goldstone mode in flocking systems is excited. This excitation facilitates the rotation of particle orientations, leading to enhanced susceptibility and, consequently, increased entropy production.

\section{Results and discussion}
\subsection{Model}
\label{sec:model}
We consider a binary mixture of circular, self-propelled particles, comprising two species $a=A,B$, which has been studied in detail in Refs.~\cite{kreienkamp_klapp_2022_clustering_flocking_chiral_active_particles_non-reciprocal_couplings,kreienkamp_klapp_2024_non-reciprocal_alignment_induces_asymmetric_clustering,kreienkamp_klapp_2024_dynamical_structures_phase_separating_nonreciprocal_polar_active_mixtures,kreienkamp_klapp_2024_synchronization_exceptional_points}. The model combines two key ingredients of active matter: steric repulsion, which drives motility-induced phase separation (MIPS) \cite{cates_tailleur_2015_mips,Bialke_2013_microscopic_theory_phase_seperation}, and alignment interactions, which can induce flocking transitions \cite{marchetti_simha_2013_hydrodynamics_soft_active_matter,vicsek_1995_novel_type_phase_transition,gregoire_chate_2004_collective_motion}. Specifically, the dynamics of each particle $\alpha (= i_a)$ in species $a$ are governed by overdamped Langevin equations for their positions $\bm{r}_{\alpha}$ and orientations $\theta_{\alpha}$. The heading vector of particle $\alpha$ is given by $\bm{p}_{\alpha}=(\cos\theta_{\alpha},\sin\theta_{\alpha})^{\rm T}$. The evolution equations read
\begin{subequations} \label{eq:Langevin_eq}
	\begin{align}
		\dot{\bm{r}}_{\alpha} &= v_0\,\bm{p}_{\alpha} + \mu_{r} \sum_{\beta\neq\alpha} \bm{F}_{\rm{rep}}^{\alpha}(\bm{r}_{\alpha},\bm{r}_{\beta}) + \sqrt{2\,D'_{\rm t}} \, \bm{\xi}_{\alpha} \label{eq:Langevin_r}\\
		\dot{\theta}_{\alpha} &= \mu_{\theta} \sum_{\beta\neq\alpha} \mathcal{T}_{\rm al}^{\alpha}(\bm{r}_{\alpha},\bm{r}_{\beta},\theta_{\alpha},\theta_{\beta}) + \sqrt{2\,D'_{\rm r}}\, \eta_{\alpha} \label{eq:Langevin_theta}.
	\end{align}
\end{subequations}
The sums over particles $\beta=j_b$ couple the dynamics of particle $\alpha$ to the positions and orientations of all other particles from both species  $b=A,B$. 
Both species share identical self-propulsion speeds $v_0$. The particles interact via pairwise symmetric repulsive force $\bm{F}_{\rm{rep}}^{\alpha} = - \sum_{\beta\neq\alpha} \nabla_{\alpha} U(r_{\alpha\beta})$ between hard disks, derived from the Weeks-Chandler-Andersen potential \cite{weeks_1971_Weeks-Chandler-Andersen_potential} 
\begin{equation}
	\label{eq:WCA_potential}
	U(r_{\alpha\beta}) = \begin{cases}
		4\epsilon \left[\left( \sigma/r_{\alpha\beta}\right)^{12} - \left( \sigma/r_{\alpha\beta}\right)^6 + \frac{1}{4} \right], \, {\rm if} \, r_{\alpha\beta}<r_{\rm c}\\
		0, \ {\rm else} ,
	\end{cases}
\end{equation}
where $r_{\alpha\beta} = \vert \bm{r}_{\alpha\beta}\vert = \vert \bm{r}_{\alpha}-\bm{r}_{\beta} \vert$. The characteristic energy scale of the potential is given by $\epsilon$. The cut-off distance is $r_{\rm c }=2^{1/6}\,\sigma$. Further, the particles are subject to independent translational ($\bm{\xi}_{\alpha}^a$) and rotational ($\eta_{\alpha}$) Gaussian white noise, with zero mean and unit variance. The translational and rotational mobilities fulfill the Einstein relation and are connected to thermal noise strengths via $\mu_r=D'_{\rm t}/(k_{\rm B}\,T)$ and $\mu_{\theta}=D'_{\rm r}/(k_{\rm B}\,T)$, where $k_{B}$ and $T$ denote the Boltzmann constant and temperature, respectively. The two species differ solely in their Vicsek-like orientational couplings, characterized by torques 
\begin{equation}
	\label{eq:torque}
	\mathcal{T}_{\rm al}^{\alpha}(\bm{r}_{\alpha}, \bm{r}_{\beta}, \theta_{\alpha}, \theta_{\beta}) = k_{ab}\, \sin(\theta_{\beta}-\theta_{\alpha}) \, \Theta(R_{\theta}-r_{\alpha\beta})
\end{equation}
of strength $k_{ab}$. The step function $\Theta(R_{\theta}-r_{\alpha\beta})$ with $\Theta(R_{\theta}-r_{\alpha\beta})= 1$, if $r_{\alpha\beta} < R_{\theta}$, and zero otherwise, ensures that these torques only act within a neighborhood of radius $R_{\theta}$. Within the alignment radius $R_{\theta}$, particles of species $a$ tend to align with those of species $b$ if $k_{ab}>0$, and anti-align if $k_{ab}<0$. The interspecies couplings $k_{AB}$ and $k_{BA}$ may be reciprocal ($k_{AB}=k_{BA}$) or non-reciprocal ($k_{AB}\neq k_{BA}$). For more details regarding the model, see Refs.~\cite{kreienkamp_klapp_2022_clustering_flocking_chiral_active_particles_non-reciprocal_couplings,kreienkamp_klapp_2024_non-reciprocal_alignment_induces_asymmetric_clustering,kreienkamp_klapp_2024_dynamical_structures_phase_separating_nonreciprocal_polar_active_mixtures,kreienkamp_klapp_2024_synchronization_exceptional_points}.

We nondimensionalize the system using the particle diameter $\ell = \sigma$ and time scale $\tau = \sigma^2/D'_{\rm t}$. This yields the key control parameters: the particle density $\rho_0^a$, the reduced orientational coupling strength $g_{ab}=k_{ab}\,\mu_{\theta}\,\tau$, the P\'eclet number ${\rm Pe}=v_0\,\tau/\ell$, and the rotational noise strength $D_{\rm r}=D'_{\rm r}\,\tau$. We perform Brownian Dynamics (BD) simulations of $N=5000$ particles.

To isolate the role of (non-)reciprocal alignment in an otherwise symmetric system, we choose equal densities for both species, $\rho_0^A=\rho_0^B=\rho_0/2$, and identical intraspecies alignment strengths $g_{AA}=g_{BB}=g = 9$. The density $\rho_0^a = 4/(5\,\pi)$, motility ${\rm Pe}=40$, and rotational noise strength $D_{\rm r} = 3\cdot 2^{-1/3}$ are selected to ensure MIPS in the absence of alignment couplings ($g_{ab}=0 \ \forall \ ab$) \cite{kreienkamp_klapp_2024_non-reciprocal_alignment_induces_asymmetric_clustering}. The non-dimensionalized translational diffusion coefficient is $D_{\rm t}=1$. The repulsive strength is chosen to be $\epsilon^{*} = \epsilon / (k_{\rm B}\,T) = 100$, where the thermal energy is set to be the energy unit ($k_{\rm B}\,T=1$). The interaction radius is set to $R_{\theta}=10\,\ell$.

\subsubsection{Collective behavior}
\label{ssec:collective_behavior}
The system exhibits a wide range of collective behaviors. An illustration on the particle level is given by the snapshots in Fig.~\ref{fig:stability_diagram_snapshots}. For separate plots of particle species and orientations, see Appendix~\ref{app:collective_behavior_particle_simulations}.

The alignment interactions between particles can lead to globally polarized states. In our system, two distinct types of steady-state polarization arise: flocking, characterized by co-aligned motion of $A$- and $B$-particles, and antiflocking, where each species remains internally aligned but moves in opposite directions of the other species \cite{kreienkamp_klapp_2024_non-reciprocal_alignment_induces_asymmetric_clustering,kreienkamp_klapp_2024_dynamical_structures_phase_separating_nonreciprocal_polar_active_mixtures,kreienkamp_klapp_2024_synchronization_exceptional_points}. Reminiscent of results obtained for aligning particles in the context of conventional Vicsek systems \cite{vicsek_1995_novel_type_phase_transition,chate_raynaud_2008_collective_motion}, the spatial distribution of particles within the flocking state depends on the relative strength of noise and alignment (or, equivalently, particle density). The flocking state can either consist of polarized bands, as seen for weaker intraspecies alignment couplings \cite{kreienkamp_klapp_2024_dynamical_structures_phase_separating_nonreciprocal_polar_active_mixtures}, or of polarized states with a more homogeneous particle distribution for the here considered case of stronger intraspecies alignment. Nevertheless, we note that the formation of large-scale banded structures may be suppressed due to the finite system sizes accessible in our simulations \cite{chate_raynaud_2008_collective_motion}. In addition to these states with constant flock directions, one can observe chiral states, in which particles move on circular trajectories and exhibit (partial) synchronization when intraspecies couplings are sufficiently strong \cite{fruchart_2021_non-reciprocal_phase_transitions}. These chiral states emerge spontaneously due to the non-reciprocal orientational torques when intraspecies alignment strengths dominate over reorientational noise -- i.e., $g> 4\,D_{\rm r}/(R_{\theta}^2\,\rho_0)$. This condition is satisfied by the parameters considered in this study. The non-equilibrium states can be characterized by order parameters, as summarized in Appendix \ref{app:microscopic_obervables}. The size of the synchronized regions and the rotational frequency depend on the strength of non-reciprocity. The same qualitative collective dynamical states are observed across the system sizes accessible in our simulations, provided that the simulation box is sufficiently large compared to the interaction radius $R_{\theta}$, see Ref.~\cite{kreienkamp_klapp_2024_synchronization_exceptional_points}.

The steric repulsion between the motile particles leads to clustering and MIPS. Intriguingly, the non-reciprocal alignment couplings additionally induce non-trivial density behavior of particles. These include asymmetric clustering \cite{kreienkamp_klapp_2024_dynamical_structures_phase_separating_nonreciprocal_polar_active_mixtures,kreienkamp_klapp_2024_non-reciprocal_alignment_induces_asymmetric_clustering} and demixing \cite{kreienkamp_klapp_2024_synchronization_exceptional_points}.

\begin{figure}
	\includegraphics[width=1\linewidth]{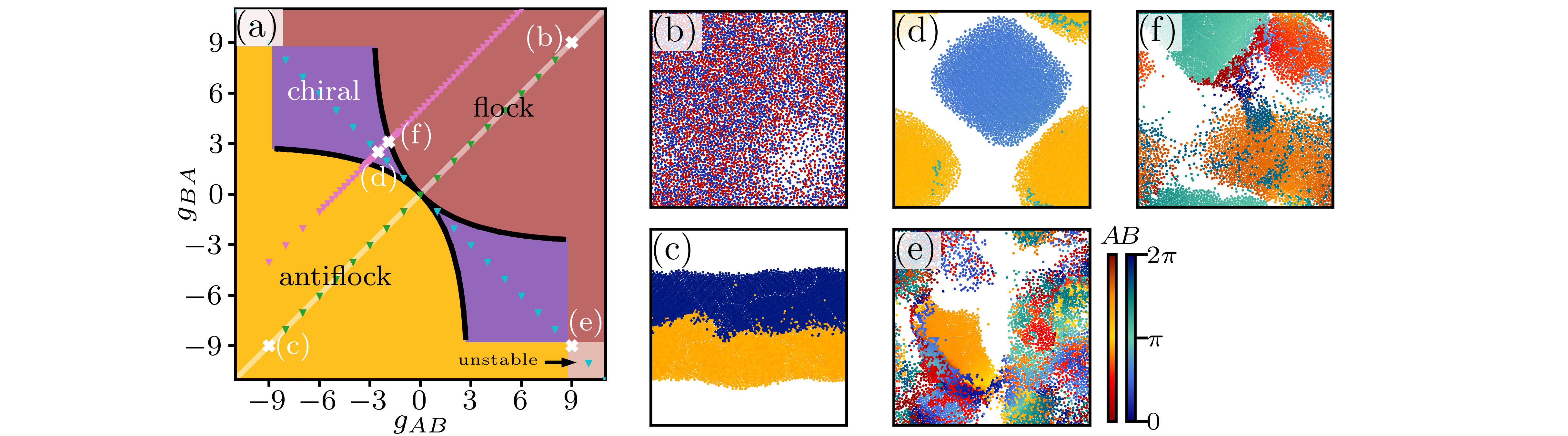}
	\caption{\label{fig:stability_diagram_snapshots}Stability diagram of the homogeneous flocking and antiflocking states against infinite-wavelength perturbations and particle simulation snapshots. The stability diagram in (a) is obtained from linear stability analyses of the continuum Eqs.~\eqref{eq:continuum_eq_density} and \eqref{eq:continuum_eq_polarization_density} for different interspecies coupling strengths $g_{AB}$ and $g_{BA}$. The intraspecies coupling strength is set to $g_{AA}=g_{BB}=9$. Exceptional points are indicated as black lines. The reciprocal system is marked with a white line. Green, cyan, and pink triangles denote data points corresponding to the particle simulations discussed in Figs.~\ref{fig:BD_pol_sus_entropy_results_reciprocal}, \ref{fig:BD_pol_sus_entropy_results_anti-symmetric}, and \ref{fig:BD_pol_sus_entropy_results_fixed_dif-5_species}, respectively. The Brownian Dynamics simulation snapshots correspond to (b) $g_{AB}=g_{BA}=9$, (c) $g_{AB}=g_{BA}=-9$, (d) $g_{AB}=-g_{BA}=-2.5$, (e) $g_{AB}=-g_{BA}=9$, and (f) $g_{AB}=g_{BA}-5=-1.9$. The color code for all snapshots is provided in (f): Particles are colored according to their species ($A$ and $B$ in yellowish-red and bluish colors, respectively), and the color gradient indicates particle orientation. Other parameters are specified in text. The stability diagram and snapshots are consistent with Ref.~\cite{kreienkamp_klapp_2024_synchronization_exceptional_points}.}
\end{figure}

\subsubsection{Continuum description}
\label{sssec:model_continuum}
The main characteristics of the collective behavior of particles can be captured on a mean-field continuum level. The continuum model is coarse-grained from the microscopic model \eqref{eq:Langevin_r} and \eqref{eq:Langevin_theta} \cite{kreienkamp_klapp_2022_clustering_flocking_chiral_active_particles_non-reciprocal_couplings,kreienkamp_klapp_2024_dynamical_structures_phase_separating_nonreciprocal_polar_active_mixtures,kreienkamp_klapp_2024_non-reciprocal_alignment_induces_asymmetric_clustering,kreienkamp_klapp_2024_synchronization_exceptional_points}. The deterministic (non-dimensionalized) continuum model consists of time evolution equations for the density fields 
$\rho^a(\bm{r},t)$,
\begin{equation}
	\label{eq:continuum_eq_density}
	\partial_t \rho^a = -  \nabla \cdot \big(v^{\rm eff}(\rho)\, \bm{w}^a -  D_{\rm t}\,\nabla\,\rho^a\big),
\end{equation}
and polarization densities $\bm{w}^a(\bm{r},t)$,
\begin{equation}
	\label{eq:continuum_eq_polarization_density}
	\partial_t \bm{w}^a = \bm{\mathcal{F}}^a[\rho^a, \rho^b, \bm{w}^a, \bm{w}^b],
\end{equation}
of species $a=A,B$, where $b\neq a$ and
\begin{equation}
	\begin{split}
		\bm{\mathcal{F}}^a[\rho^a, \rho^b,& \bm{w}^a, \bm{w}^b]
		= \sum_b g'_{ab} \, \rho^a\, \bm{w}^b - \sum_{b,c} \frac{ g'_{ab}\,g'_{ac}}{2\,D_{\rm r}} \, \bm{w}^a \, (\bm{w}^b \cdot \bm{w}^c) - D_{\rm r }\, \bm{w}^{a} + \mathcal{O}(\nabla\cdot \bm{w})+ \mathcal{O}(\nabla\rho) .
	\end{split}
\end{equation}
The polarization density measures the overall orientation at a certain position via $\bm{w}^a/\rho^a$. 
In the continuum description, the effect of steric repulsion is captured by the effective density-dependent velocity $v^{\rm eff}(\rho)={\rm Pe}- z\,\rho$ with $\rho=\sum_b\rho^b$, describing the slowing down of particles in crowded situations. The full functional $\bm{\mathcal{F}}^a$ is given in Eq.~\eqref{eq:polarization_density_functional} in Appendix~\ref{app:continuum_model_full}.

In the strong-coupling regime, i.e., for $g> 4\,D_{\rm r}/(R_{\theta}^2\,\rho_0)$, a linear stability analysis of the homogeneous flocking and antiflocking states with respect to infinite-wavelength perturbations in the mean-field continuum Eqs.~\eqref{eq:continuum_eq_density} and \eqref{eq:continuum_eq_polarization_density} yields the stability diagram shown in Fig.~\ref{fig:stability_diagram_snapshots}(a). The `chiral' state corresponds to a state, where flocking and antiflocking are predicted to be stable. Further details of the linear stability analysis can be found in Ref.~\cite{kreienkamp_klapp_2024_synchronization_exceptional_points}.

\subsubsection{Exceptional points in continuum model}
\label{sssec:model_exceptional_points}
In the present study, we are interested in the entropy production rate in the strong-coupling regime, where the dynamics are dominated by polarization. In particular, we focus on the chiral states and the transition to these states. In the continuum description, these transitions are marked by so-called exceptional points (EPs) \cite{fruchart_2021_non-reciprocal_phase_transitions}. Generally, EPs arise in non-Hermitian field theories -- including non-reciprocal field theories -- and correspond to parameter values where two eigenvalues of the linear stability matrix coalesce and their eigenvectors become parallel \cite{el-ganainy_demetrios_2018_non-Hermitian_physics_PT-symmetry_breaking,You_Baskaran_Marchetti_2020_pnas,fruchart_2021_non-reciprocal_phase_transitions,suchanek_loos_2023_time-reversal_PT_symmetry_breaking_non-Hermitian_field_theories}. 

In our system, we focus on so-called critical EPs separating the chiral states from flocking and antiflocking states, as seen in the linear stability diagram in Fig.~\ref{fig:stability_diagram_snapshots}. They occur via the coalescence of a damped flocking (or antiflocking) mode with a Goldstone mode and appear only for infinite-wavelength perturbations to the homogeneous flocking and antiflocking states. A detailed analysis of these EPs in the mean-field continuum model can be found in Refs.~\cite{fruchart_2021_non-reciprocal_phase_transitions,kreienkamp_klapp_2024_synchronization_exceptional_points}.

A physical interpretation of critical EPs from a dynamical-systems perspective is as follows. In flocking (or antiflocking) systems, the continuous rotational symmetry is spontaneously broken by the selection of a flocking direction. As a consequence, a Goldstone mode emerges, corresponding to fluctuations perpendicular to the flocking direction and reflecting the fact that global rotations of the flocking direction do not cost energy. In a linear stability analysis, this Goldstone mode therefore has a zero growth rate. At a critical EP, a previously damped mode coalesces with the Goldstone mode. As a result, the Goldstone mode is effectively actuated, which mediates the transition to chiral dynamics characterized by a persistent rotation of the flocking direction.

In Secs.~\ref{ssec:framework_entropy_production_particles} and \ref{ssec:numerical_results_entropy_production_microscopic}, we first examine the entropy production rate at the microscopic (particle) level, demonstrating that these field-theoretical EPs -- recently linked to enhanced rotational motion of particles \cite{kreienkamp_klapp_2024_synchronization_exceptional_points} -- also manifest in the entropy production rate of particles. In Sec.~\ref{ssec:entropy_production_rate_field_theory_perturbations_k-0}, we then return to the continuum framework to derive an analytical expression for the entropy production rate at the field level.

\subsection{Framework for the entropy production rate on microscopic level}
\label{ssec:framework_entropy_production_particles}
The general framework to calculate the entropy production rate of active particles is based on the formalism of stochastic thermodynamics \cite{seifert_2005_entropy_production_along_stochastic_trajectory,seifert_2012_stochastic_thermodynamics,dabelow_eichhorn_2019_irreversibility_active_matter,fodor_cates_2022_irreversibility_active_matter}. Here, we focus on the steady-state entropy production rate, defined as the time-averaged mean
\begin{equation}
	\label{eq:entropy_production_rate_definition}
	\langle \Delta \dot{S} \rangle = \lim_{\tau \rightarrow \infty} \Delta S[\vec{\bm{\varphi}}]/\tau,
\end{equation}
where $\Delta S[\vec{\bm{\varphi}}]$ is the medium entropy production associated with the fluctuating trajectory $\vec{\bm{\varphi}}$. The trajectory $\vec{\bm{\varphi}}$ characterizes the evolution of the microstate of the system. In the present context, it comprises the individual particle positions and orientations as functions of time at the particle-resolved level. At the continuum level, it corresponds to a trajectory in the space of coarse-grained fields, namely the density and polarization density fields. In the context of active matter, the quantity $\langle \Delta \dot{S} \rangle$ is referred to as the \textit{informatic} entropy production rate \cite{seifert_2012_stochastic_thermodynamics,nardini_cates_2017_entropy_production_field_theory}; it serves as a measure of the system's distance from equilibrium \cite{fodor_cates_2022_irreversibility_active_matter}.

Generally, a system's total entropy production can be decomposed into the system entropy production, associated with the system degrees of freedom, and the medium entropy production $\Delta S$, which measures the entropy produced in the environment. While the system entropy production captures the configurational changes in the system and vanishes in the non-equilibrium steady state considered here, the medium entropy production is expected to be non-zero. To calculate this quantity, we choose a trajectory that starts at a fixed $\bm{\varphi}(t_0)=\bm{\varphi}_0$ and ends at the final $\bm{\varphi}(\tau)=\bm{\varphi}_{\tau}$. The medium entropy production is then given by the log-ratio of the forward path probability $\mathfrak{p}[\vec{\bm{\varphi}}_{\backslash t_0} \vert \bm{\varphi}(t_0)]$ and the backward path probability $\mathfrak{p^\dagger}[\vec{\bm{\varphi}}^\dagger_{\backslash t_0} \vert \bm{\epsilon}\,\bm{\varphi}(\tau)]$ \cite{seifert_2005_entropy_production_along_stochastic_trajectory,seifert_2008_stochastic_thermodynamics,spinney_ford_2012_entropy_production_full_phase_space,crosato_spinney_2019_irreversibility_active_matter},
\begin{equation}
	\label{eq:medium_entropy_production_equation}
	\Delta S[\vec{\bm{\varphi}}] = k_{\rm B}\, {\rm ln}\left( \frac{\mathfrak{p}[\vec{\bm{\varphi}}_{\backslash t_0} \vert \bm{\varphi}(t_0)]}{\mathfrak{p^\dagger}[\vec{\bm{\varphi}}^\dagger_{\backslash t_0} \vert \bm{\epsilon}\,\bm{\varphi}(\tau)]} \right)
\end{equation} 
with Boltzmann constant $k_{\rm B}$. The forward path probability quantifies the probability for the trajectory $\vec{\bm{\varphi}}_{\backslash t_0} = \{\bm{\varphi}(t) \vert t \in(t_0,\tau]\}$, which does not include the initial point $\bm{\varphi}_0$.
The backward path probability is the probability of observing the time reverse of trajectory $\vec{\bm{\varphi}}_{\backslash t_0}$, defined as $\vec{\bm{\varphi}}_{\backslash t_0}^\dagger = \{\bm{\varphi}^\dagger(t) \vert t\in(t_0,\tau]\}$ with $\bm{\varphi}^\dagger(t) = \bm{\epsilon}\,\bm{\varphi}(\tau + t_0 - t)$. The time reversal operation $\bm{\epsilon}\bm{\varphi} = (\epsilon_1\,\varphi_1, ..., \epsilon_n\,\varphi_n)$ distinguishes between variables that are even or odd under time reversal \cite{spinney_ford_2012_entropy_production_full_phase_space}. For translational variables (like positions or velocities), $\varphi_i$ even implies $\epsilon_i\, \varphi_i = \varphi_i$ and $\varphi_i$ odd implies $\epsilon_i\, \varphi_i = -\varphi_i$. For rotational variables (like orientations), $\varphi_i$ even implies $\epsilon_i \, \varphi_i = \varphi_i$, and $\varphi_i$ odd implies $\epsilon_i \, \varphi_i = \varphi_i + \pi$ \cite{crosato_spinney_2019_irreversibility_active_matter}. Generally, the probability measure $\mathfrak{p^\dagger}$ for time-reversed path may differ from $\mathfrak{p}$, since it takes into account the time-reversal operation on protocol variables, like, e.g., (odd) external magnetic fields. However, in our model, all parameters are time-independent and even under time-reversal, such that $\mathfrak{p^\dagger}[\cdot] = \mathfrak{p}[\cdot]$ \cite{crosato_spinney_2019_irreversibility_active_matter}.

Following Refs.~\cite{spinney_ford_2012_entropy_production_full_phase_space,crosato_spinney_2019_irreversibility_active_matter}, the medium entropy production can be linked to the microscopic Langevin dynamics of a system by identifying reversible and irreversible components, see Appendix \ref{ssec:framework_entropy_production_microscopic}.

Which components are reversible or irreversible depends on the choice of the time reversal operator $\bm{\epsilon}$. For the translational degree of freedom, the time reversal operator is unambiguously given by 
\begin{equation}
	\bm{\epsilon}_{\bm{r}_i} \, \bm{r}_ i = \bm{r}_i \quad \text{and} \quad  \bm{\epsilon}_{\dot{\bm{r}}_i} \, \dot{\bm{r}}_ i = -\dot{\bm{r}}_i ,
\end{equation}
since particle positions remain unchanged under time-reversal, while velocities reverse direction.
However, the interpretation of $\bm{\epsilon}$ for the active forcing, which is associated with the self-propulsion in direction $\bm{p}_i = (\cos\theta_i, \sin\theta_i)$, is less clear. 

If the direction of motion is considered to be a result of a physical asymmetry of the particles themselves, it has been argued that self-propulsion should be even under time reversal (with $\epsilon_{\theta_i} \, \theta_i = \theta_i$) \cite{shankar_marchetti_2018_hidden_entropy_production}. Contrary, it has also been argued that self-propulsion should be odd when the particles themselves are active, and even when particles move due to external forces, such as passive particles in an active bath \cite{dabelow_eichhorn_2019_irreversibility_active_matter}. Importantly, however, in the absence of steric repulsion ($\bm{F}_{\rm rep}=\bm{0}$) and without translational noise ($D'_{\rm t}=0$), the self-propulsion direction $\bm{p}_i$ \textit{must} be odd under time reversal. In this case, $v_0\,\bm{p}_i$ represents a physical velocity, which reverses under time reversal \cite{ferretti_walczak_2022_signatures_irreversibility_flocking,borthne_cates_2020_time-reversal_symmetry_violations_entropy_production_field_theories_polar_active_matter}. If $\bm{p}_i$ were incorrectly treated as even in this limit, the backward path would have zero probability, resulting in a diverging entropy production. 

To avoid these issues in the deterministic limit, we adopt the odd interpretation of active forcing with \footnote{Note that the sum of the translational contributions to the entropy production for odd and even interpretations is a constant \cite{pietzonka_seifert_2017_entropy_production_active_particles,dadhichi_ramaswamy_2018_origins_nonequilibrium_character_active_systems}. The alignment contribution is independent of the choice of time-reversal convention for self-propulsion, such that peaks in the alignment contribution to the entropy production rate with the odd interpretation also occur for the even interpretation. Nevertheless, the dependence on the choice of self-propulsion time reversal symmetry -- in combination with the fact that we calculate an informatic entropy production rate -- makes a direct interpretation of the absolute value of the entropy production rate challenging.}
\begin{equation}
	\epsilon_{\theta_i} \, \theta_i = \theta_i + \pi, \quad \epsilon_{\dot{\theta}_i} \, \dot{\theta}_i = - \dot{\theta}_i, \quad \bm{\epsilon}_{\bm{p}_i} \, \bm{p}_i = -\bm{p}_i .
\end{equation}

Applying this entropy production framework to the specific case of a non-reciprocal polar active mixture governed by the overdamped Langevin equations \eqref{eq:Langevin_r} and \eqref{eq:Langevin_theta}, as shown in Appendix \ref{ssec:framework_entropy_production_microscopic},
we see that the total informatic entropy production rate of particle $\alpha$ naturally splits into two parts,
\begin{equation}
	\langle \Delta \dot{S}_{\alpha} \rangle = \langle \Delta \dot{S}_{\alpha}^{{\rm trans}} \rangle + \langle \Delta \dot{S}_{\alpha}^{{\rm align}} \rangle ,
\end{equation}
where $\langle \cdot \rangle$ denotes a noise and time average. The contribution from translational motion is given by
\begin{equation}
	\label{eq:entropy_production_rate_contribution_translational}
	\begin{split}
		\langle \Delta \dot{S}_{\alpha}^{{\rm trans}} \rangle =\, & \frac{{k_{\rm B}}}{D'_{\rm t}}  \Big \langle \Big(\mu_{r} \sum_{\beta \neq \alpha} \bm{F}^{\alpha}_{\rm rep}(\bm{r}_{\alpha},\bm{r}_{\beta}) \Big) \circ \dot{\bm{r}}_{\alpha} -	(v_0\, \bm{p}_{\alpha}) \cdot \Big( \sum_{\beta \neq \alpha} \bm{F}^{\alpha}_{\rm rep}(\bm{r}_{\alpha},\bm{r}_{\beta}) \Big) \Big\rangle,
	\end{split}
\end{equation}
where the $\circ$ symbol denotes a product interpreted in the Stratonovich sense \footnote{To evaluate a Stratonovich product of form $f(x(t),t) \circ {\rm d}x(t)$ with stochastic variable $x(t)$, one computes $f(x_{\rm mp}, t_{\rm mp}) \cdot (x(t+{\rm d}t) - x(t))$, where $f(x)$ is evaluated at the mid points $x_{\rm mp} = (x(t)+x(t+{\rm d}t))/2$, $t_{\rm mp} = (t+t+{\rm d}t)/2$ and ${\rm d}t \rightarrow 0$ \cite{risken_1996_fokker-planck_equation}.}.
The contribution from the alignment couplings is
\begin{equation}
	\label{eq:entropy_production_rate_contribution_alignment}
	\langle \Delta \dot{S}_{\alpha}^{{\rm align}} \rangle = \frac{k_{\rm B}}{D'_{\rm r}}  \Big \langle \Big(\mu_{\theta} \sum_{\beta\neq\alpha} \mathcal{T}_{\rm al}^{\alpha}(\bm{r}_{\alpha},\bm{r}_{\beta},\theta_{\alpha},\theta_{\beta}) \Big) \circ \dot{\theta}_{\alpha} \Big\rangle.
\end{equation}
Finally, the total informatic entropy production of all particles of species $a$ is given by the sum over all individual particle contributions,
\begin{equation}
\label{eq:total_informatic_entropy_production}
	\langle \Delta \dot{S}_{a} \rangle = \sum_{\alpha} \langle \Delta \dot{S}_{\alpha} \rangle .
\end{equation}

It is important to note that Langevin equations provide a coarse-grained or effective description of a system and, as such, do not capture all dissipative degrees of freedom. In these cases, the informatic entropy production in Eq.~\eqref{eq:total_informatic_entropy_production} represents only an apparent entropy production. This is evident, for example, in overdamped models, where some entropy production contributions are hidden and only become apparent in underdamped systems \cite{shankar_marchetti_2018_hidden_entropy_production}. Moreover, Langevin equations for active particles, such as Eqs.~\eqref{eq:Langevin_r} and \eqref{eq:Langevin_theta}, are in general not thermodynamically consistent. A typical case arises when the self-propulsion speed and direction do not originate from mechanical forces but instead result from behavioral rules \cite{cates_tjhung_2022_stochastic_hydrodynamics_complex_fluids}. In such situations, the absence of a connection to energetics, and therefore the violation of the thermodynamic first law of energy conservation, does not allow for the association of the path probability ratio~\eqref{eq:medium_entropy_production_equation} with energy dissipation or heat production. Recent works have developed thermodynamically consistent formulations based on microscopic Markov jump processes for non-reciprocally interacting particles \cite{mohite_rieger_2025_stochastic_thermodynamics_non-reciprocity} and on lattice models with flocking dynamics \cite{proesmans_fodor_2025_quantifying_dissipation_flocking_dynamics}. 
Nevertheless, the informatic entropy production rate computed from overdamped microscopic models provides a lower bound on the total entropy production and serves as a meaningful measure of irreversibility in the system.

\subsection{Numerical results for the entropy production rate on microscopic level}
\label{ssec:numerical_results_entropy_production_microscopic}
To actually calculate the microscopic informatic entropy production rate, we use results of BD simulations. Our main interest lies in the entropy production in chiral states and at EPs, which indicate the transitions to chiral states in the ``strong-interspecies-coupling'' regime. To this end, we fix the intra-species alignment strength at $g=9$, which satisfies the condition $g>4\,D_{\rm r}/(R_{\theta}^2\,\rho_0)$. This ensures the emergence of non-zero polarization regardless of the values of $g_{AB}$ and $g_{BA}$. The detailed dynamical behavior in this strong-coupling regime is discussed in Ref.~\cite{kreienkamp_klapp_2024_synchronization_exceptional_points}.

Here, we characterize the collective particle behavior using two observables, which turn out to be closely related to the entropy production rate: the susceptibility $\chi^{\rm mov}(\bm{P})$, which quantifies temporal fluctuations of the global polarization vector $\bm{P}$ in a co-rotating frame, and the spontaneous chirality $\vert \Omega_{\rm s}^i (t) \vert$, defined as an average angular frequency that measures the system's collective rotational motion. The explicit definitions of these microscopic observables, which we calculate for both species separately as well as combined, are provided in Appendix~\ref{app:microscopic_obervables}.

\begin{figure}
	\includegraphics[width=1\linewidth]{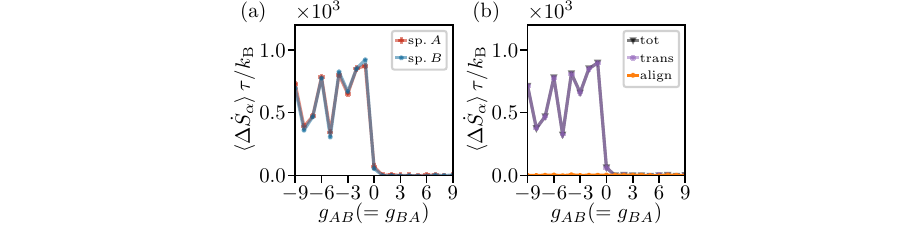}
	\caption{\label{fig:BD_pol_sus_entropy_results_reciprocal}Particle-averaged entropy production rates from particle simulations of a reciprocal system with $g_{AB} = g_{BA}=\kappa$. In (a), the entropy production rate is shown as a function of $g_{AB}$ separately for each particle species, while in (b) the contributions from translational motion and alignment couplings are plotted separately. Other parameters are as specified in Sec.~\ref{sec:model}.}
\end{figure}

\subsubsection{Reciprocal reference case}
We first consider the reciprocal system with $g_{AB}=g_{BA}=\kappa$ as a reference case. The reciprocal system exhibits flocking for $\kappa>0$ [snapshot in Fig.~\ref{fig:stability_diagram_snapshots}(b)] and antiflocking for $\kappa<0$ [snapshot in Fig.~\ref{fig:stability_diagram_snapshots}(c)]. These states share key characteristics with those in the weak-intraspecies-coupling regime with $g=3<4\,D_{\rm r}/(R_{\theta}^2\,\rho_0)$, studied in detail in Ref.~\cite{kreienkamp_klapp_2024_dynamical_structures_phase_separating_nonreciprocal_polar_active_mixtures}.

The particle-averaged informatic entropy production rate $\langle \Delta \dot{S}_{\alpha} \rangle$ is shown as a function of $\kappa$ for each species in Fig.~\ref{fig:BD_pol_sus_entropy_results_reciprocal}(a). As expected in a reciprocal system, $\langle \Delta \dot{S}_{\alpha}\rangle$ is nearly identical for both species $A$ and $B$. However, the magnitude of entropy production rate differs significantly between the flocking and antiflocking regimes. 

In the antiflocking regime, the informatic entropy production rate is clearly larger than zero,  $\langle \Delta \dot{S}_{\alpha} \rangle > 300\, k_{\rm B}\,\tau^{-1}$. In contrast, in the flocking regime, it is very small, $\langle \Delta \dot{S}_{\alpha} \rangle < 10 \, k_{\rm B}\,\tau^{-1}$. This (nearly) vanishing entropy production rate \textit{within} the flocking state is consistent with previous results for underdamped repulsive polar particles \cite{crosato_spinney_2019_irreversibility_active_matter}, overdamped non-repulsive polar particles \cite{ferretti_walczak_2022_signatures_irreversibility_flocking}, and active spins in the active Ising model \cite{yu_tu_2022_energy_cost_flocking_active_spins} \footnote{Note that, in one-species systems, the entropy production rate associated with the alignment couplings exhibits a peak at the transition from disordered to flocking motion \cite{ferretti_walczak_2022_signatures_irreversibility_flocking,yu_tu_2022_energy_cost_flocking_active_spins}. Here, we do not study the conventional ordered-disorder transition. Instead, we consider a binary mixture transitioning from antiflocking to flocking as $\kappa$ is varied.}. 

The decomposition of the entropy production rate into translational alignment contributions is plotted in Fig.~\ref{fig:BD_pol_sus_entropy_results_reciprocal}(b). In the antiflocking regime, the entropy production rate is dominated by the translational contribution, while the alignment contribution remains negligible in both regimes.

The small alignment contribution [Eq.~\eqref{eq:entropy_production_rate_contribution_alignment}] observed in both the flocking and antiflocking regimes -- i.e., in states with coherent motion in a constant direction -- can be understood from the fact that particle orientations change only weakly in time. In this case, $\dot{\theta}_{\alpha} \approx 0$, which directly leads to $\langle \Delta \dot{S}_{\alpha}^{{\rm align}} \rangle \approx 0$.
By contrast, the translational contribution $\langle \Delta \dot{S}_{\alpha}^{{\rm trans}} \rangle$ [Eq.~\eqref{eq:entropy_production_rate_contribution_translational}] depends on the steric repulsion force $\bm{F}_{\rm rep}^{\alpha}$ between particles, and is non-zero only when $\bm{F}_{\rm rep}^{\alpha} \neq \bm{0}$. This explains why the entropy production in the antiflocking state is significantly larger than in the flocking state. As seen in the snapshots in Fig.~\ref{fig:stability_diagram_snapshots}(b) and (c), in the flocking state, particles move nearly unhindered in the same direction, without much repulsive interactions, resulting in $\langle \Delta \dot{S}_{\alpha}^{{\rm trans}} \rangle \approx 0$. In contrast, in the antiflocking state, oppositely moving flocks collide. The resulting flocks slide in direction parallel to species interface with frequent collisions between different particles at the interface, leading to $\langle \Delta \dot{S}_{\alpha}^{{\rm trans}} \rangle > 0$. 

Hence, the low entropy production rate in the flocking state reflects the fact that particles have reached a configuration in which interactions do not significantly alter their motion, i.e., orientations remain constant and steric interactions are rare. Conversely, the substantially larger entropy production rate in the antiflocking state arises from persistent interactions between particles, specifically in the translational degrees of freedom.\\

\begin{figure}
	\includegraphics[width=1\linewidth]{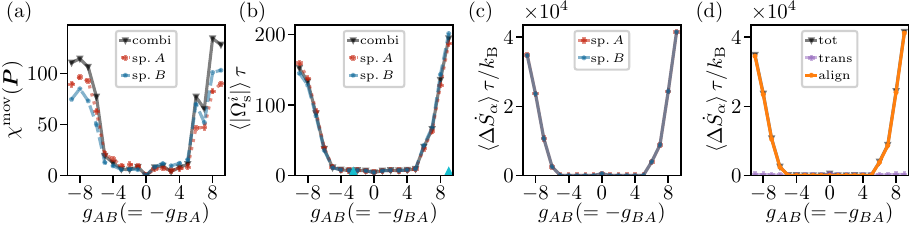}
	\caption{\label{fig:BD_pol_sus_entropy_results_anti-symmetric}Particle simulation results for a non-reciprocal, anti-symmetric system with $g_{AB} = -g_{BA} = \delta$. The (a) susceptibility, (b) spontaneous chirality, and (c),(d) particle-averaged entropy production rates are shown as functions of $g_{AB}$. The parameter values corresponding to the snapshots in Fig.~\ref{fig:stability_diagram_snapshots}(d) and (e) are indicated by cyan markers in (b). In (a), the susceptibility of the global polarization vector is calculated in a co-rotating frame, which is determined by time averaging over $0.03\,\tau$. In (c), the entropy production rate is shown separately for both particle species, while in (d) the contributions from translational motion and alignment couplings are plotted separately. Other parameters are as specified in Sec.~\ref{sec:model}. The spontaneous chirality is consistent with Ref.~\cite{kreienkamp_klapp_2024_synchronization_exceptional_points}.}
\end{figure}

\subsubsection{Non-reciprocal anti-symmetric system}    
We now turn to a non-reciprocal system with anti-symmetric couplings, $g_{AB} = -g_{BA} =\delta$. Here, particles of different species have opposing alignment goals: for $\delta>0$, $A$ wants to align with $B$-particles, while $B$-particles want to antialign with $A$-particles (and the other way round for $\delta <0$). A small (large) value of $\vert \delta \vert$ corresponds to a small (strong) degree of non-reciprocity. From a field-theoretical perspective, upon varying $\delta$ one does not cross any exceptional points. Instead, the system remains in the so-called chiral state. Microscopically, the anti-symmetric system exhibits two qualitatively distinct dynamical regimes [see also Ref.~\cite{kreienkamp_klapp_2024_synchronization_exceptional_points}].

For weak anti-symmetric couplings with $\vert \delta \vert \lesssim 5$, an almost fully demixed configuration emerges, in which each species forms a large, rotating, synchronized cluster [see snapshot in Fig.~\ref{fig:stability_diagram_snapshots}(d)]. Within each cluster, particles move collectively in nearly constant direction until encountering the cluster of the other species, triggering reorientation and resulting in global rotational motion. The clusters rotate in a phase-locked manner, maintaining a fixed phase shift between the internally synchronized $A$- and $B$-clusters. Within a single noise realization, the handedness of the cluster rotation is chosen spontaneously, and thereafter, the clusters persistently rotate either clockwise or counterclockwise~\cite{kreienkamp_klapp_2024_synchronization_exceptional_points}.
Although the susceptibility $\chi^{\rm mov}(\bm{P}_{\rm combi})$ of the combined polarization vector slightly increases with $\delta$ [Fig.~\ref{fig:BD_pol_sus_entropy_results_anti-symmetric}(a)], the single-species susceptibilities, $\chi^{\rm mov}(\bm{P}_A)$ and $\chi^{\rm mov}(\bm{P}_B)$, as well as spontaneous chirality $\langle \vert \Omega_{\rm s}^i \vert \rangle$ [Fig.~\ref{fig:BD_pol_sus_entropy_results_anti-symmetric}(b)] remain relatively small for $\vert \delta \vert \lesssim 5$. In this regime, we see that the resulting entropy production rate remains very small with $\langle \Delta \dot{S}_{\alpha} \rangle < 50 \, k_{\rm B}\,\tau^{-1}$ for $\delta \neq 0$ [Fig.~\ref{fig:BD_pol_sus_entropy_results_anti-symmetric}(c)] despite the presence of non-reciprocal couplings.

For stronger non-reciprocity ($\vert \delta \vert \gtrsim 5$), partially synchronized ``chimera-like'' states emerge. A representative snapshot of such a state is shown in Fig.~\ref{fig:stability_diagram_snapshots}(e). In this regime, the susceptibilities and the spontaneous chirality grow significantly with $\vert \delta \vert$. For example, at $\vert \delta \vert = 8$, one observes single-species and combined polarization vector susceptibilities $\chi^{\rm mov}(\bm{P}) > 80$ and spontaneous chiralities $\langle \vert \Omega_{\rm s}^i \vert \rangle > 125\,\tau^{-1}$. The informatic entropy production rate follows this trend: it increases substantially with the degree of non-reciprocity. For instance, at $\vert \delta \vert = 8$, one finds $\langle \Delta \dot{S}_{\alpha} \rangle > 2.3\cdot 10^4 \, k_{\rm B}\,\tau^{-1}$.

Despite the coupling asymmetry, the average informatic entropy production rate $\langle \dot{S}_{\alpha} \rangle$ per particles, shown in Fig.~\ref{fig:BD_pol_sus_entropy_results_anti-symmetric}(c), does not differ significantly between species.

The decomposition of the entropy production rate [Fig.~\ref{fig:BD_pol_sus_entropy_results_anti-symmetric}(d)] reveals that, in both regimes of the anti-symmetric system, the informatic entropy production rate is almost entirely due to the orientational couplings, with negligible contribution from translational motion. This observation can be explained by the nature of the collective particle motion, seen in the snapshots in Figs.~\ref{fig:stability_diagram_snapshots}(d),(e). 

In the regime of large, rotating clusters [Fig.~\ref{fig:stability_diagram_snapshots}(d)], particles of each species form flocks that move coherently until colliding with the other species' cluster. Similar to reciprocal flocking, the translational contribution remains small due to the lack of significant repulsion, i.e., $\bm{F}_{\rm rep}^{\alpha} \approx 0$. Although the clusters do rotate, spontaneous chirality, and thus $\dot{\theta}$, remains small, keeping the alignment contribution low, too.

In contrast, in the chimera-like state at $\vert \delta \vert \gtrsim 5$ [Fig.~\ref{fig:stability_diagram_snapshots}(e)], particles remain relatively loosely distributed. Thus, the translational contribution to the entropy production rate remains small, though slightly larger than at smaller $\vert \delta \vert$. However, the alignment contribution increases significantly due to rapid orientation changes ($\dot{\theta}^{\alpha}$) and stronger torques ($\mathcal{T}_{\rm al}^{\alpha}$) sustaining the rotational motion.

Thus, in the anti-symmetric regime, where non-reciprocity induces particle rotations and causes the overall polarization vector to continuously change direction, the alignment contribution to the entropy production rate becomes non-zero. Due to its intrinsic connection to changes in particle orientation [see Eq.~\ref{eq:entropy_production_rate_contribution_alignment}], the entropy production rate mirrors the trends of both the susceptibility and spontaneous chirality -- two complementary quantities that track changes in (coherent) particle orientations. However, the effect of non-reciprocity on the entropy production rate (compared to the reciprocal reference case) becomes pronounced only when the degree of non-reciprocity is sufficiently large, such that the strength of interspecies couplings is comparable to that of intraspecies alignment.

\begin{figure}
	\includegraphics[width=1\linewidth]{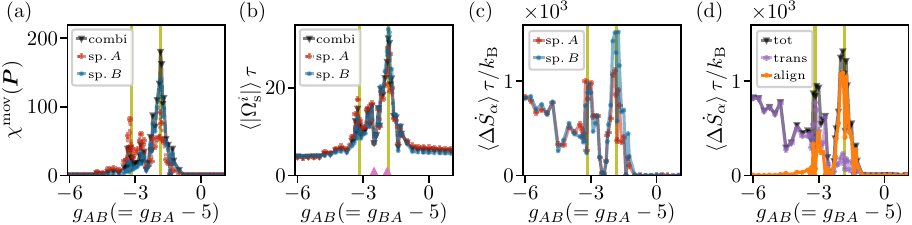}
	\caption{\label{fig:BD_pol_sus_entropy_results_fixed_dif-5_species}Particle simulation results for a non-reciprocal system with $g_{AB} = g_{BA}-d$ with $d=5$. The (a) susceptibility of the polarization vector, (b) spontaneous chirality, and (c),(d) particle-averaged entropy production rates are shown as functions of $g_{AB}$. The green vertical lines indicate the positions of critical exceptional points, respectively ($g_{AB}^{{\rm EP}>}(d=5)\approx -1.83$ and $g_{AB}^{{\rm EP}<}(d=5)\approx -3.17$). The parameter values corresponding to the snapshots in Fig.~\ref{fig:stability_diagram_snapshots}(d) and (e) are indicated by pink markers in (b). In (a), the susceptibility of the global polarization vector is calculated in a co-rotating frame, which is determined by time averaging over $0.03\,\tau$. In (c), the entropy production rate is shown separately for both particle species, while in (d) the contributions from translational motion and alignment couplings are plotted separately. Other parameters are as specified in Sec.~\ref{sec:model}. The spontaneous chirality is consistent with Ref.~\cite{kreienkamp_klapp_2024_synchronization_exceptional_points}.}
\end{figure}

\subsubsection{Non-reciprocal system crossing exceptional points}
We now consider non-reciprocal systems with couplings $g_{AB} = g_{BA}-d$, where $d \neq 0$ is fixed. Varying $g_{AB}$ in such systems leads to two crossings of exceptional points in the corresponding mean-field continuum description, which indicate the transition to the chiral state. We denote these points by $g_{AB}^{{\rm EP}>}(d)$ and $g_{AB}^{{\rm EP}<}(d)$, whereby $g_{AB}^{{\rm EP}<}(d) < g_{AB}^{{\rm EP}>}(d)$. Their locations depend on the offset parameter $d$.

In Fig.~\ref{fig:BD_pol_sus_entropy_results_fixed_dif-5_species} the susceptibility, spontaneous chirality, and entropy production rate are plotted as functions of $g_{AB}$ for $d=5$. The qualitative behavior remains similar for other values of $d$ (see Appendix \ref{sapp:different_crossings_of_exceptional_points}).

For $g_{AB} < -5$, the system is in an antiflocking state, similar to the reciprocal case shown in snapshot Fig.~\ref{fig:stability_diagram_snapshots}(c). For $g_{AB} > 0$, the system is in a flocking state, resembling the reciprocal case shown in snapshot Fig.~\ref{fig:stability_diagram_snapshots}(b). In both these limiting states, the polarization vector susceptibility and spontaneous chirality remain low. As expected, the resulting entropy production rate closely matches that of the corresponding reciprocal systems (see Fig.~\ref{fig:BD_pol_sus_entropy_results_reciprocal}).

Between these two limiting states, the system crosses the exceptional points, which mark qualitative changes in the dynamical behavior, twice. This becomes evident in the snapshots in Fig.~\ref{fig:stability_diagram_snapshots}(d),(f): while large, rotating clusters exist for anti-symmetric couplings [$g_{AB} = -g_{BA} = -2.5$ in Fig.~\ref{fig:stability_diagram_snapshots}(d)], these structures break down near the exceptional points [$g_{AB} = -g_{BA} = -1.9$ in Fig.~\ref{fig:stability_diagram_snapshots}(f)], where synchronization weakens, fluctuations of polarizations increase, and clusters dynamically form and dissolve. The dynamical behavior near exceptional points is further discussed in Ref.~\cite{kreienkamp_klapp_2024_synchronization_exceptional_points}.

The transition is reflected in the all considered particle-level observables. Near the exceptional points, both the susceptibility of the polarization vector and the spontaneous chirality display clear peaks -- more pronounced at $g_{AB}^{\rm EP>}$, closer to the flocking regime, than at $g_{AB}^{\rm EP<}$. For instance, at $g_{AB}^{\rm EP>} = -1.9$, we find $\chi^{\rm mov}(\bm{P}_{\rm combi}),\chi^{\rm mov}(\bm{P}_{B}) > 120$ and $\langle \vert \Omega_{\rm s}^i \vert \rangle > 23 \, \tau^{-1}$. These values significantly exceed the values for the anti-symmetric system with large, rotating clusters (where $\chi^{\rm mov}(\bm{P}) < 12$ and $\langle \vert \Omega_{\rm s}^i \vert \rangle < 8 \, \tau^{-1}$ at $g_{AB} = -2.5$). Notably, the susceptibility is dominated by transverse fluctuations (see Appendix~\ref{sapp:microscopic_observables_susceptibility}), consistent with our picture of the excitation of Goldstone modes close to the critical exceptional points.

Importantly, in the vicinity of the exceptional points, also the total entropy production rate exhibits pronounced peaks. These peaks are most clearly visible in the orientation-related contribution [Fig.~\ref{fig:BD_pol_sus_entropy_results_fixed_dif-5_species}(d)], which shows sharp maxima close to the positions of exceptional points identified in the continuum theory. Between these exceptional points, i.e., for intermediate anti-symmetric couplings, the entropy production rate decreases substantially.

Since the polarization vector susceptibility, spontaneous chirality, and the alignment contribution to the entropy production rate are all closely linked to changes in particle orientations, the positions of peaks and minima in these quantities coincide. Moreover, the peak heights show qualitative agreement: both the entropy production rate and susceptibility attain higher maxima near the exceptional point $g_{AB}^{\rm EP>}$, which is closer to the flocking regime.

The appearance of peaks in all three quantities near the critical EPs can be understood from a dynamical systems perspective. As discussed in Sec.~\ref{sssec:model_exceptional_points}, the critical EPs correspond to points where the Goldstone mode of broken continuous rotational symmetry is actuated, facilitating the rotation of particle orientations. However, the good agreement between the positions of the peaks -- observed in three different microscopic observables derived from particle-level simulations -- and the critical EPs predicted by the continuum theory is both surprising and not expected a priori.

\subsection{Entropy production rate in field theory}
\label{ssec:entropy_production_rate_field_theory_perturbations_k-0}
To substantiate the peaks in the entropy production rate at coupling strengths associated with exceptional points, we now turn to the continuum description of the system.

\subsubsection{Fluctuating hydrodynamic model for polarization perturbations}
At the continuum level, we calculate the entropy production rate based on path \textit{probabilities}, which requires considering \textit{stochastic} fields instead of deterministic ones (that were considered in Sec.~\ref{sssec:model_continuum}). The fluctuating hydrodynamic equations include both the deterministic contributions, obtained through coarse-graining, and additional noise terms that are added ``by hand'' \cite{marchetti_simha_2013_hydrodynamics_soft_active_matter,nardini_cates_2017_entropy_production_field_theory,toner_tu_2005_hydrodynamics_phases_of_flocks,borthne_cates_2020_time-reversal_symmetry_violations_entropy_production_field_theories_polar_active_matter,dadhichi_ramaswamy_2018_origins_nonequilibrium_character_active_systems}. The full fluctuating hydrodynamic equations are given in Appendix \ref{app:continuum_model_fluctuating}.

For the analytical calculation below, we focus on the time evolution of small perturbations $\bm{w}'^a$ around stationary solutions $\bm{w}^a_0$ of the deterministic continuum Eqs.~\eqref{eq:continuum_eq_density} and \eqref{eq:continuum_eq_polarization_density}. By focusing on the entropy production rate arising from polarization perturbations (instead of the entropy production rate of the total polarization) we are in the position to obtain analytical results. Further, we specialize on the infinite-wavelength limit with wavenumber $k=0$, where exceptional points are predicted to occur. Due to number conservation, density fluctuations vanish at $k=0$. Hence, we restrict our analysis to the polarization perturbation dynamics alone. To linear order, the mean-field polarization perturbation dynamics for species $a$ at $k=0$ is given by
\begin{equation}
	\label{eq:fluctuating_hydrodynamic_polarization_k-0_lin}
	\partial_t \bm{w}'^a = \bm{\mathcal{F}}^{a}_{k=0,{\rm lin}}[\bm{w}'^a,\bm{w}'^b] + \tilde{\bm{\eta}}_a ,
\end{equation}
where $b \neq a $, and both $\bm{w}'^a = \bm{w}'^a(t)$, $\tilde{\bm{\eta}}_a = \tilde{\bm{\eta}}_a(t)$ are homogeneous in space.
For stationary solutions of the form $\bm{w}^a_0 = (w_0^a, 0)^{\rm T}$, the deterministic contribution is described by the functional
\begin{equation}
\label{eq:continuum_functional_k-0_linear}
	\begin{split}
		\bm{\mathcal{F}}^{a}_{k=0,{\rm lin}}[\bm{w}'^a,\bm{w}'^b] = \bm{\gamma}^a_a[w_0^a, w_0^b] \cdot \bm{w}'^a +  \bm{\gamma}^a_b[w_0^a, w_0^b] \cdot  \bm{w}'^b .
	\end{split}
\end{equation}
The prefactors $\bm{\gamma}_a^a$ and $\bm{\gamma}_b^a$ can be calculated analytically and are plotted as functions of $g_{AB}$ in Fig.~\ref{fig:prefactors_continuum_functional} in Appendix \ref{app:prefactors_function_k-0}.
The alignment strength at the continuum level is given by $g'_{ab} = g_{ab}\,R_{\theta}^2\,\rho_0^b/2$. The noise term $\tilde{\bm{\eta}}_a$ in Eq.~\eqref{eq:fluctuating_hydrodynamic_polarization_k-0_lin} is of strength $D_{\rm h}$, has zero mean, and satisfies
\begin{equation}
\label{eq:noise_correlation_field_lin}
	\langle \tilde{\eta}_{a,i}(t) \, \tilde{\eta}_{b,j}(t') \rangle = 2\,D_{\rm h}\,\delta_{ab}\,\delta_{ij}\,\delta(t-t') .
\end{equation}

\subsubsection{Dynamical action}
\label{sssec:entropy_production_dynamical_action}
We begin by considering the total entropy production from polarization perturbations, defined as the log-ratio of forward ($\mathfrak{p}[\vec{\bm{w}}']$) and backward ($\mathfrak{p}[\vec{\bm{w}}'^{\dagger}]$) path probabilities \cite{seifert_2005_entropy_production_along_stochastic_trajectory,borthne_cates_2020_time-reversal_symmetry_violations_entropy_production_field_theories_polar_active_matter,dadhichi_ramaswamy_2018_origins_nonequilibrium_character_active_systems},
\begin{equation}
	\label{eq:entropy_production_path_probability}
	\Delta S[\vec{\bm{w}}'] = k_{\rm B}\,{\rm ln}\left( \frac{\mathfrak{p}[\vec{\bm{w}}']}{\mathfrak{p}[\vec{\bm{w}}'^\dagger]} \right),
\end{equation}
for the trajectory $\vec{\bm{w}}' = \{(\bm{w}'^A(t),\bm{w}'^B(t)) \vert t \in[t_0,\tau]\}$.

The expression for the probability of individual trajectories in $(\bm{w}'^A,\bm{w}'^B)$-space is provided by the standard Onsager-Machlup path integral \cite{onsager_machlup_1953_fluctuations_irreversible_processes,nardini_cates_2017_entropy_production_field_theory}
\begin{equation}
	\label{eq:path_probability_action_lin}
	\mathfrak{p}[\vec{\bm{w}}'] \propto {\rm exp}(-\mathcal{A}[\vec{\bm{w}}'])
\end{equation}
with dynamical action $\mathcal{A}[\vec{\bm{w}}']$.

For our binary mixture at $k=0$, the total dynamical action includes contributions from $\bm{w}'^A$ and $\bm{w}'^B$, such that
\begin{equation}
	\mathcal{A}[\vec{\bm{w}}'] = \mathcal{A}_{\bm{w}'^A} + \mathcal{A}_{\bm{w}'^B}. 
\end{equation}
In the $k=0$-limit, the contribution from species $a=A,B$ takes the form
\begin{equation}
	\label{eq:action_k-0_forward_path_lin}
	\begin{split}
		\mathcal{A}_{\bm{w}'^a} = & \frac{1}{4\,D_{\rm h}} \iint \big(\partial_t \bm{w}'^a - \bm{\mathcal{F}}^a_{k=0, {\rm lin}}[\bm{w}'^a,\bm{w}'^b] \big)^2 \, {\rm d}\bm{r}\,{\rm d}t + \mathcal{A}_{\bm{w}'^a}^{\rm conv},
	\end{split}
\end{equation}
where 
\begin{equation}
	\mathcal{A}_{\bm{w}'^a}^{\rm conv} = \frac{1}{2} \, \iint \frac{\delta \bm{\mathcal{F}}^a_{k=0, {\rm lin}}[\bm{w}'^a,\bm{w}'^b]}{\delta \bm{w}'^a} \, {\rm d}\bm{r}\,{\rm d}t
\end{equation}
accounts for the convention-dependent contributions to the dynamics due to the discretization scheme with respect to noise. Here, we use Stratonovich (midpoint) discretization.

\subsubsection{Backwards probability}
\label{sssec:entropy_production_backwards_probability}
In addition to the forward path probability \eqref{eq:path_probability_action_lin}, we also require the probability for the time-reversed backward path, given by
\begin{equation}
	\label{eq:path_probability_backwards_action_lin}
	\mathfrak{p}[\vec{\bm{w}}'^{\dagger}] \propto {\rm exp}(-\mathcal{A}[\vec{\bm{w}}'^{\dagger}]) = {\rm exp}(-\overleftarrow{\mathcal{A}}_{\bm{w}'^A} + \overleftarrow{\mathcal{A}}_{\bm{w}'^B}),
\end{equation}
which measures the probability of observing the trajectory $\vec{\bm{w}}'^{\dagger}$ under the same forward time evolution equation \eqref{eq:fluctuating_hydrodynamic_polarization_k-0_lin}.
For Stratonovich convention, the normalization factors are identical for the forward and backward path probabilities in Eqs.~\eqref{eq:path_probability_action_lin} and \eqref{eq:path_probability_backwards_action_lin} \cite{cates_tjhung_2022_stochastic_hydrodynamics_complex_fluids,cugliandolo_lecomte_2017_rules_calculus_path_intergral_representation_langevin_equations}.

To define the backward path, we must specify a time-reversal protocol. Consistent with our particle-level considerations, we impose a polarity flip under time reversal, such that
\begin{equation}
	\bm{w}'^{a\dagger}(\bm{r},t) = - \bm{w}'^{a}(\bm{r}, \tau + t_0 - t) .
\end{equation}
This polarity flip is indeed necessary when the density dynamics are purely deterministic \cite{borthne_cates_2020_time-reversal_symmetry_violations_entropy_production_field_theories_polar_active_matter}.

The $a$-species action of the time-reversed trajectory is
\begin{equation}
	\label{eq:action_k-0_backward_path_lin}
	\begin{split}
		\overleftarrow{\mathcal{A}}_{\bm{w}'^a} = &\frac{1}{4\,D_{\rm h}} \iint \big(\partial_t \bm{w}'^{a\dagger} - \bm{\mathcal{F}}^{a}_{k=0, {\rm lin}}[\bm{w}'^{a\dagger},\bm{w}'^{b\dagger}] \big)^2 \, {\rm d}\bm{r}\,{\rm d}t + \overleftarrow{\mathcal{A}}_{\bm{w}'^a}^{\rm conv}
	\end{split}
\end{equation}
with
\begin{equation}
	\overleftarrow{\mathcal{A}}_{\bm{w}'^a}^{\rm conv} = \frac{1}{2} \, \iint \frac{\delta \bm{\mathcal{F}}^a_{k=0, {\rm lin}}[\bm{w}'^{a\dagger},\bm{w}'^{b\dagger}]}{\delta \bm{w}'^{a\dagger}} \, {\rm d}\bm{r}\,{\rm d}t .
\end{equation}
The functional $\bm{\mathcal{F}}^{a}_{k=0}$ is anti-symmetric under time reversal, i.e., 
\begin{equation}
	\bm{\mathcal{F}}^{a}_{k=0, {\rm lin}}[\bm{w}'^{a\dagger},\bm{w}'^{b\dagger}] = -\bm{\mathcal{F}}^{a}_{k=0, {\rm lin}}[\bm{w}'^{a},\bm{w}'^{b}] .
\end{equation} 
Hence, the action of the time-reversed trajectory, expressed in terms of the forward trajectory, is 
\begin{equation}
	\begin{split}
		\overleftarrow{\mathcal{A}}_{\bm{w}'^a} = &\frac{1}{4\,D_{\rm h}} \iint \big(\partial_t \bm{w}'^{a} + \bm{\mathcal{F}}^{a}_{k=0, {\rm lin}}[\bm{w}'^{a},\bm{w}'^{b}] \big)^2 \, {\rm d}\bm{r}\,{\rm d}t + \mathcal{A}_{\bm{w}'^a}^{\rm conv} .
	\end{split}
\end{equation}

\subsubsection{Entropy production} 
From Eq.~\eqref{eq:entropy_production_path_probability}, it follows that the entropy production for species $a$ along the trajectory $\vec{\bm{w}}'$ is given by
\begin{equation}
	\label{eq:entropy_production_actions_lin}
	\Delta S^{a} = k_{\rm B} \, \big(\overleftarrow{\mathcal{A}}_{\bm{w}'^a} - \mathcal{A}_{\bm{w}'^a} \big) .
\end{equation}
Inserting the expressions \eqref{eq:action_k-0_forward_path_lin} and \eqref{eq:action_k-0_backward_path_lin} for the actions, we obtain \footnote{Note that $\mathcal{A}_{\bm{w}^a}^{\rm conv}$ and $\overleftarrow{\mathcal{A}}_{\bm{w}^a}^{\rm conv}$ cancel when the functional $\mathcal{F}^a_{k=0}$ is fully anti-symmetric in time (as in this case). This is not generally the case.}
\begin{equation}
	\label{eq:entropy_production_continuum_k-0_lin}
	\Delta S^{a} = - \frac{k_{\rm B}}{D_{\rm h}} \iint \big(- (\partial_t \bm{w}'^a) \circ \bm{\mathcal{F}}^{a}_{k=0, {\rm lin}}[\bm{w}'^{a},\bm{w}'^{b}] \big)\,{\rm d}\bm{r}\,{\rm d}t,
\end{equation}
where $\circ$ indicates a product interpreted in the Stratonovich sense.

We now compute the entropy production rate $\langle \Delta \dot{S} \rangle = \langle \Delta \dot{S}^A \rangle + \langle \Delta \dot{S}^B \rangle$ as defined in Eq.~\eqref{eq:entropy_production_rate_definition}. To this end, we first split $\bm{\mathcal{F}}^a_{k=0, {\rm lin}}$ into a component depending on $\bm{w}'^a$,
\begin{equation}
	\begin{split}
		&\bm{\mathcal{F}}^{a,\bm{w}'^a}_{k=0, {\rm lin}}[\bm{w}'^{a}] = \bm{\gamma}^a_a[w_0^a, w_0^b] \cdot \bm{w}'^a,
	\end{split}
\end{equation}
and one depending on $\bm{w}'^b$ with $b\neq a$,
\begin{equation}
	\begin{split}
		& \bm{\mathcal{F}}^{a,\bm{w}'^b}_{k=0, {\rm lin}}[\bm{w}'^{b}] = \bm{\gamma}^a_b[w_0^a, w_0^b] \cdot \bm{w}'^b.
	\end{split}
\end{equation}
Accordingly, the entropy production $\Delta S^a$ splits into two parts: $\Delta S^a = \Delta S^a_1 + \Delta S^a_2$. Since the integral in Eq.~\eqref{eq:entropy_production_continuum_k-0_lin} is interpreted in the Stratonovich sense, the (common) chain rule applies. The first part, involving $\bm{\mathcal{F}}^{a,\bm{w}'^a}_{k=0, {\rm lin}}$, becomes
\begin{equation}
	\label{eq:entropy_production_first_part_lin}
	\begin{split}
		\Delta S^{a}_1 & = \frac{k_{\rm B}}{D_{\rm h}} \iint (\partial_t \bm{w}'^a) \circ \bm{\mathcal{F}}^{a, \bm{w}'_a}_{k=0, {\rm lin}}[\bm{w}'^{a}] \,{\rm d}\bm{r}\,{\rm d}t\\
		& = \frac{k_{\rm B}}{D_{\rm h}} \iint \big( \bm{\mathcal{F}}^{a,\bm{w}'^a}_{k=0, {\rm lin}}[\bm{w}'^{a}] \big)\,{\rm d}\bm{r}\,{\rm d}\bm{w}'^a \\
		& = \frac{k_{\rm B}}{D_{\rm h}} \int \left( \frac{\bm{\gamma}^a_a}{2} \cdot \bm{w}'^a\right) \cdot \bm{w}'^a \, {\rm d}\bm{r},
	\end{split}
\end{equation}
where we used direct integration in the last step.
For the entropy production \textit{rate} [Eq.~\eqref{eq:entropy_production_rate_definition}], the term $\Delta S_1/\tau \rightarrow 0$ as $\tau \rightarrow \infty$, because all moments of $\bm{w}'^a$ remain finite in the steady state.

The second part, involving $\bm{\mathcal{F}}^{a,\bm{w}'^b}_{k=0, {\rm lin}}$, is split into two contributions by inserting Eq.~\eqref{eq:fluctuating_hydrodynamic_polarization_k-0_lin} for $\partial_t\bm{w}'^a$,
\begin{equation}
	\begin{split}
		\Delta S^a_{2} & = \frac{k_{\rm B}}{D_{\rm h}} \iint (\partial_t \bm{w}'^a) \circ \bm{\mathcal{F}}^{a, \bm{w}'_b}_{k=0, {\rm lin}}[\bm{w}'^{b}] \,{\rm d}\bm{r}\,{\rm d}t\\
		& = \Delta S^a_{2.1} + \Delta S^a_{2.2}
	\end{split}
\end{equation}
with
\begin{equation}
	\begin{split}
		\Delta S^a_{2.1} = &\frac{k_{\rm B}}{D_{\rm h}} \iint	\bm{\mathcal{F}}^{a}_{k=0, {\rm lin}}	\cdot \bm{\mathcal{F}}^{a,\bm{w}'^b}_{k=0, {\rm lin}} {\rm d}\bm{r}\,{\rm d}t
	\end{split}
\end{equation}
and
\begin{equation}
\label{eq:entropy_production_second_part_stratonobvich_lin}
	\begin{split}
		\Delta S^a_{2.2} = &\frac{k_{\rm B}}{D_{\rm h}} \iint \tilde{\bm{\eta}}_a \circ \bm{\mathcal{F}}^{a,\bm{w}'^b}_{k=0, {\rm lin}} \, {\rm d}\bm{r}\,{\rm d}t.
	\end{split}
\end{equation}
While $\Delta S_{2.1}$ is noise-independent and directly yields a contribution to the entropy production rate upon division by $\tau$ and taking the limit $\tau \rightarrow \infty$, $\Delta S_{2.2}$ must be calculated explicitly.

To treat the Stratonovich product in Eq.~\eqref{eq:entropy_production_second_part_stratonobvich_lin}, we convert it into an It\^o product. The advantage is that in It\^o calculus, equal-time correlations like $\langle \tilde{\bm{\eta}}^a \cdot \bm{w}'^c \rangle$ (with $c = a$ or $c \neq a$) vanish \cite{gardiner_1985_handbook_stochastic_methods}\footnote{In It\^o calculus, the delta-correlated noise $\tilde{\bm{\eta}}^a$ only affects $\bm{w}^c$ at later time steps, making $\tilde{\bm{\eta}}^a$ and $\bm{w}^c$ uncorrelated at equal times.}.
The relevant conversions between Stratonovich and It\^o are of form \cite{gardiner_1985_handbook_stochastic_methods,borthne_cates_2020_time-reversal_symmetry_violations_entropy_production_field_theories_polar_active_matter,cates_tjhung_2022_stochastic_hydrodynamics_complex_fluids}
\begin{equation}
	\label{eq:entropy_production_conversion_Ito_Stratonovich_noise_function_main}
	\begin{split}
		\langle h(\bm{w}'^c(\bm{r},&t)) \circ \tilde{\eta}_{a,i}(\bm{r}',t) \rangle
		=\, \langle h(\bm{w}'^c(\bm{r},t)) \cdot \tilde{\eta}_{a,i}(\bm{r}',t) \rangle + D_{\rm h} \, \Big \langle \frac{\partial h(\bm{w}'^c(\bm{r},t))}{\partial w'^b_{j}(\bm{r}',t)} \Big \rangle \, \delta(\bm{r}-\bm{r}')\, \delta_{ab}\, \delta_{ij} ,
	\end{split}
\end{equation}
where $h$ is a function of $\bm{w}'$, and $\cdot$ denotes the It\^o product \footnote{This conversion is commonly presented for particles and unit-variance noise, e.g., \cite{gardiner_1985_handbook_stochastic_methods,cates_tjhung_2022_stochastic_hydrodynamics_complex_fluids}: for a Langevin equation
\begin{equation}
	\frac{{\rm d}x_i}{{\rm d}t} = f_i(\{x_i\}) + g_{ij}(\{x_i\})\,\eta_j(t)
\end{equation}	
for $\{x_i(t)\}$, where $i=1,2,...,d$, and $\langle \eta_i(t)\rangle = 0$, $\langle \eta_i(t)\,\eta_j(t') \rangle = \delta_{ij}\,\delta(t-t')$, one has \cite{cates_tjhung_2022_stochastic_hydrodynamics_complex_fluids}
\begin{equation}
	\begin{split}
		\int_{t_0}^{t_N} h_{i}(\{x_i\}) \circ \eta_j(t) \, {\rm d}t = & \int_{t_0}^{t_N} h_{i}(\{x_i\}) \cdot \eta_j(t) \, {\rm d}t + \frac{1}{2}\, \int_{t_0}^{t_N} \frac{\partial h_{i}(\{x_i\})}{\partial x_k} \, g_{kj} \, {\rm d}t,
	\end{split}
\end{equation}
where $h_i$ is a function of $\{x_i(t)\}$. In our field-theoretical case, the noise $\tilde{\bm{\eta}}^b$ has no unit-variance but `includes' the $g_{jk}$. Therefore, the additional prefactor $2\,D_{\rm h}\,\delta(\bm{r}-\bm{r}')$ needs to be taken into account.}. Since the It\^o product vanishes in the equal-time expectation, only the conversion contribution, i.e., the second contribution on the r.h.s.~of Eq.~\eqref{eq:entropy_production_conversion_Ito_Stratonovich_noise_function_main}, remains to be evaluated. In our case, $\bm{\mathcal{F}}^{a,\bm{w}'^b}_{k=0, {\rm lin}}$ is independent of $\bm{w}'^a$, so also the conversion term vanishes, and hence $\Delta S^a_{2.2} = 0$. 

Therefore, the only contribution to the entropy production rate for polarization fluctuations of species $a$ arises from $\Delta S^a_{2.1}$, such that
\begin{equation}
	\begin{split}
		\langle \Delta \dot{S}^a \rangle = \frac{ k_{\rm B}}{D_{\rm h}} \left\langle \int	\bm{\mathcal{F}}^{a}_{k=0, {\rm lin}}	\cdot \bm{\mathcal{F}}^{a,\bm{w}'^b}_{k=0, {\rm lin}} {\rm d}\bm{r} \right\rangle
	\end{split}
\end{equation}
Substituting the explicit forms of $\bm{\mathcal{F}}^{a}_{k=0, {\rm lin}}$ and $\bm{\mathcal{F}}^{a,\bm{w}'^b}_{k=0, {\rm lin}}$ yields
\begin{equation}
\label{eq:entropy_production_rate_k-0_lin_susceptibilities}
	\begin{split}
		\langle \Delta \dot{S}^a \rangle = \frac{ k_{\rm B}}{D_{\rm h}} \Big(& \gamma^a_{a,\ell}\,\gamma^a_{b,\ell} \, \chi_{{\rm f},\ell}^{ab} + (\gamma^a_{b,\ell})^2 \, \chi_{{\rm f},\ell}^{bb} + \gamma^a_{a,t}\,\gamma^a_{b,t} \, \chi_{{\rm f},t}^{ab} + (\gamma^a_{b,t})^2 \, \chi_{{\rm f},t}^{bb} \Big)
	\end{split}
\end{equation}
with longitudinal and transversal susceptibilities of the polarization fluctuations defined as
\begin{equation}
	\chi_{{\rm f},\ell}^{aa} = \left\langle \int w'^a_x \, w'^a_x \, {\rm d}\bm{r}\right\rangle, \ \  \chi_{{\rm f},\ell}^{ab} = \left\langle \int w'^a_x \, w'^b_x \, {\rm d}\bm{r}\right\rangle,
\end{equation}
and
\begin{equation}
	\chi_{{\rm f},t}^{aa} = \left\langle \int w'^a_y \, w'^a_y \, {\rm d}\bm{r}\right\rangle, \ \  \chi_{{\rm f},t}^{ab} = \left\langle \int w'^a_y \, w'^b_y \, {\rm d}\bm{r}\right\rangle.
\end{equation}
In the $k=0$-limit, the field $\bm{w}'^a = \bm{w}'^a(t)$ is spatially uniform, so spatial integrals reduce to multiplication by the system size.

The key result, given by Eq.~\eqref{eq:entropy_production_rate_k-0_lin_susceptibilities}, is that, in the infinite-wavelength limit, the entropy production rate of polarization perturbations scales with their susceptibilities.
According to the prefactors to the susceptibilities (see Fig.~\ref{fig:prefactors_continuum_functional} in Appendix \ref{app:prefactors_function_k-0}), the longitudinal mixed-species susceptibilities, $\chi^{AB}_{{\rm f},\ell}$ and $\chi^{BA}_{{\rm f},\ell}$, contribute most significantly. However, the transversal and same-species contributions also remain nonzero.

Taken together, we have shown that both the particle and continuum entropy production rates are closely related to orientational susceptibilities. Despite this qualitative agreement, we note that the numerical (particle-based) and analytical (field-theoretical) quantities differ in detail. On the particle level, the entropy production rate behaves similarly to the temporal fluctuations of the global polarization vector. On the other hand, the analytical approach considers $k=0$-perturbations around homogeneous base states, using perturbations $\bm{w}'^a = \bm{w}^a - \bm{w}^a_0$. The entropy production at the field level scales with susceptibilities defined through these perturbations, which are not directly accessible on the particle level. Nevertheless, the analytical field-theoretical results provide a qualitative explanation for the observed correlation between susceptibility and informatic entropy production rate on the particle level.

While we focused here only on the contribution of the polarization perturbations alone to the entropy production rate, the analytical calculation of the entropy production rate for the full polarization dynamics in the limit of $k=0$ is presented in Appendix \ref{app:entropy_production_k-0}. To go beyond the infinite-wavelength limit, we also provide the full entropy production rate of the fluctuating hydrodynamic density and polarization density fields [see Eqs.~\eqref{eq:continuum_eq_density_fluctuating}-\eqref{eq:continuum_eq_polarization_density_fluctuating}] in Appendix \ref{app:total_continuum_entropy_production_rate}. Numerical continuum simulations reveal that the overall dependence of the entropy production rate on the coupling strength resembles the one obtained from particle simulations, including, in particular, the peaks near exceptional points.

\section{Conclusion}
We have studied the informatic entropy production rate at phase transitions associated to non-reciprocity-induced collective behavior in active matter systems. As a measure of distance from equilibrium, the entropy production rate captures the twofold non-equilibrium character arising from both activity and non-reciprocal orientational couplings. Moreover, it reflects phase transitions through pronounced signatures at critical exceptional points.

Specifically, we have focused on a binary mixture of polar particles, where non-reciprocal orientational couplings lead to spontaneous rotational motion and (partial) synchronization.

As a first main result, our particle simulations demonstrate that non-reciprocal orientational couplings can leave a clear signature in the entropy production rate -- also in active systems consisting of self-propelling polar particles. For sufficiently strong non-reciprocal couplings, the informatic entropy production rate increases with the strength of non-reciprocity. However, this is not always the case: when non-reciprocity is weak compared to the intraspecies alignment and the non-reciprocal couplings are anti-symmetric, the entropy production rate remains relatively low -- comparable to that of conventional reciprocal flocking states.

Interestingly, even in the regime of weak non-reciprocity, the entropy production rate exhibits pronounced peaks when the coupling strengths are not anti-symmetric but correspond to exceptional points in the associated field theory. These findings are consistent with field-theoretical studies of scalar Cahn-Hilliard systems, where transitions to traveling states are also accompanied by peaks in entropy production \cite{suchanek_loos_2023_irreversible_fluctuations_emergence_dynamical_phases,suchanek_loos_2023_entropy_production_nonreciprocal_Cahn-Hilliard}. This suggests that, as in scalar field theories, active polar particle systems exhibit non-trivial entropy production behavior at exceptional points.

At the particle level, we find a strong correspondence between the behavior of the polarization vector susceptibility, spontaneous chirality, and entropy production rate as functions of non-reciprocal couplings.

As the second key result, we substantiate this correspondence through field-theoretical analytical calculations. In the analytically tractable infinite-wavelength limit, we show that the entropy production rate of polarization perturbations scales with their susceptibilities, providing a qualitative explanation for the particle-based findings.

The close relationship between susceptibility and entropy production rate allows us to explain the observed peaks in entropy production near critical EPs. From a dynamical systems perspective, these EPs mark points where the Goldstone mode in flocking systems is excited. This excitation facilitates rotations of particle orientations, leading to enhanced susceptibility and, consequently, increased entropy production.

The combination of our particle- and continuum-level results suggests that the susceptibility of the polarization vector, which could be measured in experiments, can serve as an estimate of the informatic entropy production rate of active particles. Indeed, the entropy production has recently gained importance as a key quantity in optimal control strategies of, e.g., active particles \cite{soriani_markovich_2025_control_active_field_theories,garcia-millan_loos_2024_optimal_closed-loop_control_active_particles} or passive particles in complex environments \cite{loos_bechinger_2024_universal_symmetry_optimal_control_microscale}. It is also emerging as an important tool for gaining insight into biological processes \cite{li_fakhri_2024_measuring_irreversibility_learned_representations_biological_patterns,tan_fakhri_2021_scale-dependent_irreversibility_living_matter,seifert_2019_thermodynamic_inference}. An intriguing direction for future research is to explore the connection between entropy production rate and universal scaling behaviors near critical exceptional points \cite{liu_littlewood_2025_universal_scaling_non-reciprocity}.

\begin{acknowledgments}
This work was funded by the Deutsche Forschungsgemeinschaft (DFG, German Research
Foundation) -- Projektnummer 163436311 (SFB 910) and Projektnummer 517665044.
\end{acknowledgments}

\appendix

\section{Snapshots of particle simulations}
\label{app:collective_behavior_particle_simulations}
\begin{figure}
	\includegraphics[width=0.85\linewidth]{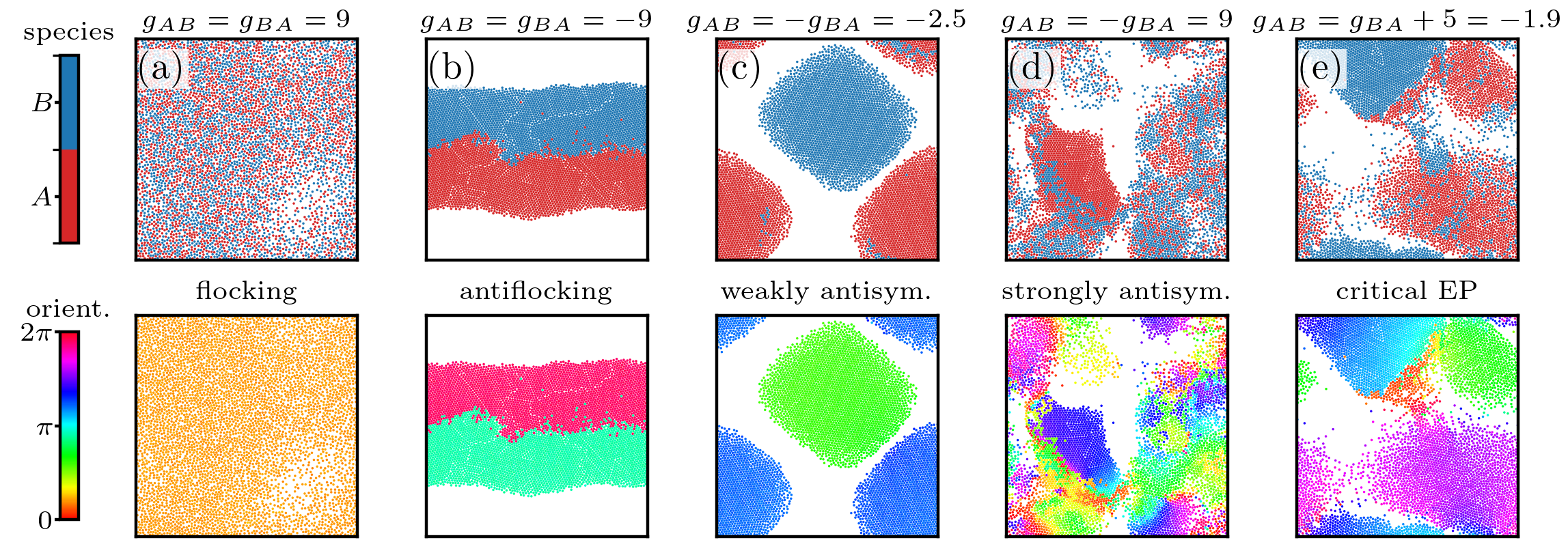}
	\caption{\label{fig:additional_snapshots}Brownian Dynamics simulation snapshots of the Langevin Eqs.~\eqref{eq:Langevin_r} and \eqref{eq:Langevin_theta}. The upper row shows particle species, and the lower row shows particle orientations, using separate color coding for clarity. The snapshots correspond to the parameter combinations (a) $g_{AB}=g_{BA}=9$, (b) $g_{AB}=g_{BA}=-9$, (c) $g_{AB}=-g_{BA}=-2.5$, (d) $g_{AB}=-g_{BA}=9$, and (e) $g_{AB}=g_{BA}-5=-1.9$. The intraspecies coupling strength is set to $g_{AA}=g_{BB}=9$. Other parameters are specified in text.}
\end{figure}

In Fig.~\ref{fig:additional_snapshots}, we show snapshots of various collective behaviors obtained from Brownian Dynamics simulations of the Langevin Eqs.~\eqref{eq:Langevin_r} and \eqref{eq:Langevin_theta}, described in Sec.~\ref{ssec:collective_behavior}. The snapshots correspond to the same parameter combinations as in Fig.~\ref{fig:stability_diagram_snapshots}, but here we indicate particle species and particle orientation in two separate panels.

\section{Full deterministic continuum equations}
\label{app:continuum_model_full}
The continuum equation for the particle density is the continuity Eq.~\eqref{eq:continuum_eq_density}. The evolution equation for the polarization density, Eq.~\eqref{eq:continuum_eq_polarization_density}, contains the functional  
\begin{equation}
\label{eq:polarization_density_functional}
	\begin{split}
		\bm{\mathcal{F}}^a[\rho^a, \rho^b, \bm{w}^a, \bm{w}^b]
		=& - \frac{1}{2} \, \nabla \,\big(v^{\rm eff}(\rho)\, \rho^a\big)  - D_{\rm r }\, \bm{w}^{a} + \sum_b g'_{ab} \, \rho^a\, \bm{w}^b +  D_{\rm t}\,\nabla^2\,\bm{w}^a + \frac{v^{\rm eff}(\rho)}{16\,D_{\rm r}} \, \nabla^2\,\Big(v^{\rm eff}(\rho)\,\bm{w}^a \Big) \\
		&- \sum_{b,c} \frac{ g'_{ab}\,g'_{ac}}{2\,D_{\rm r}} \, \bm{w}^a \, (\bm{w}^b \cdot \bm{w}^c) - \frac{z}{16\,D_{\rm r}} \, \nabla\rho \cdot \big[\nabla \big(v^{\rm eff}(\rho)\,\bm{w}^a\big) - \nabla^* \big(v^{\rm eff}(\rho)\,\bm{w}^{a*} \big) \big]\\
		&+ \sum_b \frac{g'_{ab}}{8 \,D_{\rm r}} \Big[ \bm{w}^b \cdot \nabla \big(v^{\rm eff}(\rho)\, \bm{w}^a\big) + \bm{w}^{b*} \cdot \nabla \big(v^{\rm eff}(\rho)\, \bm{w}^{a*}\big) -2 \, \Big\{ v^{\rm eff}(\rho)\,\bm{w}^a \cdot \nabla \bm{w}^b \\
		&\qquad   +\bm{w}^b \, \nabla \cdot \big(v^{\rm eff}(\rho)\,\bm{w}^a\big) - v^{\rm eff}(\rho)\, \bm{w}^{a*} \cdot \nabla \bm{w}^{b*} - \bm{w}^{b*}\,  \nabla \cdot \big( v^{\rm eff}(\rho)\, \bm{w}^{a*}\big) \Big \}  \Big] ,
	\end{split}
\end{equation}
where $\bm{w}^*=(w_y, -w_x)^{\rm T}$ and $\nabla^*=(\partial_y, -\partial_x)^{\rm T}$. The continuum equations are non-dimensionalized using a characteristic time scale $\tau$ and length scale $\ell$. The particle and polarization densities of species $a$ are scaled by the average particle density $\rho_0^a$. The remaining five dimensionless control parameters are the P\'eclet number ${\rm Pe}=v_0\,\tau/\ell$, $z = \zeta\,\rho^a_0\,\tau/\ell$ measuring the particle velocity-reduction due to the environment, the translational diffusion coefficient $D_{\rm t}=D'_{\rm t}\,\tau/\ell^2$, the rotational diffusion coefficient $D_{\rm r}=D'_{\rm r}\,\tau$, and the orientational coupling parameter $g'_{ab} = k_{ab}\,\mu_{\theta}\,R_{\theta}^2\,\rho_0^b\,\tau/2$. (For the parameters used in this study, the continuum ($g'_{ab}$) and particle ($g_{ab}$) orientational coupling parameters are related via $g'_{ab} = 12.73\,g_{ab}$.)

The continuum parameters are chosen to match the corresponding particle-level parameters. While most of these can be directly adopted, the velocity reduction parameter $z$ needs to be determined from the pair distribution function of particles. As described in Ref.~\cite{kreienkamp_klapp_2024_dynamical_structures_phase_separating_nonreciprocal_polar_active_mixtures}, it is given by $z = \zeta \, \rho^a_0 \, \tau/\ell = 57.63 \, \rho^a_0 \, \tau/\ell = 0.37 \, {\rm Pe}/\rho_0^{\rm con}$ with $\rho_0^{\rm con} = 1$. Furthermore, we adopt an ad-hoc choice of $D_{\rm t}=9$ \cite{kreienkamp_klapp_2024_dynamical_structures_phase_separating_nonreciprocal_polar_active_mixtures}.

\section{Calculation of the entropy production rate on the microscopic level}
\label{ssec:framework_entropy_production_microscopic}
Following Refs.~\cite{spinney_ford_2012_entropy_production_full_phase_space,crosato_spinney_2019_irreversibility_active_matter}, the medium entropy production can be linked to the microscopic Langevin dynamics of a system $\bm{\varphi} = (\phi_1, ..., \phi_n)$, consisting of $n$ coupled stochastic differential equations of form 
\begin{equation}
	\label{eq:Langevin_odd_even}
	{\rm d}\phi_i = A^{\rm rev}_{\phi_i}[\bm{\varphi}(t),t]\,{\rm d}t + A^{\rm ir}_{\phi_i}[\bm{\varphi}(t),t]\,{\rm d}t + B_{\phi_i} \, {\rm d}W_i,
\end{equation}
where $W_i$ are Wiener processes with ${\rm d}W_i\,{\rm d}W_j=\delta_{ij}\,{\rm d}t$ and $B_{\phi_i}$ are scalar. The reversible and irreversible components of the deterministic dynamics satisfy \cite{spinney_ford_2012_entropy_production_full_phase_space,risken_1996_fokker-planck_equation} 
\begin{equation}
	\label{eq:reversible_part_definition}
	A^{\rm rev}_{\phi_i}[\bm{\varphi}(t),t] = -\epsilon_{\phi_i} \, A^{\rm rev}_{\phi_i}[\bm{\epsilon}\bm{\varphi}(t),t]
\end{equation}
and
\begin{equation}
	\label{eq:irreversible_part_definition}
	A^{\rm ir}_{\phi_i}[\bm{\varphi}(t),t] = \epsilon_{\phi_i} \, A^{\rm ir}_{\phi_i}[\bm{\epsilon}\bm{\varphi}(t),t],
\end{equation}
where $\epsilon_{\phi_i}\in\{-1,+1\}$ for $\phi_i$ odd and even, respectively \cite{crosato_spinney_2019_irreversibility_active_matter}.

To calculate the entropy production from path probabilities, one considers the Fokker-Planck equation that corresponds to the Langevin equation \eqref{eq:Langevin_odd_even} and its solution, the short time Green's function or ``short time propagator'' $\mathfrak{p}(\bm{\varphi}',t+{\rm d}t\vert \bm{\varphi},t)$. The latter is the conditional probability of a displacement $d\bm{\varphi} = \bm{\varphi}'- \bm{\varphi}$ in a time ${\rm d}t$ given an initial $\delta$-function. The medium entropy production along the path $\vec{\bm{\varphi}}$ is then given by the path integral \cite{spinney_ford_2012_entropy_production_full_phase_space,seifert_2005_entropy_production_along_stochastic_trajectory,crosato_spinney_2019_irreversibility_active_matter}
\begin{equation}
	\Delta S[\vec{\bm{\varphi}}] = \int_{t_0}^{\tau} {\rm d}\Delta S(t)
\end{equation}
with increments
\begin{equation}
	\label{eq:increment_entropy_production_general}
	\begin{split}
		{\rm d}\Delta S(t) &= k_{\rm B}\,{\rm ln} \left( \frac{\mathfrak{p}(\bm{\varphi}',t+{\rm d}t\vert \bm{\varphi},t)}{\mathfrak{p}(\bm{\epsilon}\bm{\varphi}',t+{\rm d}t\vert \bm{\epsilon}\bm{\varphi},t)}\right)\\
		&= k_{\rm B} \sum_{\phi_i \in \bm{\varphi}} \Big( \frac{A^{\rm ir}_{\phi_i}(\bm{\varphi}(t))}{D_{\phi_i}} \circ {\rm d}\phi_i - \frac{A^{\rm ir}_{\phi_i}(\bm{\varphi}(t))\,A^{\rm rev}_{\phi_i}(\bm{\varphi}(t))}{D_{\phi_i}}\, {\rm d}t \Big),
	\end{split}
\end{equation}
where $D_{\phi_i}=B_{\phi_i}^2/2$ denotes the diffusion coefficient. For the propagators in Eq.~\eqref{eq:increment_entropy_production_general} Stratonovich evaluation was chosen, such that the $\circ$ symbol denotes a product interpreted in the Stratonovich sense.

We now apply the entropy production framework to the specific case of a non-reciprocal polar active mixture governed by the overdamped Langevin equations \eqref{eq:Langevin_r} and \eqref{eq:Langevin_theta}. With the convention of odd active forcing, we decompose the non-dimensionalized microscopic dynamics into reversible and irreversible parts [see Eqs.~\eqref{eq:reversible_part_definition} and \eqref{eq:irreversible_part_definition}], that are
\begin{subequations} 
	\begin{align}
		\bm{A}^{\rm rev}_{\bm{r}_{\alpha}} &= v_0\, \bm{p}_{\alpha}\\
		\bm{A}^{\rm ir}_{\bm{r}_{\alpha}} &= \mu_{r} \sum_{\beta \neq \alpha} \bm{F}^{\alpha}_{\rm rep}(\bm{r}_{\alpha},\bm{r}_{\beta})\\
		A^{\rm rev}_{\theta_{\alpha}} &= 0\\
		A^{\rm ir}_{\theta_{\alpha}} &= \mu_{\theta} \sum_{\beta\neq\alpha} \mathcal{T}_{\rm al}^{\alpha}(\bm{r}_{\alpha},\bm{r}_{\beta},\theta_{\alpha},\theta_{\beta}).
	\end{align}
\end{subequations}
From the general formula for the entropy production, Eq.~\eqref{eq:increment_entropy_production_general}, we can compute the entropy produced by each particle.

\section{Numerical methods}
\subsection{Particle simulations}
We carry out numerical Brownian Dynamics (BD) simulations based on the Langevin Eqs.~\eqref{eq:Langevin_r} and \eqref{eq:Langevin_theta}, using a square simulation box of size $L \times L$ with periodic boundary conditions. The simulations are conducted with the following dimensionless parameters. The total area fraction is fixed at $\Phi = (\rho^A_0 + \rho^B_0) \,\pi\,\ell^2 / 4 = 0.4$, where $\rho_0^a = N_a / L$ defines the number density. Each species contains an equal number of particles, i.e., $\Phi_A = \Phi_B = \Phi/2$. We choose a P\'eclet number of ${\rm Pe}=40$, and set the repulsion strength to $\epsilon^*=100$. The diffusion coefficients are $D_{\rm t}' = 1\, \ell^2/\tau$ for translational and $D_{\rm r}' = 3 \cdot 2^{-1/3} / \tau$ for rotational motion. Orientational coupling strengths are defined as $g_{ab}= k_{ab}\,\mu_{\theta}\,\tau$. We focus on strong intraspecies couplings by fixing $g_{AA}=g_{BB}=9$, while the interspecies couplings $g_{AB}$, $g_{BA}$ are varied independently. In all simulations, the cut-off radius for orientational couplings is set to $R_{\theta} = 10\,\ell$. We use a total of $N=5000$ particles, corresponding to a box size of $L=99\,\ell$. As long as the systems contains a sufficient number of particles, its dynamical behavior is not affected by the particle number \cite{kreienkamp_klapp_2024_synchronization_exceptional_points}. 

The system is initialized in a random configuration. We integrate the equations of motion using an Euler-Mayurama algorithm with a timestep of $\delta t = 10^{-5}\,\tau$. Before data evaluation, we let the simulations evolve for $150\,\tau$ to reach a steady state. Typically, we consider three independent noise realizations. Time-averaged quantities characterizing collective behavior (i.e., the susceptibility of the polarization vector and spontaneous chirality) are calculated between $150\,\tau$ and $170\,\tau$ after initialization.

To compute the entropy production rate, we begin from the non-equilibrium steady state reached at $170\,\tau$ and integrate the system for an additional $5\,\tau$ using the same Euler-Mayurama scheme with a timestep of $\delta t = 10^{-5}\,\tau$. The entropy production rates are computed over this $5\,\tau$ interval. During this time interval, the systems remain in the non-equilibrium steady states discussed in the main text. The time derivatives $\dot{\bm{r}}$ and $\dot{\theta}$, required for the calculation of the entropy production rate [see Eqs.~\eqref{eq:entropy_production_rate_contribution_translational} and \eqref{eq:entropy_production_rate_contribution_alignment}], are approximated using finite differences over $t_{s} = 100$ timesteps. 

Concretely, to evaluate the alignment contribution to the entropy production rate, given in Eq.~\eqref{eq:entropy_production_rate_contribution_alignment}, we discretize time as $t = i\,t_{\rm s}\,\delta t$, where $i=0,1,...,N$. At the midpoint time $t_{i+\frac{1}{2}}$, we compute
\begin{equation}
	\Delta \dot{S}_{\alpha}^{{\rm align}}(t_{i+\frac{1}{2}}) = \frac{k_{\rm B}}{D'_{\rm r}}  \Big(\mu_{\theta} \sum_{\beta\neq\alpha} \mathcal{T}_{\rm al}^{\alpha}(\bm{r}_{\alpha}(t_{i+\frac{1}{2}}),\bm{r}_{\beta}(t_{i+\frac{1}{2}}),\theta_{\alpha}(t_{i+\frac{1}{2}}),\theta_{\beta}(t_{i+\frac{1}{2}})) \Big) \circ \dot{\theta}_{\alpha}(t_{i+\frac{1}{2}}),
\end{equation}
where $\bm{r}_{\alpha}(t_{i+\frac{1}{2}}) = (\bm{r}_{\alpha}(t) + \bm{r}_{\alpha}(t+1))/2$, $\theta_{\alpha}(t_{i+\frac{1}{2}}) = (\theta_{\alpha}(t) + \theta_{\alpha}(t+1))/2$, and 
\begin{equation}
	\dot{\theta}_{\alpha}(t_{i+\frac{1}{2}}) = \frac{\theta_{\alpha}(t+1) - \theta_{\alpha}(t)}{t_{\rm s}\,\delta t} .
\end{equation}
We then take the time and ensemble average of $\Delta \dot{S}_{\alpha}^{{\rm align}}(t)$ to obtain $\langle \Delta \dot{S}_{\alpha}^{{\rm align}}(t) \rangle$. The same procedure is applied to determine $\langle \Delta \dot{S}_{\alpha}^{{\rm trans}}(t) \rangle$.

\subsection{Continuum simulations}
\label{sapp:numerical_methods_continuum}
We also compute the entropy production rate from the full fluctuating continuum model (see Appendix \ref{app:continuum_model_fluctuating}) via numerical simulations in two-dimensional periodic systems. The parameters are chosen as described in Appendix \ref{app:continuum_model_full}. The noise strength is set to $D_{\rm h}^{\rho} = D_{\rm h}^{\bm{w}} = D_{\rm h}= 1\cdot 10^{-5}$. 

To solve the stochastic partial differential continuum equations numerically, we employ a pseudo-spectral method combined with an operator splitting technique to accurately treat the linear terms. The pseudo-spectral code determines derivatives in Fourier space and nonlinear terms in real space \cite{canuto_zang_2007_spectral_methods}. For the time integration, we use the Euler-Mayurama algorithm, analogous to the stochastic particle-level simulations \cite{kloeden_platen_2011_numerical_solution_SDE}. The initial state is a slightly perturbed disordered configuration, with zero polarization $\bm{w}(\bm{r},0)=\bm{0}$ and a constant density $\rho^a(\bm{r},0) = 1$. The simulation box of size $50\,\ell \times 50\,\ell$ is discretized into $126 \times 126$ grid points. The timestep is set to $\delta t = 10^{-5}\,\tau$. The entropy production rate is evaluated between $330\,\tau$ and $350\,\tau$ after initialization. To compute derivatives we use the difference over 1000 timesteps.

\section{Microscopic observables}
\label{app:microscopic_obervables}
To understand how the entropy production rate relates to the different non-equilibrium states of the system, we identify observables that characterize the collective behavior. 

\subsection{Polarization}
The first such observable is the global polarization vector of the entire system (including both particle species), defined as
\begin{equation}
	\bm{P}_{\rm combi}(t) =  \frac{1}{N} \sum_{i=1}^N  \bm{p}_i(t).
\end{equation}
Its modulus presents the global polarization order parameter $P_{\rm combi}(t) = \vert \bm{P}_{\rm combi}(t) \vert$.
Its time and ensemble average is denoted $\langle P_{\rm combi} \rangle$. A value of $\langle P_{\rm combi} \rangle=1$ corresponds to perfect collective motion, where all particles align in the same direction, while \mbox{$\langle P_{\rm combi} \rangle=0$} reflects disorder or mutual cancellation of particle orientations. 

In addition to the combined polarization $\bm{P}_{\rm combi}(t)$ of $A$- and $B$-particles, we also consider the polarization of each species separately. For particles of species $a$, the polarization vector is
\begin{equation}
\label{eq:polarization_vector_species}
	\bm{P}_a(t) = \frac{1}{N_a} \sum_{\alpha=1}^{N_a}  \bm{p}_{\alpha}(t),
\end{equation}
where $N_a$ is the number of $a$-particles and $\alpha$ denotes a particle of species $a$. Its modulus yields the single-species polarization order parameter, $P_a(t) = \vert \bm{P}_a \vert$. Again, we define the time and ensemble average as $\langle P_a \rangle$. A perfect flocking state is characterized by $\langle P_A \rangle = \langle P_B \rangle = \langle P_{\rm combi} \rangle = 1$, while a perfect antiflocking state has $\langle P_A \rangle = \langle P_B \rangle = 1$ and $\langle P_{\rm combi} \rangle = 0$.

\subsection{Susceptibility}
\label{sapp:microscopic_observables_susceptibility}
Temporal fluctuations of the scalar global order parameters $P_{\rm combi}$ and $P_a$ are quantified by the susceptibility $\chi(P)$, defined as \cite{martin_gomez_pagonabarraga_2018_collective_motion_ABP_alignment_volume_exclusion,Baglietto_Albano_2008_Finite-size_scaling_analysis_self-driven_indiviudals,Adhikary_Santra_2022_pattern_formation_phase_transition_binary_mixture}
\begin{equation}
	\begin{split}
		\chi&(P_{\rm combi}) = N\,{\rm Var}(P_{\rm combi}) = \frac{1}{N} \Big( \Big\langle \Big\vert \sum_{i=1}^N \bm{p}_i \Big\vert \, \Big\vert \sum_{j=1}^N \bm{p}_j \Big\vert \Big\rangle - \Big\langle \Big\vert \sum_{i=1}^N \bm{p}_i \Big\vert \Big\rangle^2 \Big),
	\end{split}
\end{equation}
where ${\rm Var}(P) =\langle (P - \langle P \rangle)^2 \rangle$ is the variance. The susceptibility of species $a$ is given by $\chi_a(P_a) = N_a\, {\rm Var}(P_a)$ \cite{kreienkamp_klapp_2024_dynamical_structures_phase_separating_nonreciprocal_polar_active_mixtures}.

In this study, we focus primarily on the fluctuations of the heading vectors of individual particles $i$. These are defined as the deviation of the individual particle heading vectors from the respective single-species polarization, i.e.,
\begin{equation}
	\delta\bm{p}^{\rm av}_{\alpha}(t) = \bm{p}_{\alpha}(t) - \langle \bm{P}_a \rangle.
\end{equation}
Here, the superscript $\cdot^{\rm av}$ denotes that $\delta\bm{p}^{\rm av}_{\alpha}(t)$ measures the deviation of individual particles of species $a$ from the (time and ensemble) averaged total $a$-species polarization $\bm{P}_a$. The susceptibilities associated with the polarization vectors $\bm{P}_{\rm combi}$ and $\bm{P}_a$ are
\begin{equation}
	\begin{split}
		\chi(\bm{P}_{a}) &= N_a\,{\rm Var}(\bm{P}_a) = \frac{1}{N_a} \Big( \sum_{\alpha=1}^{N_a} \sum_{\alpha'=1}^{N_a} \langle \bm{p}_{\alpha} \cdot \bm{p}_{\alpha'} \rangle - \Big\langle \sum_{i=\alpha}^{N_a} \bm{p}_{\alpha} \Big\rangle^2 \Big)= \frac{1}{N_a} \sum_{\alpha=1}^{N_a} \sum_{\alpha'=1}^{N_a} \langle \delta\bm{p}^{\rm av}_{\alpha} \cdot \delta\bm{p}^{\rm av}_{\alpha'} \rangle
	\end{split}
\end{equation}
and
\begin{equation}
	\begin{split}
		\chi&(\bm{P}_{\rm combi}) = N\,{\rm Var}(\bm{P}_{\rm combi}) = \frac{1}{N} \sum_{i=1}^{N} \sum_{j=1}^{N} \langle \delta\bm{p}^{\rm av}_{i} \cdot \delta\bm{p}^{\rm av}_{j} \rangle
	\end{split} .
\end{equation}

The susceptibility $\chi(\bm{P}_a)$ is related to infinite-wavelength correlations of polarization fluctuations, expressed in terms of the connected correlation function \cite{hansen_mcDonald_2013_theory_simple_liquids,cavagna_grigera_2018_physics_of_flocking_correlations}
\begin{equation}
	C_{\rm c}^a(\bm{r}) = \frac{V}{N_a^2} \sum_{\alpha=1}^{N_a} \sum_{\alpha'=1}^{N_a} \langle \delta\bm{p}^{\rm av}_{\alpha} \cdot \delta\bm{p}^{\rm av}_{\alpha'} \, \delta(\bm{r}-\bm{r}_{\alpha\alpha'}) \rangle,
\end{equation}
where $V$ denotes the system area and $\bm{r}_{\alpha\alpha'} = \bm{r}_{\alpha} - \bm{r}_{\alpha'}$. In Fourier space (denoted by $\hat{\cdot}$), the relation is given by 
\begin{equation}
\label{eq:susceptibility_connected_correlation_relation}
	\hat{C}_{\rm c}^a(k = 0) = \frac{V}{N}\,\chi(\bm{P}_a) . 
\end{equation}
Note that the Fourier transform at $k=0$ corresponds to a spatial integral.

\subsubsection{Instantaneous susceptibility}
The definition of the heading vector deviation $\delta \bm{p}_{\alpha}^{\rm av}$ relies on the time- and ensemble-averaged polarization $\langle \bm{P}_a \rangle$. This quantity can be determined straightforwardly in systems with (more or less) constant polarization direction. However, for rotating states, $\langle \bm{P}_a \rangle$ is not meaningful. In this case, the instantaneous polarization vector $\bm{P}_a(t)$ can have (approximately) constant magnitude, while its direction rotates in time. Hence, $\langle \bm{P}_a \rangle$ crucially depends on the time frame over which the average is performed.

To circumvent this issue, it is \textit{not} sufficient to consider particle heading vector deviations relative to the \textit{instantaneous} average polarization $\bm{P}_{\alpha}$, defined as \cite{cavagna_grigera_2018_physics_of_flocking_correlations}
\begin{equation}
	\delta \bm{p}_{\alpha}^{\rm ins}(t) = \bm{p}_{\alpha}(t) - \bm{P}_a(t) .
\end{equation}
The reason is that, since the sum over all instantaneous deviation is always zero, i.e.,
\begin{equation}
	\sum_{\alpha} \delta \bm{p}_{\alpha}^{\rm ins}(t) = 0,
\end{equation}
the corresponding susceptibility $\chi^{\rm ins}(\bm{P}_a)$ trivially vanishes:
\begin{equation}
\label{eq:susceptibility_instantaneous_constraint}
	\chi^{\rm ins}(\bm{P}_a) = \frac{1}{N_a} \sum_{\alpha=1}^{N_a} \sum_{\alpha'=1}^{N_a} \langle \delta\bm{p}^{\rm ins}_{\alpha} \cdot \delta\bm{p}^{\rm ins}_{\alpha'} \rangle = \Bigg\langle \frac{1}{N_a} \left( \sum_{\alpha=1}^{N_a} \delta\bm{p}^{\rm ins}_{\alpha} \right) \cdot \left( \sum_{\alpha'=1}^{N_a} \delta\bm{p}^{\rm ins}_{\alpha'} \right) \Bigg\rangle= 0.
\end{equation}

\begin{figure}
	\includegraphics[width=1\linewidth]{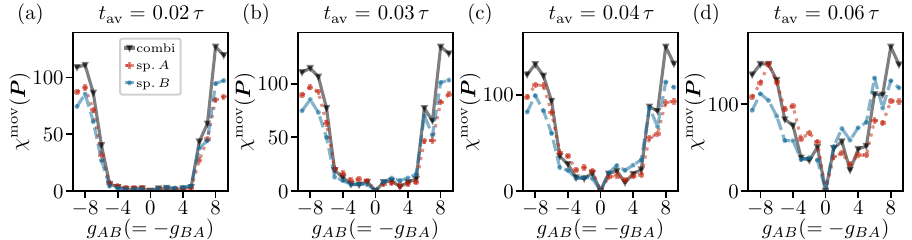}
	\caption{\label{fig:susceptibilities_comparison_anti-symmetric}Susceptibility of the polarization vector from particle simulations of a non-reciprocal, anti-symmetric system with $g_{AB} = -g_{BA}$. The susceptibilities are calculated from deviations of particle heading vectors relative to a co-rotating mean polarization direction. Panels (a)-(d) display the results for different time intervals $t_{\rm av}$ used to compute the mean polarization direction.}
\end{figure}

\begin{figure}
	\includegraphics[width=1\linewidth]{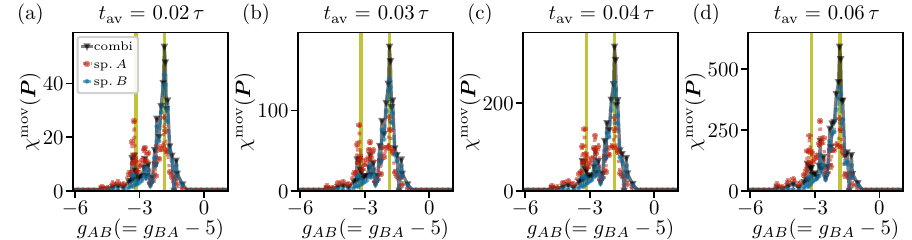}
	\caption{\label{fig:susceptibilities_comparison_fd-5}Susceptibility of the polarization vector obtained from particle simulations of a non-reciprocal system with $g_{AB} = g_{BA}-5$. The susceptibilities are calculated from deviations of particle heading vectors relative to a co-rotating mean polarization direction. Panels (a)-(d) display the results for different time intervals $t_{\rm av}$ used to compute the mean polarization direction. The green vertical lines indicate the positions of critical exceptional points, respectively.}
\end{figure}

\subsubsection{Moving mean susceptibility}
To overcome this issue, we propose an adapted calculation of the susceptibility that involves a moving mean of the polarization vector. To this end, we calculate the average polarization vector $\overline{\bm{P}}_a(t)$ at each time $t$ as the average over the polarization vectors $\bm{P}_a(t)$ [defined in Eq.~\eqref{eq:polarization_vector_species}] within the preceding time interval $t_{\rm av}$. This yields the moving mean polarization vector
\begin{equation}
	\overline{\bm{P}}_a(t) = \frac{1}{t_{\rm av}} \, \int_{t-t_{\rm av}}^{t_{\rm av}} \bm{P}_a(t') \, {\rm d}t'.
\end{equation}
The deviations of individual particle orientations relative to this moving mean polarization vector of the respective particle species are
\begin{equation}
	\delta \bm{p}_{\alpha}^{\rm mov}(t) = \bm{p}_{\alpha}(t) - \overline{\bm{P}}_a(t) .
\end{equation}
The corresponding moving mean susceptibility is
\begin{equation}
	\label{eq:susceptibility_moving_mean}
	\chi^{\rm mov}(\bm{P}_a) = \frac{1}{N_a} \sum_{\alpha=1}^{N_a} \sum_{\alpha'=1}^{N_a} \langle \delta\bm{p}^{\rm mov}_{\alpha} \cdot \delta\bm{p}^{\rm mov}_{\alpha'} \rangle = \Bigg\langle \frac{1}{N_a} \left( \sum_{\alpha=1}^{N_a} \delta\bm{p}^{\rm mov}_{\alpha} \right) \cdot \left( \sum_{\alpha'=1}^{N_a} \delta\bm{p}^{\rm mov}_{\alpha'} \right) \Bigg\rangle.
\end{equation}
If both particle species are taken into account, we calculate
\begin{equation}
	\chi^{\rm mov}(\bm{P}_{\rm combi}) = \frac{1}{N} \sum_{i=1}^{N} \sum_{j=1}^{N} \langle \delta\bm{p}^{\rm mov}_{i} \cdot \delta\bm{p}^{\rm mov}_{j} \rangle .
\end{equation}
One can understand this susceptibility as a measure of fluctuations of the polarization vector in a co-rotating reference frame, which becomes necessary in states where the polarization direction rotates.

The moving mean susceptibility depends on the choice of $t_{\rm av}$. However, in Figs.~\ref{fig:susceptibilities_comparison_anti-symmetric} and \ref{fig:susceptibilities_comparison_fd-5}, we show that for two different paths in the $g_{AB}-g_{BA}$ plane, the specific choice of $t_{\rm av}$ has little qualitative impact when averaging over small time intervals. The qualitative behavior as a function of alignment strength, including the peaks shown in Fig.~\ref{fig:susceptibilities_comparison_fd-5}, remains similar, though absolute values may differ. Since we focus on qualitative behavior in the main text, this moving mean susceptibility is an appropriate quantity for our purposes. In Figs.~\ref{fig:BD_pol_sus_entropy_results_anti-symmetric}, \ref{fig:BD_pol_sus_entropy_results_fixed_dif-5_species}, \ref{fig:entropy_production_particle_fixed_dif--3_species}, and \ref{fig:entropy_production_particle_fixed_dif--7_species}, we set $t_{\rm av}=0.03\,\tau$.

\subsubsection{Longitudinal and transverse susceptibility}
One can further distinguish longitudinal and transverse susceptibilities by considering corresponding particle heading deviations relative to the moving mean polarization vector \cite{kardar_2007_statistical_physics_fields}. The transverse susceptibilities are typically associated with the Goldstone modes of the system.

Let $\hat{\bm{e}}_1(t)$ denote the unit vector along the instantaneous mean polarization direction and $\hat{\bm{e}}_2(t)$ a perpendicular unit vector, i.e.,
\begin{equation}
	\hat{\bm{e}}_1(t) = \frac{\overline{\bm{P}}_a(t)}{\vert \overline{\bm{P}}_a(t) \vert} \quad {\rm and} \quad \hat{\bm{e}}_1(t) \cdot \hat{\bm{e}}_2(t) = 0 .
\end{equation}
Longitudinal and transverse fluctuations are then defined as \cite{kardar_2007_statistical_physics_fields}
\begin{equation}
	\delta p_{\alpha,\ell}^{\rm mov}(t) = \big( \bm{p}_{\alpha}(t) \cdot \hat{\bm{e}}_1(t) \big) - \vert \overline{\bm{P}}_a(t) \vert \quad {\rm and} \quad \delta p_{\alpha,{\rm t}}^{\rm mov}(t) = \big( \bm{p}_{\alpha}(t) \cdot \hat{\bm{e}}_2(t) \big) .
\end{equation}
The corresponding susceptibilities are given by
\begin{equation}
	\chi_{\ell}^{\rm mov}(\bm{P}_a) = \frac{1}{N_a} \sum_{\alpha=1}^{N_a} \big\langle \big(\delta p^{\rm mov}_{\alpha,\ell}\big)^2 \big\rangle \quad {\rm and} \quad \chi_{\rm t}^{\rm mov}(\bm{P}_a) = \frac{1}{N_a} \sum_{\alpha=1}^{N_a} \big\langle \big(\delta p^{\rm mov}_{\alpha,{\rm t}}\big)^2 \big\rangle .
\end{equation}

\begin{figure}
	\includegraphics[width=1\linewidth]{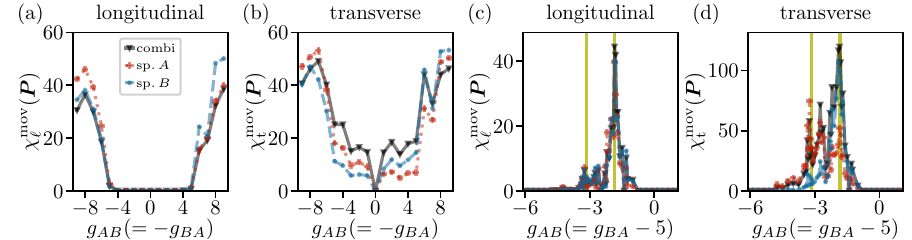}
	\caption{\label{fig:susceptibilities_long_trans}Longitudinal and transverse susceptibilities of the polarization vector obtained from particle simulations of non-reciprocal systems with (a),(b) $g_{AB}=-g_{BA}$ and (c),(d) $g_{AB}=g_{BA}-5$. The susceptibilities are calculated from deviations of particle heading vectors relative to a co-rotating mean polarization direction, which is determined by time averaging over $0.03,\tau$. The green vertical lines in (c),(d) indicate the locations of the critical exceptional points.}
\end{figure}

In Fig.~\ref{fig:susceptibilities_long_trans}, we plot the longitudinal and transverse susceptibilities for two different paths in the $g_{AB}$-$g_{BA}$ plane.

For the anti-symmetric system with $g_{AB}=-g_{BA}$, shown in Fig.~\ref{fig:susceptibilities_long_trans}(a) and (b), the longitudinal susceptibility becomes non-negligible only beyond a certain threshold of non-reciprocity and subsequently increases with increasing non-reciprocity. The transverse susceptibility is non-zero for all values of non-reciprocity and also increases with its strength. For sufficiently large non-reciprocity, both longitudinal and transverse susceptibilities become comparable in magnitude.

The longitudinal and transverse susceptibilities of the non-reciprocal system with $g_{AB}=g_{BA}-5$, which crosses exceptional points twice, are shown in Fig.~\ref{fig:susceptibilities_long_trans}(c) and (d). Both susceptibilities are non-negligible, particularly in the vicinity of the critical exceptional points, where pronounced peaks are observed. From a dynamical systems perspective, these critical exceptional points correspond to parameter values at which the Goldstone mode is excited. Accordingly, the transverse susceptibility becomes especially large close to these points.

Overall, the susceptibility associated with the global polarization vector, $\chi^{\rm mov}(\bm{P})$, shown in the main text, generally contains contributions from both longitudinal and transverse fluctuations, both of which are relevant. However, as expected, in the vicinity of critical exceptional points, the transverse susceptibility exhibits particularly strong peaks.

\subsection{Spontaneous chirality}
We further characterize collective rotational motion through a quantity called spontaneous chirality. It is defined as the absolute value of the average phase difference rate \cite{kreienkamp_klapp_2024_synchronization_exceptional_points},
\begin{equation}
	\vert \Omega_{\rm s}^i (t) \vert = \Big \vert \frac{\theta_i(t) - \theta_i(t-\Delta t)}{\Delta t} \Big \vert,
\end{equation}
where we set $\Delta t = 0.01 \, \tau$. Taking the absolute value avoids cancellation of clockwise and counterclockwise rotations when averaging. The time- and ensemble-averaged spontaneous chirality is denoted $\langle \vert \Omega_{\rm s}^i \vert \rangle$. 
In the absence of alignment couplings (i.e., $g_{ab} = 0 \, \forall \, a,b$), rotational motion results purely from rotational diffusion. In this case, the spontaneous chirality is $\langle \vert \Omega_{\rm s}^{\rm no \, align}\vert \rangle = 2\,\sqrt{D'_{\rm r}/(\pi\,\Delta t)}$, which is \mbox{$17.41 \, \tau^{-1}$} for the parameters used here \cite{kreienkamp_klapp_2024_synchronization_exceptional_points}. For polarized states in constant direction, i.e., flocking and antiflocking, the spontaneous chirality is small. It increases in non-reciprocal systems, where non-reciprocity induces chiral motion of particles.

While BD simulation results for polarization and spontaneous chirality have previously been used to characterize collective states in this system (see Ref.~\cite{kreienkamp_klapp_2024_synchronization_exceptional_points}), the focus here is on relating these collective states to the entropy production rate.

\section{Additional particle simulation results: different crossings of exceptional points}
\label{sapp:different_crossings_of_exceptional_points}
\begin{figure}
	\includegraphics[width=1\linewidth]{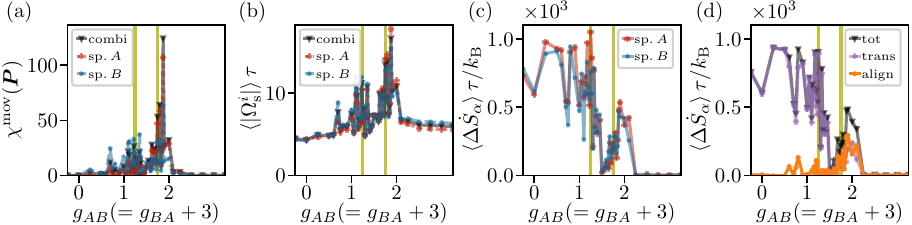}
	\caption{\label{fig:entropy_production_particle_fixed_dif--3_species}Particle simulation results for a non-reciprocal system with $g_{AB} = g_{BA}-d$ with $d=-3$. The (a) susceptibility of the polarization vector, (b) spontaneous chirality, and (c),(d) particle-averaged entropy production rates are shown as functions of $g_{AB}$. The green vertical lines indicate the positions of critical exceptional points, respectively ($g_{AB}^{{\rm EP}>}(d=-3)\approx 1.75$ and $g_{AB}^{{\rm EP}<}(d=-3)\approx 1.25$). In (a), the susceptibility of the global polarization vector is calculated in a co-rotating frame, which is determined by time averaging over $0.03\,\tau$. In (c), the entropy production rate is shown separately for both particle species, while in (d) the contributions from translational motion and alignment couplings are plotted separately. Other parameters are as specified in Sec.~\ref{sec:model}. The spontaneous chirality is consistent with Ref.~\cite{kreienkamp_klapp_2024_synchronization_exceptional_points}.}
\end{figure}

\begin{figure}
	\includegraphics[width=1\linewidth]{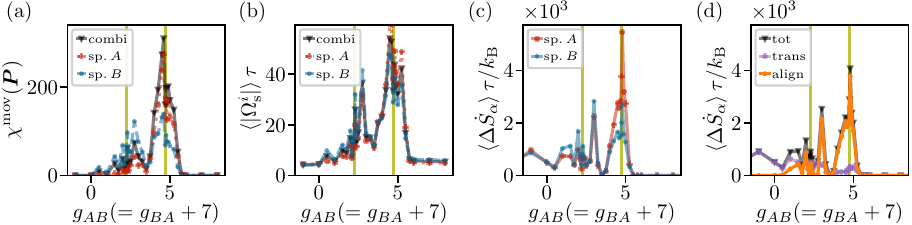}
	\caption{\label{fig:entropy_production_particle_fixed_dif--7_species}Particle simulation results for a non-reciprocal system with $g_{AB} = g_{BA}-d$ with $d=-7$. The (a) susceptibility of the polarization vector, (b) spontaneous chirality, and (c),(d) particle-averaged entropy production rates are shown as functions of $g_{AB}$. The green vertical lines indicate the positions of critical exceptional points, respectively ($g_{AB}^{{\rm EP}>}(d=-7)\approx 4.74$ and $g_{AB}^{{\rm EP}<}(d=-7)\approx 2.26$). In (a), the susceptibility of the global polarization vector is calculated in a co-rotating frame, which is determined by time averaging over $0.03\,\tau$. In (c), the entropy production rate is shown separately for both particle species, while in (d) the contributions from translational motion and alignment couplings are plotted separately. Other parameters are as specified in Sec.~\ref{sec:model}.}
\end{figure}

In the main text, we show in Fig.~\ref{fig:BD_pol_sus_entropy_results_fixed_dif-5_species} BD simulation results for a non-reciprocal system, where $g_{AB} = g_{BA}-d$ with $d=5$, which crosses exceptional points twice as $g_{AB}$ is varied. Here, in Figs.~\ref{fig:entropy_production_particle_fixed_dif--3_species} and \ref{fig:entropy_production_particle_fixed_dif--7_species}, we show that qualitatively similar behavior is observed for $g_{AB} = g_{BA} +3$ and $g_{AB} = g_{BA} + 7$. 

For these cases (with $d<0$), the system exhibits antiflocking for $g_{AB} < 0$, and flocking for $g_{AB} > -d$. At the exceptional points $g_{AB}^{\rm EP <}$ and $g_{AB}^{\rm EP >}$, the polarization vector susceptibilities, spontaneous chiralities, and informatic entropy production rates exhibit peaks.

Note that the peaks in the alignment coupling contributions to the informatic entropy production rate become more pronounced as the difference $\vert d \vert$ between $g_{AB}$ and $g_{BA}$ increases. This effect arises due to the growing dominance of the alignment-related contribution relative to the translational part as the non-reciprocity becomes larger (see Fig.~\ref{fig:BD_pol_sus_entropy_results_anti-symmetric}).\\

\section{Prefactors for polarization perturbations in entropy production rate in the infinite-wavelength limit}
\label{app:prefactors_function_k-0}
In this appendix, we show the analytical prefactors $\gamma$ and $\gamma^2$ in the functional \eqref{eq:continuum_functional_k-0_linear}, which governs the deterministic contribution to the time evolution of polarization perturbations in the infinite-wavelength limit, and of the corresponding entropy production rate \eqref{eq:entropy_production_rate_k-0_lin_susceptibilities}. 

\begin{figure}
	\includegraphics[width=1\linewidth]{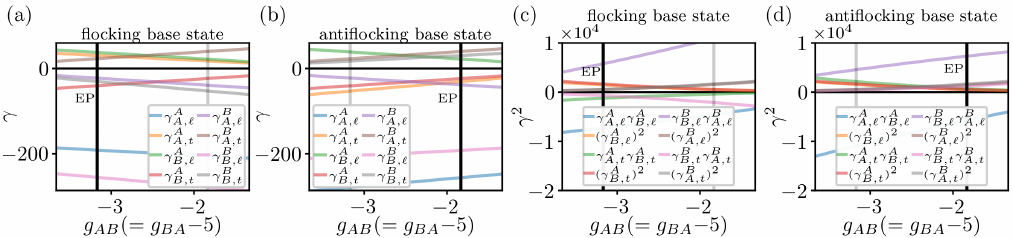}
	\caption{\label{fig:prefactors_continuum_functional}Analytical prefactors in field-theoretical description of polarization perturbations and the corresponding entropy production rate. The prefactors to the functional \eqref{eq:continuum_functional_k-0_linear} describing the deterministic contribution to the time evolution of polarization perturbations around the homogeneous (a) flocking and (b) antiflocking base states in the infinite-wavelength limit are shown as functions of $g_{AB} = g_{BA} -5$. Panels (c) and (d) show the corresponding prefactors of the entropy production rate \eqref{eq:entropy_production_rate_k-0_lin_susceptibilities}. The vertical black lines indicate the exceptional point associated with the respective base state, while the vertical grey lines mark the exceptional point of the other base state. Note that in (c), the line for $(\gamma_{B,\ell}^A)^2$ (orange) is covered by the line for $(\gamma_{B,t}^A)^2$ (red), and the line for $(\gamma_{A,\ell}^B)^2$ (brown) is covered by the line for $(\gamma_{A,t}^B)^2$ (grey). In (d), the lines for $(\gamma_{B,\ell}^A)^2$ (orange), $\gamma_{A,t}^A\,\gamma_{B,t}^A$ (green) and $(\gamma_{B,t}^A)^2$ (red) nearly overlap, as do the lines for $(\gamma_{A,\ell}^B)^2$ (brown), $\gamma_{B,t}^B\,\gamma_{A,t}^B$ (pink) and $(\gamma_{A,t}^B)^2$ (grey).}
\end{figure}

The prefactors in the functional \eqref{eq:continuum_functional_k-0_linear} are
\begin{equation}
	\begin{split}
		\bm{\gamma}^a_a[w_0^a, w_0^b] &= \begin{pmatrix}
			\gamma^a_{a,\ell} & 0\\ 0 & \gamma^a_{a,t}
		\end{pmatrix} \\
		&= (g' - D_{\rm r}) \begin{pmatrix}
			1 & 0\\ 0 & 1
		\end{pmatrix} - \frac{1}{2\,D_{\rm r}} \Big[(g'\,\bm{w}_0^a)^2 \, \begin{pmatrix}
			3 & 0\\ 0 & 1
		\end{pmatrix} + (g_{ab}'\,\bm{w}_0^b)^2 \begin{pmatrix}
			1 & 0\\ 0 & 1
		\end{pmatrix} + 2\,g'\,g_{ab}'\, \begin{pmatrix}
			2 & 0\\ 0 & 1
		\end{pmatrix} \, \bm{w}_0^a \cdot \bm{w}_0^b \Big]
	\end{split}
\end{equation}
and
\begin{equation}
	\begin{split}
		\bm{\gamma}^a_b [w_0^a, w_0^b] &= \begin{pmatrix}
			\gamma^a_{b,\ell} & 0\\ 0 & \gamma^a_{b,t}
		\end{pmatrix} \\
		& = g_{ab}' \begin{pmatrix}
			1 & 0\\ 0 & 1
		\end{pmatrix}  - \frac{1}{D_{\rm r}}  \begin{pmatrix}
			1 & 0\\ 0 & 0
		\end{pmatrix} \Big[g'\,g_{ab}'\,  (\bm{w}_0^a)^2 + g_{ab}'^2  \, \bm{w}_0^a \cdot \bm{w}_0^b \Big].
	\end{split}
\end{equation}
They are plotted as functions of $g_{AB} = g_{BA} - d$ with $d=5$ in Fig.~\ref{fig:prefactors_continuum_functional}. They do not exhibit special behavior at exceptional points. Nevertheless, they indicate the relevance of different contributions.

Near the exceptional points, certain prefactors of both the flocking [Fig.~\ref{fig:prefactors_continuum_functional}(a)] and antiflocking [Fig.~\ref{fig:prefactors_continuum_functional}(b)] base states -- specifically $\gamma^A_{B,\ell}$, $\gamma^B_{A,t}$ -- are positive, while their species-inverted counterparts ($\gamma^B_{A,\ell}$, $\gamma^A_{B,t}$) are negative but of similar magnitude. This indicates that, in the given parameter regime, longitudinal perturbations in species $B(A)$ enhance (dampen) longitudinal perturbations in $A(B)$. For transversal perturbations, the effect is reversed: transversal perturbations in species $A(B)$ enhance (dampen) transversal perturbations in $B(A)$. 

The transversal self-coupling terms, $\gamma^A_{A,t}$ and $\gamma^B_{B,t}$, have opposite signs and change sign between the flocking to antiflocking base states. This indicates that transversal $A$-perturbations to the flocking state are amplified, whereas transversal $B$-perturbations to the antiflocking state are amplified. In contrast, longitudinal same-species perturbations ($\gamma^A_{A,\ell}$ and $\gamma^B_{B,\ell}$) are always strongly damped.

Figs.~\ref{fig:prefactors_continuum_functional}(c) and (d) display the prefactors of the susceptibilities ($\gamma^2$) in the entropy production rate \eqref{eq:entropy_production_rate_k-0_lin_susceptibilities} for $g_{AB} = g_{BA}-5$ for flocking [Fig.~\ref{fig:prefactors_continuum_functional}(c)] and antiflocking base states [Fig.~\ref{fig:prefactors_continuum_functional}(d)]. The largest magnitude contributions arise from the terms $\gamma^A_{A,\ell}\,\gamma^A_{B,\ell}$ and $\gamma^B_{B,\ell}\,\gamma^B_{A\,\ell}$, pertaining to longitudinal mixed-species susceptibilities along the direction of polarization. However, other contributions, including transversal ones, are also nonzero and remain relevant.

For $d>0$, at the exceptional point associated to the flocking base state, $g_{AB}^{{\rm EP}<}$ (with $g_{AB}^{{\rm EP}<} < g_{AB}^{{\rm EP}>}$), the transversal susceptibility contributions to the entropy production of species $A$ -- namely $\gamma^A_{A,t}\,\gamma^A_{B,t}$ and $(\gamma^A_{B,t})^2$ -- are larger in magnitude than those for species $B$ ($\gamma^B_{B,t}\,\gamma^B_{A,t}$ and $(\gamma^B_{A,t})^2$). In contrast, at the exceptional point associated with the antiflocking base state, $g_{AB}^{{\rm EP}>}$, this behavior is reversed: the transversal susceptibility contributions to the entropy production rate of species $B$ are larger than those of species $A$.

This may explain why the entropy production rate of species $A$ peaks at $g_{AB}^{{\rm EP}<}$, while that of species $B$ peaks at $g_{AB}^{{\rm EP}>}$ -- both at the particle level (Fig.~\ref{fig:BD_pol_sus_entropy_results_fixed_dif-5_species}) and in the continuum model (Fig.~\ref{fig:entropy_production_continuum_fixed_dif}) for $g_{AB} = g_{BA}-5$. Note that this behavior is reversed for $d<0$.

\section{Full fluctuating continuum equations}
\label{app:continuum_model_fluctuating}
In the main text, we consider only the fluctuating hydrodynamic equations for infinite-wavelength perturbations of the polarization density with respect to homogeneous (anti)flocking base states, as given in Eq.~\eqref{eq:fluctuating_hydrodynamic_polarization_k-0_lin}. However, one can also write down the fluctuating hydrodynamic equations corresponding to the complete continuum description -- including spatial variations in the particle density and polarization density.

To this end, we start from the full deterministic continuum Eqs.~\eqref{eq:continuum_eq_polarization_density} and \eqref{eq:continuum_eq_density} for the particle density and polarization density. Adding noise terms to the (conserved) density and (non-conserved) polarization equations yields the full fluctuating continuum equations, given by
\begin{equation}
	\label{eq:continuum_eq_density_fluctuating}
	\partial_t \rho^a = -  \nabla \cdot \big(\bm{j}^a[\rho^a, \rho^b,\bm{w}^a] + \bm{\xi}^{\rm f}_{a}\big),
\end{equation}
with flux
\begin{equation}
	\bm{j}^a[\rho^a, \rho^b,\bm{w}^a] = v^{\rm eff}(\rho)\, \bm{w}^a -  D_{\rm t}\,\nabla\,\rho^a
\end{equation}
and
\begin{equation}
	\label{eq:continuum_eq_polarization_density_fluctuating}
	\partial_t \bm{w}^a = \bm{\mathcal{F}}^a[\rho^a, \rho^b, \bm{w}^a, \bm{w}^b] + \bm{\eta}^{\rm f}_{a},
\end{equation}
with functional $\bm{\mathcal{F}}^a[\rho^a, \rho^b, \bm{w}^a, \bm{w}^b]$ as given in Eq.~\eqref{eq:polarization_density_functional},
for species $a=A,B$ with $b\neq a$. The noise fields $\bm{\xi}^{\rm f}_{a}(\bm{r},t)$ and $\bm{\eta}^{\rm f}_{a}(\bm{r},t)$ have zero mean. Their variances are
\begin{equation}
	\begin{split}
		\langle \xi_{a,i}^{\rm f}(\bm{r},t)&\, \xi_{b,j}^{\rm f}(\bm{r}',t') \rangle = 2\,D_{\rm h}^{\rho} \, \delta_{ab}\, \delta_{ij}\, \delta(\bm{r}-\bm{r}')\,\delta(t-t')
	\end{split}
\end{equation}
and
\begin{equation}
	\begin{split}
		\langle \eta_{a,i}^{\rm f}(\bm{r},t)&\, \eta_{b,j}^{\rm f}(\bm{r}',t') \rangle = 2\,D_{\rm h}^{\bm{w}} \, \delta_{ab}\, \delta_{ij}\, \delta(\bm{r}-\bm{r}')\,\delta(t-t'),
	\end{split}
\end{equation}
respectively. For simplicity, we set $D_{\rm h}^{\rho} = D_{\rm h}^{\bm{w}} = D_{\rm h}$. In our numerical continuum simulations, we set $D_{\rm h} = 1\cdot 10^{-5}$.

\section{Full entropy production rate in infinite-wavelength limit}
\label{app:entropy_production_k-0}
In the main text, we analyze the contributions to the entropy production rate arising from perturbations of the polarization field around homogeneous base states in the infinite-wavelength limit. In this appendix, we also consider the infinite-wavelength limit, but provide the analytical calculation of the entropy production rate associated with the full polarization field (not just its perturbations), which involves higher-order correlations. Since the analysis remains restricted to infinite wavelengths, the density field is constant and only the polarization field is taken into account.

The mean-field $k=0$-polarization dynamics for species $a$ is given by \cite{fruchart_2021_non-reciprocal_phase_transitions,kreienkamp_klapp_2024_non-reciprocal_alignment_induces_asymmetric_clustering}
\begin{equation}
	\label{eq:fluctuating_hydrodynamic_polarization_k-0}
	\partial_t \bm{w}^a = \bm{\mathcal{F}}^{a}_{k=0}[\bm{w}^a,\bm{w}^b] + \tilde{\bm{\eta}}_a
\end{equation}
with the deterministic contribution given by the functional
\begin{equation}
	\begin{split}
		&\bm{\mathcal{F}}^{a}_{k=0}[\bm{w}^a,\bm{w}^b] =   (g' - D_{\rm r} - \frac{(g'\,\bm{w}^a + g'_{ab} \, \bm{w}^b)^2}{2\,D_{\rm r}}  ) \cdot \bm{w}^a + g'_{ab} \, \bm{w}^b ,
	\end{split}
\end{equation}
where $b \neq a $ and $\bm{w}^a = \bm{w}^a(\bm{r},t)$, $\tilde{\bm{\eta}}_a = \tilde{\bm{\eta}}_a(\bm{r},t)$. The alignment strength on the continuum level is given by $g'_{ab} = g_{ab}\,R_{\theta}^2\,\rho_0^b/2$. The noise term of strength $D_{\rm h}$ has zero mean and variance
\begin{equation}
	\begin{split}
	\label{eq:noise_correlation_field}
		\langle \tilde{\eta}_{a,i}(\bm{r},t) & \, \tilde{\eta}_{b,j}(\bm{r}', t') \rangle = 2\,D_{\rm h}\,\delta_{ab}\,\delta_{ij}\,\delta(\bm{r}-\bm{r}')\,\delta(t-t') .
	\end{split}
\end{equation}

The dynamical action of the forward and backward paths are of the same form as given in Secs.~\ref{sssec:entropy_production_dynamical_action} and \ref{sssec:entropy_production_backwards_probability}, just with $\bm{\mathcal{F}}^{a}_{k=0, {\rm lin}}[\bm{w}'^{a},\bm{w}'^{b}]$ replaced by $\bm{\mathcal{F}}^a_{k=0}[\bm{w}^a,\bm{w}^b]$. The resulting expression for the informatic entropy production rate is
\begin{equation}
	\label{eq:entropy_production_continuum_k-0}
	\Delta S^{a} = - \frac{k_{\rm B}}{D_{\rm h}} \iint \big(- (\partial_t \bm{w}^a) \circ \bm{\mathcal{F}}^{a}_{k=0}[\bm{w}^{a},\bm{w}^{b}] \big)\,{\rm d}\bm{r}\,{\rm d}t.
\end{equation}
Since the integral in Eq.~\eqref{eq:entropy_production_continuum_k-0} is a Stratonovich integral, the chain rule applies, and we can rewrite it as
\begin{equation}
	\label{eq:entropy_production_split}
	\begin{split}
		\Delta S^{a} = &\frac{k_{\rm B}}{D_{\rm h}} \Bigg( \iint \big( \bm{\mathcal{F}}^{a,{\rm only \ }\bm{w}^a}_{k=0}[\bm{w}^{a}] \big)\,{\rm d}\bm{r}\,{\rm d}\bm{w}^a +\iint \big((\partial_t \bm{w}^a) \circ \bm{\mathcal{F}}^{a,{\rm also \ }\bm{w}^b}_{k=0}[\bm{w}^{a},\bm{w}^{b}] \big)\,{\rm d}\bm{r}\,{\rm d}t \Bigg),
	\end{split}
\end{equation}
where we split $\mathcal{F}^a_{k=0}$ into a part that only explicitly depends on $\bm{w}^a$,
\begin{equation}
	\bm{\mathcal{F}}^{a,{\rm only \, }\bm{w}^a}_{k=0}[\bm{w}^{a}] = \big(g' - D_{\rm r} - \frac{g'^2}{2\,D_{\rm r}} (\bm{w}^a)^2\big) \, \bm{w}^a,
\end{equation}
and a second part that depends (also) on $\bm{w}^b$ with $b\neq a$,
\begin{equation}
	\begin{split}
		\bm{\mathcal{F}}^{a,{\rm also \, }\bm{w}^b}_{k=0}[\bm{w}^{a},\bm{w}^{b}] = &\ g_{ab}' \, \bm{w}^b - \frac{g'^2_{ab}}{2\,D_{\rm r}} (\bm{w}^b \cdot \bm{w}^b) \, \bm{w}^a  -\frac{g' \, g_{ab}'}{D_{\rm r}} \, (\bm{w}^a \cdot \bm{w}^b) \, \bm{w}^a.
	\end{split}
\end{equation}
The first part (involving $\bm{\mathcal{F}}^{a,{\rm only \, }\bm{w}^a}_{k=0}$) can be integrated directly and yields
\begin{equation}
	\begin{split}
		\Delta S_1 = &\frac{k_{\rm B}}{D_{\rm h}} \int \bigg( \frac{g'-D_{\rm r}}{2} \, (\bm{w}^a)^2 - \frac{g'^2}{8\,D_{\rm r}} \, (\bm{w}^a)^4 \bigg) \, {\rm d}\bm{r}.
	\end{split}
\end{equation}
For the entropy production \textit{rate} [see Eq.~\eqref{eq:entropy_production_rate_definition}], we take the limit $\tau \rightarrow \infty$. This yields $\Delta S_1/\tau \rightarrow 0$, since all moments of $\bm{w}^a$ are finite in the steady state. 

This leaves us with the second part of the r.h.s.~of Eq.~\eqref{eq:entropy_production_split}. Inserting Eq.~\eqref{eq:fluctuating_hydrodynamic_polarization_k-0} for $\partial_t\bm{w}^a$ yields $\Delta S_2 + \Delta S_3$, with 
\begin{equation}
	\begin{split}
		\Delta S_{2} = &\frac{k_{\rm B}}{D_{\rm h}} \iint	\bm{\mathcal{F}}^{a}_{k=0}	\cdot \bm{\mathcal{F}}^{a,{\rm also \, }\bm{w}^b}_{k=0} {\rm d}\bm{r}\,{\rm d}t
	\end{split}
\end{equation}
and
\begin{equation}
	\begin{split}
		\Delta S_{3} = &\frac{k_{\rm B}}{D_{\rm h}} \iint \tilde{\bm{\eta}}_a \circ \bm{\mathcal{F}}^{a,{\rm also \, }\bm{w}^b}_{k=0} \, {\rm d}\bm{r}\,{\rm d}t.
	\end{split}
\end{equation}
While $\Delta S_2$ is already independent of the noise and directly yields an entropy production rate upon division by $\tau$ and then taking the limit $\tau \rightarrow \infty$, $\Delta S_3$ needs to be calculated explicitly.

We transform the Stratonovich product in $\Delta S_{3}$ into an It\^o product, following the transformation rule \eqref{eq:entropy_production_conversion_Ito_Stratonovich_noise_function_main}. Since the It\^o product vanishes in the equal-time expectation, only the conversion contribution [second contribution on r.h.s.~of Eq.~\eqref{eq:entropy_production_conversion_Ito_Stratonovich_noise_function_main}] remains to be evaluated.

To calculate the conversion contributions, we consider Fourier space with the Fourier transform defined as
\begin{equation}
	\hat{f}(\bm{k}) = \int f(\bm{r}) \, {\rm e}^{-2\pi i \bm{r}\cdot \bm{k}} \, {\rm d}\bm{r}
\end{equation}
and inverse
\begin{equation}
	f(\bm{r}) = \int \hat{f}(\bm{k}) \, {\rm e}^{2\pi i \bm{r}\cdot \bm{k}} \, {\rm d}\bm{r} .
\end{equation}
In Fourier space the noise correlation $\langle \tilde{\eta}_{\alpha}(\bm{r},t) \, \tilde{\eta}_{\beta}(\bm{r}',t') \rangle$, given in Eq.~\eqref{eq:noise_correlation_field}, transforms into
\begin{equation}
	\begin{split}
		\frac{\langle \hat{\tilde{\eta}}_{\alpha}(\bm{k},t) \, \hat{\tilde{\eta}}_{\beta}(\bm{k}',t') \rangle}{2\,D_{\rm h}\, \delta_{\alpha\beta}\, \delta(t-t')} &= \iint \delta(\bm{r}-\bm{r}') \, {\rm e}^{-2\pi i (\bm{r} \cdot \bm{k} + \bm{r}' \cdot \bm{k}')} \, {\rm d}{\bm{r}}\, {\rm d}\bm{r}' \\
		&= \iint \, {\rm e}^{-2\pi i \bm{r} \cdot (\bm{k} + \bm{k}')}\, {\rm d}{\bm{r}}\\
		&= \delta(\bm{k} + \bm{k}') .
	\end{split}
\end{equation}

We now go through the individual terms in $\Delta S_3$ step-by-step. The first term is zero, i.e.,
\begin{equation}
	\begin{split}
		\Delta S_{3.1} = & \frac{k_{\rm B}}{D_{\rm h}} \, g_{ab}' \iint \bm{w}^b \circ \tilde{\bm{\eta}}_a \, {\rm d}\bm{r}\,{\rm d}t = 0,
	\end{split}
\end{equation}
because, here, the Stratonovich product coincides with the corresponding It\^o product. For the second term,
\begin{equation}
	\begin{split}
		\Delta S_{3.2} = & - \frac{k_{\rm B}}{D_{\rm h}} \, \frac{g_{ab}'^2}{2\,D_{\rm r}} \iint (\bm{w}^b \cdot \bm{w}^b) \, \bm{w}^a \circ \tilde{\bm{\eta}}_a \, {\rm d}\bm{r}\,{\rm d}t,
	\end{split}
\end{equation}
we assume that we can replace temporal averages by averages over noise realizations. For the entropy production \textit{rate}, this implies
\begin{equation}
	\begin{split}
		\Delta \dot{S}_{3.2} &= - \frac{k_{\rm B}}{D_{\rm h}} \, \frac{g_{ab}'^2}{2\,D_{\rm r}} \left \langle \int (\bm{w}^b \cdot \bm{w}^b) \, \bm{w}^a \circ \tilde{\bm{\eta}}_a \, {\rm d}\bm{r} \right \rangle\\
		& = - k_{\rm B} \, \frac{g_{ab}'^2}{D_{\rm r}}  \, \delta(\bm{0})  \left \langle \int (\bm{w}^b \cdot \bm{w}^b) \, {\rm d}\bm{r} \right \rangle
	\end{split}
\end{equation}
Note that the two dimensions lead to the factor of $2$. The $\delta(\bm{0})$ appears due to the equal-position product [i.e., $\bm{r} = \bm{r}'$ in Eq.~\eqref{eq:entropy_production_conversion_Ito_Stratonovich_noise_function_main}].
The third term is 
\begin{equation}
	\begin{split}
		\Delta S_{3.3} & = - \frac{k_{\rm B}}{D_{\rm h}} \, \frac{g' \, g_{ab}'}{D_{\rm r}} \left\langle \int (\bm{w}^a \cdot \bm{w}^b) \, \bm{w}^a \circ \tilde{\bm{\eta}}_a \, {\rm d}\bm{r} \right\rangle \\
		& = - k_{\rm B} \, \frac{g' \, g_{ab}'}{D_{\rm r}} \, \delta(\bm{0}) \, \left\langle \int 3\, (\bm{w}^a \cdot \bm{w}^b)\, {\rm d}\bm{r} \right\rangle
	\end{split}
\end{equation}
This entropy production contribution depends on the system size due to the $\delta(\bm{0})= \int {\rm e}^{-2\pi i \bm{r}\cdot \bm{k}}\vert_{\bm{k}=\bm{0}}\,{\rm d}\bm{r} = L^2$.

The resulting entropy production rate reads
\begin{equation}
	\label{eq:entropy_production_rate_final_expression_k-0}
	\begin{split}
		\Delta \dot{S} =\, & k_{\rm B} \, \int \Bigg ( \frac{1}{D_{\rm h}} \left\langle 	\bm{\mathcal{F}}^{a}_{k=0}	\cdot \bm{\mathcal{F}}^{a,{\rm also \, }\bm{w}^b}_{k=0} \right\rangle - \frac{\delta(\bm{0})}{D_{\rm r}} \, \Big \langle g_{ab}'^2 \, (\bm{w}^b)^2 + 3\, g'\,g'_{ab} \, \bm{w}^a \cdot \bm{w}^b  \Big \rangle \Bigg)\,{\rm d}\bm{r} .
	\end{split}
\end{equation}
This result can, in principle, be used to investigate analytically the scaling of the entropy production with the noise strength $D_{\rm h}$, as done, e.g., for phase separating \cite{nardini_cates_2017_entropy_production_field_theory}, flocking \cite{borthne_cates_2020_time-reversal_symmetry_violations_entropy_production_field_theories_polar_active_matter,dadhichi_ramaswamy_2018_origins_nonequilibrium_character_active_systems}, and non-reciprocal scalar systems \cite{suchanek_loos_2023_time-reversal_PT_symmetry_breaking_non-Hermitian_field_theories}, to quantify the nature of time-reversal symmetry breaking \cite{fodor_cates_2022_irreversibility_active_matter}. To this end, one would need to define proper ground state trajectories to obtain meaningful scalings regarding the transition to, e.g., chiral states.

\section{Full entropy production rate at arbitrary wavelengths from numerical continuum simulations}
\label{app:total_continuum_entropy_production_rate}
The entropy production rate associated with infinite-wavelength polarization fluctuations considered in the main text allows for analytical insights. However, to obtain a more complete picture, one can also consider the entropy production rate derived from the full continuum description, as given in Appendix~\ref{app:continuum_model_fluctuating}, which involves both the particle density and polarization density.

The dynamical action of the trajectory $\vec{\bm{\varphi}} = \{(\rho^A(t), \rho^B(t),\bm{w}^A(t), \bm{w}^B(t)) \vert t \in [t_0,\tau]\}$ contains contributions from both the densities and polarization densities of both species, i.e.,
\begin{equation}
	\mathcal{A}[\vec{\bm{\varphi}}] = \mathcal{A}_{\rho^A} + \mathcal{A}_{\rho^B} + \mathcal{A}_{\bm{w}^A} + \mathcal{A}_{\bm{w}^B} .
\end{equation}
The density contributions are given by \cite{nardini_cates_2017_entropy_production_field_theory}
\begin{equation}
	\begin{split}
		\mathcal{A}_{\rho^a} =& \frac{1}{4\,D_{\rm h}} \iint \Big (\nabla^{-1} (\partial_t \rho^a + \nabla \cdot \bm{j}^a[\rho^a, \rho^b,\bm{w}^a]) \Big )^2 \, {\rm d}\bm{r}\,{\rm d}t + \mathcal{A}_{\rho^a}^{\rm conv},
	\end{split}
\end{equation}
where $\nabla^{-1} X = \nabla \, \nabla^{-2} X$ for a scalar $X$, and $\nabla^{-2}$ is the functional
inverse to the Laplacian \cite{nardini_cates_2017_entropy_production_field_theory} \footnote{In two dimensions: $\nabla^{-2} f(\bm{r}) = \int_{\mathbb{R}^2} \frac{1}{2\,\pi} {\rm ln}(\vert \bm{r} - \bm{r}' \vert) \, f(\bm{r}') \, {\rm d}^2\bm{r}'$.}. The polarization contributions are, similarly to what is discussed in the main text,
\begin{equation}
	\begin{split}
		\mathcal{A}_{\bm{w}^a} =& \frac{1}{4\,D_{\rm h}} \iint \Big (\partial_t \bm{w}^a - \bm{\mathcal{F}}^a[\rho^a, \rho^b, \bm{w}^a, \bm{w}^b] \Big )^2 \, {\rm d}\bm{r}\,{\rm d}t + \mathcal{A}_{\bm{w}^a}^{\rm conv}.
	\end{split}
\end{equation}
For Stratonovich discretization, the terms that depend explicitly on the choice of time discretization with respect to noise, are
\begin{equation}
	\mathcal{A}_{\rho^a}^{\rm conv} = -\dfrac{1}{2} \iint \frac{\delta \nabla \cdot \bm{j}_a[\rho^a, \rho^b,\bm{w}^a]}{\delta \rho^a} \, {\rm d}\bm{r}\,{\rm d}t
\end{equation}
and
\begin{equation}
	\mathcal{A}_{\bm{w}^a}^{\rm conv} = \dfrac{1}{2} \iint \frac{\delta \bm{\mathcal{F}}^a[\rho^a, \rho^b,\bm{w}^a,\bm{w}^b]}{\delta \bm{w}^a} \, {\rm d}\bm{r}\,{\rm d}t.
\end{equation}

To calculate the entropy production rate as the ratio between forward and backward path probabilities, we must consider the time-reversed density, which is even under time-reversal, i.e.,
\begin{equation}
	\rho^{a\dagger}(\bm{r},t) = \rho^a(\bm{r},\tau + t_0 - t)
\end{equation}
and the time-reversed polarization density, which we consider odd under time-reversal, i.e.,
\begin{equation}
	\bm{w}^{a\dagger}(\bm{r},t) = - \bm{w}^{a}(\bm{r}, \tau + t_0 - t) .
\end{equation}
The functional $\bm{\mathcal{F}}^a[\rho^a, \rho^b,\bm{w}^a,\bm{w}^b]$, which governs the time evolution of $\bm{w}^a$, can be split into time-symmetric and time-anti-symmetric contributions, which satisfy
\begin{equation}
	\bm{\mathcal{F}}^{a}_{\rm sym}[\rho^{a\dagger}, \rho^{b\dagger},\bm{w}^{a\dagger},\bm{w}^{b\dagger}] = \bm{\mathcal{F}}^{a}_{\rm sym}[\rho^{a}, \rho^{b},\bm{w}^{a},\bm{w}^{b}]
\end{equation}
and
\begin{equation}
	\bm{\mathcal{F}}^{a}_{\rm asym}[\rho^{a\dagger}, \rho^{b\dagger},\bm{w}^{a\dagger},\bm{w}^{b\dagger}] = - \bm{\mathcal{F}}^{a}_{\rm asym}[\rho^{a}, \rho^{b},\bm{w}^{a},\bm{w}^{b}] .
\end{equation}
They are given by
\begin{equation}
	\begin{split}
		\bm{\mathcal{F}}^a_{\rm sym}[\rho^a, \rho^b, \bm{w}^a, \bm{w}^b] =& - \frac{1}{2} \, \nabla \,\big(v^{\rm eff}(\rho)\, \rho^a\big)+ \sum_b \frac{g'_{ab}}{8 \,D_{\rm r}} \Big[ \bm{w}^b \cdot \nabla \big(v^{\rm eff}(\rho)\, \bm{w}^a\big) + \bm{w}^{b*} \cdot \nabla \big(v^{\rm eff}(\rho)\, \bm{w}^{a*}\big) \\
		&  -2 \, \Big\{ v^{\rm eff}(\rho)\,\bm{w}^a \cdot \nabla \bm{w}^b   +\bm{w}^b \, \nabla \cdot \big(v^{\rm eff}(\rho)\,\bm{w}^a\big) - v^{\rm eff}(\rho)\, \bm{w}^{a*} \cdot \nabla \bm{w}^{b*} - \bm{w}^{b*}\,  \nabla \cdot \big( v^{\rm eff}(\rho)\, \bm{w}^{a*}\big) \Big \}  \Big] ,
	\end{split}
\end{equation}
and
\begin{equation}
	\begin{split}
		\bm{\mathcal{F}}^a_{\rm asym}[\rho^a, \rho^b, \bm{w}^a, \bm{w}^b] =& - D_{\rm r }\, \bm{w}^{a} + \sum_b g'_{ab} \, \rho^a\, \bm{w}^b +  D_{\rm t}\,\nabla^2\,\bm{w}^a + \frac{v^{\rm eff}(\rho)}{16\,D_{\rm r}} \, \nabla^2\,\Big(v^{\rm eff}(\rho)\,\bm{w}^a \Big) \\
		&- \sum_{b,c} \frac{ g'_{ab}\,g'_{ac}}{2\,D_{\rm r}} \, \bm{w}^a \, (\bm{w}^b \cdot \bm{w}^c) - \frac{z}{16\,D_{\rm r}} \, \nabla\rho \cdot \big[\nabla \big(v^{\rm eff}(\rho)\,\bm{w}^a\big) - \nabla^* \big(v^{\rm eff}(\rho)\,\bm{w}^{a*} \big) \big].
	\end{split}
\end{equation}
The action of the time-reversed trajectory is given by
\begin{equation}
	\mathcal{A}[\vec{\bm{\varphi}}'^{\dagger}] = \overleftarrow{\mathcal{A}}_{\rho^A} + \overleftarrow{\mathcal{A}}_{\rho^B} + \overleftarrow{\mathcal{A}}_{\bm{w}^A} + \overleftarrow{\mathcal{A}}_{\bm{w}^B}
\end{equation}
with
\begin{equation}
	\begin{split}
		\overleftarrow{\mathcal{A}}_{\rho^a} = &\iint \Big (\nabla^{-1} (\partial_t \rho^{a\dagger} + \nabla \cdot \bm{j}^a[\rho^{a\dagger}, \rho^{b\dagger},\bm{w}^{a\dagger}]) \Big )^2 \, {\rm d}\bm{r}\,{\rm d}t + \overleftarrow{\mathcal{A}}_{\rho^a}^{\rm conv},
	\end{split}
\end{equation}
where
\begin{equation}
	\overleftarrow{\mathcal{A}}_{\rho^a}^{\rm conv} = -\dfrac{1}{2} \iint \frac{\delta \nabla \cdot \bm{j}_a[\rho^{a\dagger}, \rho^{b\dagger},\bm{w}^{a\dagger}]}{\delta \rho^{a\dagger}} \, {\rm d}\bm{r}\,{\rm d}t,
\end{equation}
and
\begin{equation}
	\begin{split}
		\overleftarrow{\mathcal{A}}_{\bm{w}^a} = &\frac{1}{4\,D_{\rm h}} \iint \big(\partial_t \bm{w}^{a\dagger} - \bm{\mathcal{F}}^{a}[\rho^{a\dagger}, \rho^{b\dagger},\bm{w}^{a\dagger},\bm{w}^{b\dagger}] \big)^2 \, {\rm d}\bm{r}\,{\rm d}t + \overleftarrow{\mathcal{A}}_{\bm{w}^a}^{\rm conv},
	\end{split}
\end{equation}
where 
\begin{equation}
	\overleftarrow{\mathcal{A}}_{\bm{w}^a}^{\rm conv} = \dfrac{1}{2} \iint \frac{\delta \bm{\mathcal{F}}^a[\rho^{a\dagger}, \rho^{b\dagger},\bm{w}^{a\dagger},\bm{w}^{b\dagger}]}{\delta \bm{w}^{a\dagger}} \, {\rm d}\bm{r}\,{\rm d}t.
\end{equation}

Hence, the entropy production related to the density and polarization dynamics of species $a$ are
\begin{equation}
	\begin{split}
		\Delta S^a_{\rho^a} =& k_{\rm B} \, \left(\overleftarrow{\mathcal{A}}_{\rho^a} - \mathcal{A}_{\rho^a} \right)\\
		=& -\frac{k_{\rm B}}{D_{\rm h}} \iint \Big( \nabla^{-2} \Big (-(\partial_t \rho^a) \circ (D_{\rm t}\,\nabla^2\rho^a)\Big ) - (v^{\rm eff}(\rho) \, \bm{w}^a))\cdot (D_{\rm t}\,\nabla \rho^a) \Big)\, {\rm d}\bm{r} \, {\rm d}t\\
		& + \iint \frac{\delta \nabla \cdot (v^{\rm eff}(\rho) \, \bm{w}^a)}{\delta \rho^a} \, {\rm d}\bm{r}\,{\rm d}t
	\end{split}
\end{equation}
and
\begin{equation}
	\label{eq:full_entropy_production_polarization_density}
	\begin{split}
		\Delta S^a_{\bm{w}^a} &= k_{\rm B} \, \left(\overleftarrow{\mathcal{A}}_{\bm{w}^a} - \mathcal{A}_{\bm{w}^a} \right) \\
		&= \frac{k_{\rm B}}{D_{\rm h}} \iint \Big ((\partial_t \bm{w}^a) \circ \bm{\mathcal{F}}^{a}_{\rm asym} - \bm{\mathcal{F}}^{a}_{\rm asym} \cdot \bm{\mathcal{F}}^{a}_{\rm sym} \Big ) \, {\rm d}\bm{r} \, {\rm d}t + \iint \frac{\delta \bm{\mathcal{F}}_{\rm sym}^a}{\delta \bm{w}^a} {\rm d}\bm{r}\,{\rm d}t,
	\end{split}
\end{equation}
respectively. The total entropy production \textit{rate}, $\Delta \dot{S}^a = \Delta \dot{S}^a_{\rho^a} + \Delta \dot{S}^a_{\bm{w}^a}$, is then calculated as the time-averaged mean, defined in Eq.~\eqref{eq:entropy_production_rate_definition}.

Following the line of calculations shown in Appendix \ref{app:entropy_production_k-0} for the (simpler) infinite-wavelength limit, where density dynamics and gradient terms are neglected, one can, in principle, obtain an analytical expression for the full entropy production rate involving spatially varying density and polarization fields. However, as already indicated by the final expression in Appendix \ref{app:entropy_production_k-0}, given in Eq.~\eqref{eq:entropy_production_rate_final_expression_k-0}, the resulting formula is lengthy, depends on the system size, and cannot be easily interpreted to yield meaningful insights. 

Therefore, we turn to numerical continuum simulations (for details on the continuum simulation methods, see Appendix \ref{sapp:numerical_methods_continuum}). Fig.~\ref{fig:entropy_production_continuum_fixed_dif} shows the entropy production rate obtained from continuum simulations of the full fluctuating hydrodynamic Eqs.~\eqref{eq:continuum_eq_density_fluctuating}-\eqref{eq:continuum_eq_polarization_density_fluctuating} for the case $g_{AB} = g_{BA} - d$ with $d=5$ and $d=-7$. Generally, these results are not expected to quantitatively match the particle-level results presented in the main text. One reason is the choice of the noise strength $D_{\rm h}$, which lacks a clear microscopic counterpart and is set, ad hoc, equal for both particle density and polarization density. Nevertheless, the general dependence of the entropy production rate on the coupling strength shows qualitative similarities to the one obtained from particle simulations shown in Fig.~\ref{fig:BD_pol_sus_entropy_results_fixed_dif-5_species}. 

In particular, for $d=5$ [Fig.~\ref{fig:entropy_production_continuum_fixed_dif}(a)], the total entropy production rate in the continuum system (as the sum of $A$- and $B$-species contributions) also exhibits peaks at the exceptional points. As suggested by the prefactors of the susceptibilities in the analytical expression \eqref{eq:entropy_production_rate_k-0_lin_susceptibilities} for the entropy production rate (see Appendix \ref{app:prefactors_function_k-0}), the $A$-species peak occurs near the exceptional point $g_{AB}^{{\rm EP}<}$ (with $g_{AB}^{{\rm EP}<} < g_{AB}^{{\rm EP}>}$), while the $B$-species peak appears near $g_{AB}^{{\rm EP}>}$. This behavior is consistent with the particle simulation results (in Fig.~\ref{fig:BD_pol_sus_entropy_results_fixed_dif-5_species}). 

An apparent difference between the continuum and particle results concerns the translational and orientational contributions to the entropy production rate [Fig.~\ref{fig:entropy_production_continuum_fixed_dif}(b)]. While at the particle level, the translational dynamics significantly contribute to the entropy production rate in the antiflocking regime, at the continuum level (with equal noise strengths for particle density and polarization density), the entropy production is almost exclusively governed by the polarization density dynamics within all regimes. However, these two levels of description are not directly comparable: on the particle level, the orientational contributions [Eq.~\eqref{eq:entropy_production_rate_contribution_alignment}] account solely for particle orientations, whereas the polarization density field combines the director (unit vector) and the particle density, thus blending different aspects [in Eq.~\eqref{eq:full_entropy_production_polarization_density}].

To demonstrate that this behavior is consistent for different $d$, we show the entropy production rate of the continuum system for $g_{AB}=g_{BA}+7$ in Fig.~\ref{fig:entropy_production_continuum_fixed_dif}(c) and (d). This non-reciprocal system also exhibits peaks near the exceptional points, with the small shift of the peak at $g_{AB}^{{\rm EP}<}$ observed at both the particle and continuum levels. Notably, as for the particle system (Fig.~\ref{fig:entropy_production_particle_fixed_dif--7_species}), the peaks of $A$- and $B$-species entropy production are reversed compared to the case of $g_{AB} = g_{BA} - 5$.

\begin{figure}
	\includegraphics[width=1\linewidth]{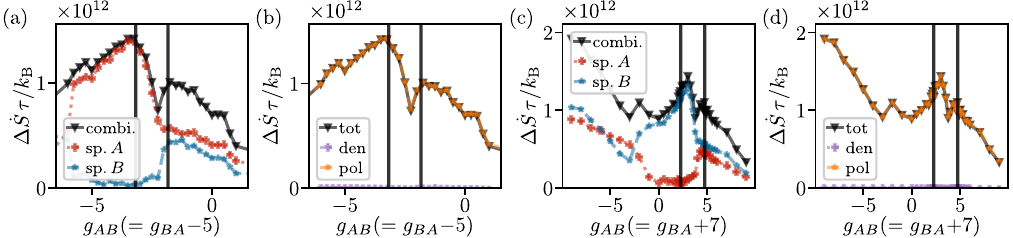}
	\caption{\label{fig:entropy_production_continuum_fixed_dif}Continuum-level entropy production rate of non-reciprocal systems obtained from continuum simulations with (a),(b) $g_{AB} = g_{BA} - 5$ and (c),(d) $g_{AB} = g_{BA} + 7$. In (a),(c), the contributions from individual species are shown in red and blue, while the black data points indicates their sum. In (b),(d), the contributions from the polarization and density are shown in orange and purple, respectively, while their sum is shown in black. Black vertical lines indicate parameters associated with exceptional points. The data shows simulation results between $330\,\tau$ and $350\,\tau$ after initialization.}
\end{figure}

An interesting question is how the noise amplitude, or different noise strengths in the density and polarization dynamics, affect the entropy production at the field level. We leave this question for further investigation.

%\bibliography{bibliography}% Produces the bibliography via BibTeX.

\begin{thebibliography}{83}%
\makeatletter
\providecommand \@ifxundefined [1]{%
 \@ifx{#1\undefined}
}%
\providecommand \@ifnum [1]{%
 \ifnum #1\expandafter \@firstoftwo
 \else \expandafter \@secondoftwo
 \fi
}%
\providecommand \@ifx [1]{%
 \ifx #1\expandafter \@firstoftwo
 \else \expandafter \@secondoftwo
 \fi
}%
\providecommand \natexlab [1]{#1}%
\providecommand \enquote  [1]{``#1''}%
\providecommand \bibnamefont  [1]{#1}%
\providecommand \bibfnamefont [1]{#1}%
\providecommand \citenamefont [1]{#1}%
\providecommand \href@noop [0]{\@secondoftwo}%
\providecommand \href [0]{\begingroup \@sanitize@url \@href}%
\providecommand \@href[1]{\@@startlink{#1}\@@href}%
\providecommand \@@href[1]{\endgroup#1\@@endlink}%
\providecommand \@sanitize@url [0]{\catcode `\\12\catcode `\$12\catcode
  `\&12\catcode `\#12\catcode `\^12\catcode `\_12\catcode `\%12\relax}%
\providecommand \@@startlink[1]{}%
\providecommand \@@endlink[0]{}%
\providecommand \url  [0]{\begingroup\@sanitize@url \@url }%
\providecommand \@url [1]{\endgroup\@href {#1}{\urlprefix }}%
\providecommand \urlprefix  [0]{URL }%
\providecommand \Eprint [0]{\href }%
\providecommand \doibase [0]{http://dx.doi.org/}%
\providecommand \selectlanguage [0]{\@gobble}%
\providecommand \bibinfo  [0]{\@secondoftwo}%
\providecommand \bibfield  [0]{\@secondoftwo}%
\providecommand \translation [1]{[#1]}%
\providecommand \BibitemOpen [0]{}%
\providecommand \bibitemStop [0]{}%
\providecommand \bibitemNoStop [0]{.\EOS\space}%
\providecommand \EOS [0]{\spacefactor3000\relax}%
\providecommand \BibitemShut  [1]{\csname bibitem#1\endcsname}%
\let\auto@bib@innerbib\@empty
%</preamble>
\bibitem [{\citenamefont
  {Seifert}(2012)}]{seifert_2012_stochastic_thermodynamics}%
  \BibitemOpen
  \bibfield  {author} {\bibinfo {author} {\bibfnamefont {Udo}\ \bibnamefont
  {Seifert}},\ }\bibfield  {title} {\enquote {\bibinfo {title} {Stochastic
  thermodynamics, fluctuation theorems and molecular machines},}\ }\href@noop
  {} {\bibfield  {journal} {\bibinfo  {journal} {Rep. Prog. Phys.}\ }\textbf
  {\bibinfo {volume} {75}},\ \bibinfo {pages} {126001} (\bibinfo {year}
  {2012})}\BibitemShut {NoStop}%
\bibitem [{\citenamefont {Peliti}\ and\ \citenamefont
  {Pigolotti}(2021)}]{peliti_pigolotti_2021_stochastic_thermodynamics_introduction}%
  \BibitemOpen
  \bibfield  {author} {\bibinfo {author} {\bibfnamefont {Luca}\ \bibnamefont
  {Peliti}}\ and\ \bibinfo {author} {\bibfnamefont {Simone}\ \bibnamefont
  {Pigolotti}},\ }\href@noop {} {\emph {\bibinfo {title} {Stochastic
  thermodynamics: an introduction}}}\ (\bibinfo  {publisher} {Princeton
  University Press},\ \bibinfo {year} {2021})\BibitemShut {NoStop}%
\bibitem [{\citenamefont {Fodor}\ \emph {et~al.}(2022)\citenamefont {Fodor},
  \citenamefont {Jack},\ and\ \citenamefont
  {Cates}}]{fodor_cates_2022_irreversibility_active_matter}%
  \BibitemOpen
  \bibfield  {author} {\bibinfo {author} {\bibfnamefont {{\'E}tienne}\
  \bibnamefont {Fodor}}, \bibinfo {author} {\bibfnamefont {Robert~L}\
  \bibnamefont {Jack}}, \ and\ \bibinfo {author} {\bibfnamefont {Michael~E}\
  \bibnamefont {Cates}},\ }\bibfield  {title} {\enquote {\bibinfo {title}
  {Irreversibility and biased ensembles in active matter: Insights from
  stochastic thermodynamics},}\ }\href@noop {} {\bibfield  {journal} {\bibinfo
  {journal} {Annu. Rev. Condens. Matter Phys.}\ }\textbf {\bibinfo {volume}
  {13}},\ \bibinfo {pages} {215--238} (\bibinfo {year} {2022})}\BibitemShut
  {NoStop}%
\bibitem [{\citenamefont {O’Byrne}\ \emph {et~al.}(2022)\citenamefont
  {O’Byrne}, \citenamefont {Kafri}, \citenamefont {Tailleur},\ and\
  \citenamefont {van
  Wijland}}]{OByrne_vanWijland_2022_time_irreversibility_active_matter}%
  \BibitemOpen
  \bibfield  {author} {\bibinfo {author} {\bibfnamefont {J{\'e}r{\'e}my}\
  \bibnamefont {O’Byrne}}, \bibinfo {author} {\bibfnamefont {Yariv}\
  \bibnamefont {Kafri}}, \bibinfo {author} {\bibfnamefont {Julien}\
  \bibnamefont {Tailleur}}, \ and\ \bibinfo {author} {\bibfnamefont
  {Fr{\'e}d{\'e}ric}\ \bibnamefont {van Wijland}},\ }\bibfield  {title}
  {\enquote {\bibinfo {title} {Time irreversibility in active matter, from
  micro to macro},}\ }\href@noop {} {\bibfield  {journal} {\bibinfo  {journal}
  {Nature Reviews Physics}\ }\textbf {\bibinfo {volume} {4}},\ \bibinfo {pages}
  {167--183} (\bibinfo {year} {2022})}\BibitemShut {NoStop}%
\bibitem [{\citenamefont {Dabelow}\ \emph {et~al.}(2019)\citenamefont
  {Dabelow}, \citenamefont {Bo},\ and\ \citenamefont
  {Eichhorn}}]{dabelow_eichhorn_2019_irreversibility_active_matter}%
  \BibitemOpen
  \bibfield  {author} {\bibinfo {author} {\bibfnamefont {Lennart}\ \bibnamefont
  {Dabelow}}, \bibinfo {author} {\bibfnamefont {Stefano}\ \bibnamefont {Bo}}, \
  and\ \bibinfo {author} {\bibfnamefont {Ralf}\ \bibnamefont {Eichhorn}},\
  }\bibfield  {title} {\enquote {\bibinfo {title} {Irreversibility in active
  matter systems: Fluctuation theorem and mutual information},}\ }\href@noop {}
  {\bibfield  {journal} {\bibinfo  {journal} {Phys. Rev. X}\ }\textbf {\bibinfo
  {volume} {9}},\ \bibinfo {pages} {021009} (\bibinfo {year}
  {2019})}\BibitemShut {NoStop}%
\bibitem [{\citenamefont {Pietzonka}\ \emph {et~al.}(2019)\citenamefont
  {Pietzonka}, \citenamefont {Fodor}, \citenamefont {Lohrmann}, \citenamefont
  {Cates},\ and\ \citenamefont
  {Seifert}}]{pietzonka_seifert_2019_autonomous_engines_driven_active_matter}%
  \BibitemOpen
  \bibfield  {author} {\bibinfo {author} {\bibfnamefont {Patrick}\ \bibnamefont
  {Pietzonka}}, \bibinfo {author} {\bibfnamefont {{\'E}tienne}\ \bibnamefont
  {Fodor}}, \bibinfo {author} {\bibfnamefont {Christoph}\ \bibnamefont
  {Lohrmann}}, \bibinfo {author} {\bibfnamefont {Michael~E}\ \bibnamefont
  {Cates}}, \ and\ \bibinfo {author} {\bibfnamefont {Udo}\ \bibnamefont
  {Seifert}},\ }\bibfield  {title} {\enquote {\bibinfo {title} {Autonomous
  engines driven by active matter: Energetics and design principles},}\
  }\href@noop {} {\bibfield  {journal} {\bibinfo  {journal} {Phys. Rev. X}\
  }\textbf {\bibinfo {volume} {9}},\ \bibinfo {pages} {041032} (\bibinfo {year}
  {2019})}\BibitemShut {NoStop}%
\bibitem [{\citenamefont {Bebon}\ \emph {et~al.}(2025)\citenamefont {Bebon},
  \citenamefont {Robinson},\ and\ \citenamefont
  {Speck}}]{bebon_speck_2025_thermodynamics_active_matter_across_scales}%
  \BibitemOpen
  \bibfield  {author} {\bibinfo {author} {\bibfnamefont {Robin}\ \bibnamefont
  {Bebon}}, \bibinfo {author} {\bibfnamefont {Joshua~F}\ \bibnamefont
  {Robinson}}, \ and\ \bibinfo {author} {\bibfnamefont {Thomas}\ \bibnamefont
  {Speck}},\ }\bibfield  {title} {\enquote {\bibinfo {title} {Thermodynamics of
  active matter: Tracking dissipation across scales},}\ }\href@noop {}
  {\bibfield  {journal} {\bibinfo  {journal} {Phys. Rev. X}\ }\textbf {\bibinfo
  {volume} {15}},\ \bibinfo {pages} {021050} (\bibinfo {year}
  {2025})}\BibitemShut {NoStop}%
\bibitem [{\citenamefont {Loos}\ and\ \citenamefont
  {Klapp}(2020)}]{loos_klapp_2020_irreversibility_heat_non-reciprocal_interactions}%
  \BibitemOpen
  \bibfield  {author} {\bibinfo {author} {\bibfnamefont {Sarah A~M}\
  \bibnamefont {Loos}}\ and\ \bibinfo {author} {\bibfnamefont {Sabine H~L}\
  \bibnamefont {Klapp}},\ }\bibfield  {title} {\enquote {\bibinfo {title}
  {Irreversibility, heat and information flows induced by non-reciprocal
  interactions},}\ }\href@noop {} {\bibfield  {journal} {\bibinfo  {journal}
  {New J. Phys.}\ }\textbf {\bibinfo {volume} {22}},\ \bibinfo {pages} {123051}
  (\bibinfo {year} {2020})}\BibitemShut {NoStop}%
\bibitem [{\citenamefont {Loos}\ \emph {et~al.}(2021)\citenamefont {Loos},
  \citenamefont {Hermann},\ and\ \citenamefont
  {Klapp}}]{loos_klapp_2021_medium_entropy_recution_delay}%
  \BibitemOpen
  \bibfield  {author} {\bibinfo {author} {\bibfnamefont {Sarah A~M}\
  \bibnamefont {Loos}}, \bibinfo {author} {\bibfnamefont {Simon}\ \bibnamefont
  {Hermann}}, \ and\ \bibinfo {author} {\bibfnamefont {Sabine H~L}\
  \bibnamefont {Klapp}},\ }\bibfield  {title} {\enquote {\bibinfo {title}
  {Medium entropy reduction and instability in stochastic systems with
  distributed delay},}\ }\href@noop {} {\bibfield  {journal} {\bibinfo
  {journal} {Entropy}\ }\textbf {\bibinfo {volume} {23}},\ \bibinfo {pages}
  {696} (\bibinfo {year} {2021})}\BibitemShut {NoStop}%
\bibitem [{\citenamefont {Saha}\ \emph {et~al.}(2020)\citenamefont {Saha},
  \citenamefont {Agudo-Canalejo},\ and\ \citenamefont
  {Golestanian}}]{saha_scalar_active_mixtures_2020}%
  \BibitemOpen
  \bibfield  {author} {\bibinfo {author} {\bibfnamefont {Suropriya}\
  \bibnamefont {Saha}}, \bibinfo {author} {\bibfnamefont {Jaime}\ \bibnamefont
  {Agudo-Canalejo}}, \ and\ \bibinfo {author} {\bibfnamefont {Ramin}\
  \bibnamefont {Golestanian}},\ }\bibfield  {title} {\enquote {\bibinfo {title}
  {Scalar active mixtures: The nonreciprocal {C}ahn-{H}illiard model},}\
  }\href@noop {} {\bibfield  {journal} {\bibinfo  {journal} {Phys. Rev. X}\
  }\textbf {\bibinfo {volume} {10}},\ \bibinfo {pages} {041009} (\bibinfo
  {year} {2020})}\BibitemShut {NoStop}%
\bibitem [{\citenamefont {Ivlev}\ \emph {et~al.}(2015)\citenamefont {Ivlev},
  \citenamefont {Bartnick}, \citenamefont {Heinen}, \citenamefont {Du},
  \citenamefont {Nosenko},\ and\ \citenamefont
  {L{\"o}wen}}]{Ivlev_2015_statistical_mechanics_where_newtons_third_law_is_broken}%
  \BibitemOpen
  \bibfield  {author} {\bibinfo {author} {\bibfnamefont {Alexei~V}\
  \bibnamefont {Ivlev}}, \bibinfo {author} {\bibfnamefont {J{\"o}rg}\
  \bibnamefont {Bartnick}}, \bibinfo {author} {\bibfnamefont {Marco}\
  \bibnamefont {Heinen}}, \bibinfo {author} {\bibfnamefont {C-R}\ \bibnamefont
  {Du}}, \bibinfo {author} {\bibfnamefont {V}~\bibnamefont {Nosenko}}, \ and\
  \bibinfo {author} {\bibfnamefont {Hartmut}\ \bibnamefont {L{\"o}wen}},\
  }\bibfield  {title} {\enquote {\bibinfo {title} {Statistical mechanics where
  {N}ewton’s third law is broken},}\ }\href@noop {} {\bibfield  {journal}
  {\bibinfo  {journal} {Phys. Rev. X}\ }\textbf {\bibinfo {volume} {5}},\
  \bibinfo {pages} {011035} (\bibinfo {year} {2015})}\BibitemShut {NoStop}%
\bibitem [{\citenamefont {Scheibner}\ \emph {et~al.}(2020)\citenamefont
  {Scheibner}, \citenamefont {Souslov}, \citenamefont {Banerjee}, \citenamefont
  {Sur{\'o}wka}, \citenamefont {Irvine},\ and\ \citenamefont
  {Vitelli}}]{scheibner_vitelli_2020_odd_elasticity}%
  \BibitemOpen
  \bibfield  {author} {\bibinfo {author} {\bibfnamefont {Colin}\ \bibnamefont
  {Scheibner}}, \bibinfo {author} {\bibfnamefont {Anton}\ \bibnamefont
  {Souslov}}, \bibinfo {author} {\bibfnamefont {Debarghya}\ \bibnamefont
  {Banerjee}}, \bibinfo {author} {\bibfnamefont {Piotr}\ \bibnamefont
  {Sur{\'o}wka}}, \bibinfo {author} {\bibfnamefont {William~TM}\ \bibnamefont
  {Irvine}}, \ and\ \bibinfo {author} {\bibfnamefont {Vincenzo}\ \bibnamefont
  {Vitelli}},\ }\bibfield  {title} {\enquote {\bibinfo {title} {Odd
  elasticity},}\ }\href@noop {} {\bibfield  {journal} {\bibinfo  {journal}
  {Nat. Phys.}\ }\textbf {\bibinfo {volume} {16}},\ \bibinfo {pages} {475--480}
  (\bibinfo {year} {2020})}\BibitemShut {NoStop}%
\bibitem [{\citenamefont {Bowick}\ \emph {et~al.}(2022)\citenamefont {Bowick},
  \citenamefont {Fakhri}, \citenamefont {Marchetti},\ and\ \citenamefont
  {Ramaswamy}}]{Bowick_2022_symmetry_thermodynamcs_topology_active_matter}%
  \BibitemOpen
  \bibfield  {author} {\bibinfo {author} {\bibfnamefont {Mark~J.}\ \bibnamefont
  {Bowick}}, \bibinfo {author} {\bibfnamefont {Nikta}\ \bibnamefont {Fakhri}},
  \bibinfo {author} {\bibfnamefont {M.~Cristina}\ \bibnamefont {Marchetti}}, \
  and\ \bibinfo {author} {\bibfnamefont {Sriram}\ \bibnamefont {Ramaswamy}},\
  }\bibfield  {title} {\enquote {\bibinfo {title} {Symmetry, thermodynamics,
  and topology in active matter},}\ }\href@noop {} {\bibfield  {journal}
  {\bibinfo  {journal} {Phys. Rev. X}\ }\textbf {\bibinfo {volume} {12}},\
  \bibinfo {pages} {010501} (\bibinfo {year} {2022})}\BibitemShut {NoStop}%
\bibitem [{\citenamefont {Lotka}(1920)}]{lotka_1920_lotka_volterra_model}%
  \BibitemOpen
  \bibfield  {author} {\bibinfo {author} {\bibfnamefont {Alfred~J}\
  \bibnamefont {Lotka}},\ }\bibfield  {title} {\enquote {\bibinfo {title}
  {Analytical note on certain rhythmic relations in organic systems},}\
  }\href@noop {} {\bibfield  {journal} {\bibinfo  {journal} {Proc. Natl. Acad.
  Sci. U.S.A.}\ }\textbf {\bibinfo {volume} {6}},\ \bibinfo {pages} {410--415}
  (\bibinfo {year} {1920})}\BibitemShut {NoStop}%
\bibitem [{\citenamefont
  {Volterra}(1926)}]{volterra_1926_lotka_volterra_model}%
  \BibitemOpen
  \bibfield  {author} {\bibinfo {author} {\bibfnamefont {Vito}\ \bibnamefont
  {Volterra}},\ }\bibfield  {title} {\enquote {\bibinfo {title} {Fluctuations
  in the abundance of a species considered mathematically},}\ }\href@noop {}
  {\bibfield  {journal} {\bibinfo  {journal} {Nature}\ }\textbf {\bibinfo
  {volume} {118}},\ \bibinfo {pages} {558--560} (\bibinfo {year}
  {1926})}\BibitemShut {NoStop}%
\bibitem [{\citenamefont {Xiong}\ \emph {et~al.}(2020)\citenamefont {Xiong},
  \citenamefont {Cao}, \citenamefont {Cooper}, \citenamefont {Rappel},
  \citenamefont {Hasty},\ and\ \citenamefont
  {Tsimring}}]{xiong_tsimring_2020_flower-like_patterns_multi-species_bacteria}%
  \BibitemOpen
  \bibfield  {author} {\bibinfo {author} {\bibfnamefont {Liyang}\ \bibnamefont
  {Xiong}}, \bibinfo {author} {\bibfnamefont {Yuansheng}\ \bibnamefont {Cao}},
  \bibinfo {author} {\bibfnamefont {Robert}\ \bibnamefont {Cooper}}, \bibinfo
  {author} {\bibfnamefont {Wouter-Jan}\ \bibnamefont {Rappel}}, \bibinfo
  {author} {\bibfnamefont {Jeff}\ \bibnamefont {Hasty}}, \ and\ \bibinfo
  {author} {\bibfnamefont {Lev}\ \bibnamefont {Tsimring}},\ }\bibfield  {title}
  {\enquote {\bibinfo {title} {Flower-like patterns in multi-species bacterial
  colonies},}\ }\href@noop {} {\bibfield  {journal} {\bibinfo  {journal}
  {{e}{L}ife}\ }\textbf {\bibinfo {volume} {9}},\ \bibinfo {pages} {e48885}
  (\bibinfo {year} {2020})}\BibitemShut {NoStop}%
\bibitem [{\citenamefont {Pruessner}\ and\ \citenamefont
  {Garcia-Millan}(2022)}]{pruessner_garcia-millan_2022_field_theories_active_matter_entropy_production}%
  \BibitemOpen
  \bibfield  {author} {\bibinfo {author} {\bibfnamefont {Gunnar}\ \bibnamefont
  {Pruessner}}\ and\ \bibinfo {author} {\bibfnamefont {Rosalba}\ \bibnamefont
  {Garcia-Millan}},\ }\bibfield  {title} {\enquote {\bibinfo {title} {Field
  theories of active particle systems and their entropy production},}\
  }\href@noop {} {\bibfield  {journal} {\bibinfo  {journal} {arXiv preprint
  arXiv:2211.11906}\ } (\bibinfo {year} {2022})}\BibitemShut {NoStop}%
\bibitem [{\citenamefont {Brossollet}\ and\ \citenamefont
  {Biroli}(2025)}]{brossollet_biroli_2025_entropy_production_field_theories_interacting_particles}%
  \BibitemOpen
  \bibfield  {author} {\bibinfo {author} {\bibfnamefont {Antonin}\ \bibnamefont
  {Brossollet}}\ and\ \bibinfo {author} {\bibfnamefont {Giulio}\ \bibnamefont
  {Biroli}},\ }\bibfield  {title} {\enquote {\bibinfo {title} {Entropy
  production from density field theories for interacting particles systems},}\
  }\href@noop {} {\bibfield  {journal} {\bibinfo  {journal} {arXiv preprint
  arXiv:2507.15131}\ } (\bibinfo {year} {2025})}\BibitemShut {NoStop}%
\bibitem [{\citenamefont {Nardini}\ \emph {et~al.}(2017)\citenamefont
  {Nardini}, \citenamefont {Fodor}, \citenamefont {Tjhung}, \citenamefont
  {Van~Wijland}, \citenamefont {Tailleur},\ and\ \citenamefont
  {Cates}}]{nardini_cates_2017_entropy_production_field_theory}%
  \BibitemOpen
  \bibfield  {author} {\bibinfo {author} {\bibfnamefont {Cesare}\ \bibnamefont
  {Nardini}}, \bibinfo {author} {\bibfnamefont {{\'E}tienne}\ \bibnamefont
  {Fodor}}, \bibinfo {author} {\bibfnamefont {Elsen}\ \bibnamefont {Tjhung}},
  \bibinfo {author} {\bibfnamefont {Fr{\'e}d{\'e}ric}\ \bibnamefont
  {Van~Wijland}}, \bibinfo {author} {\bibfnamefont {Julien}\ \bibnamefont
  {Tailleur}}, \ and\ \bibinfo {author} {\bibfnamefont {Michael~E}\
  \bibnamefont {Cates}},\ }\bibfield  {title} {\enquote {\bibinfo {title}
  {Entropy production in field theories without time-reversal symmetry:
  quantifying the non-equilibrium character of active matter},}\ }\href@noop {}
  {\bibfield  {journal} {\bibinfo  {journal} {Phys. Rev. X}\ }\textbf {\bibinfo
  {volume} {7}},\ \bibinfo {pages} {021007} (\bibinfo {year}
  {2017})}\BibitemShut {NoStop}%
\bibitem [{\citenamefont {Borthne}\ \emph {et~al.}(2020)\citenamefont
  {Borthne}, \citenamefont {Fodor},\ and\ \citenamefont
  {Cates}}]{borthne_cates_2020_time-reversal_symmetry_violations_entropy_production_field_theories_polar_active_matter}%
  \BibitemOpen
  \bibfield  {author} {\bibinfo {author} {\bibfnamefont {{\O}yvind~L}\
  \bibnamefont {Borthne}}, \bibinfo {author} {\bibfnamefont {{\'E}tienne}\
  \bibnamefont {Fodor}}, \ and\ \bibinfo {author} {\bibfnamefont {Michael~E}\
  \bibnamefont {Cates}},\ }\bibfield  {title} {\enquote {\bibinfo {title}
  {Time-reversal symmetry violations and entropy production in field theories
  of polar active matter},}\ }\href@noop {} {\bibfield  {journal} {\bibinfo
  {journal} {New J. Phys.}\ }\textbf {\bibinfo {volume} {22}},\ \bibinfo
  {pages} {123012} (\bibinfo {year} {2020})}\BibitemShut {NoStop}%
\bibitem [{\citenamefont {Ferretti}\ \emph {et~al.}(2022)\citenamefont
  {Ferretti}, \citenamefont {Grosse-Holz}, \citenamefont {Holmes},
  \citenamefont {Shivers}, \citenamefont {Giardina}, \citenamefont {Mora},\
  and\ \citenamefont
  {Walczak}}]{ferretti_walczak_2022_signatures_irreversibility_flocking}%
  \BibitemOpen
  \bibfield  {author} {\bibinfo {author} {\bibfnamefont {Federica}\
  \bibnamefont {Ferretti}}, \bibinfo {author} {\bibfnamefont {Simon}\
  \bibnamefont {Grosse-Holz}}, \bibinfo {author} {\bibfnamefont {Caroline}\
  \bibnamefont {Holmes}}, \bibinfo {author} {\bibfnamefont {Jordan~L}\
  \bibnamefont {Shivers}}, \bibinfo {author} {\bibfnamefont {Irene}\
  \bibnamefont {Giardina}}, \bibinfo {author} {\bibfnamefont {Thierry}\
  \bibnamefont {Mora}}, \ and\ \bibinfo {author} {\bibfnamefont {Aleksandra~M}\
  \bibnamefont {Walczak}},\ }\bibfield  {title} {\enquote {\bibinfo {title}
  {Signatures of irreversibility in microscopic models of flocking},}\
  }\href@noop {} {\bibfield  {journal} {\bibinfo  {journal} {Phys. Rev. E}\
  }\textbf {\bibinfo {volume} {106}},\ \bibinfo {pages} {034608} (\bibinfo
  {year} {2022})}\BibitemShut {NoStop}%
\bibitem [{\citenamefont {Proesmans}\ \emph {et~al.}(2025)\citenamefont
  {Proesmans}, \citenamefont {Falasco}, \citenamefont {Mohite}, \citenamefont
  {Esposito},\ and\ \citenamefont
  {Fodor}}]{proesmans_fodor_2025_quantifying_dissipation_flocking_dynamics}%
  \BibitemOpen
  \bibfield  {author} {\bibinfo {author} {\bibfnamefont {Karel}\ \bibnamefont
  {Proesmans}}, \bibinfo {author} {\bibfnamefont {Gianmaria}\ \bibnamefont
  {Falasco}}, \bibinfo {author} {\bibfnamefont {Atul~Tanaji}\ \bibnamefont
  {Mohite}}, \bibinfo {author} {\bibfnamefont {Massimiliano}\ \bibnamefont
  {Esposito}}, \ and\ \bibinfo {author} {\bibfnamefont {{\'E}tienne}\
  \bibnamefont {Fodor}},\ }\bibfield  {title} {\enquote {\bibinfo {title}
  {Quantifying dissipation in flocking dynamics: When tracking internal states
  matter},}\ }\href@noop {} {\bibfield  {journal} {\bibinfo  {journal} {arXiv
  preprint arXiv:2505.13113}\ } (\bibinfo {year} {2025})}\BibitemShut {NoStop}%
\bibitem [{\citenamefont {Martynec}\ \emph {et~al.}(2020)\citenamefont
  {Martynec}, \citenamefont {Klapp},\ and\ \citenamefont
  {Loos}}]{martynec_loos_2020_entropy_production_criticality_nonequilibrium_Potts}%
  \BibitemOpen
  \bibfield  {author} {\bibinfo {author} {\bibfnamefont {Thomas}\ \bibnamefont
  {Martynec}}, \bibinfo {author} {\bibfnamefont {Sabine~HL}\ \bibnamefont
  {Klapp}}, \ and\ \bibinfo {author} {\bibfnamefont {Sarah~AM}\ \bibnamefont
  {Loos}},\ }\bibfield  {title} {\enquote {\bibinfo {title} {Entropy production
  at criticality in a nonequilibrium potts model},}\ }\href@noop {} {\bibfield
  {journal} {\bibinfo  {journal} {New Journal of Physics}\ }\textbf {\bibinfo
  {volume} {22}},\ \bibinfo {pages} {093069} (\bibinfo {year}
  {2020})}\BibitemShut {NoStop}%
\bibitem [{\citenamefont {Herpich}\ \emph {et~al.}(2018)\citenamefont
  {Herpich}, \citenamefont {Thingna},\ and\ \citenamefont
  {Esposito}}]{herpich_esposito_2018_collective_power}%
  \BibitemOpen
  \bibfield  {author} {\bibinfo {author} {\bibfnamefont {Tim}\ \bibnamefont
  {Herpich}}, \bibinfo {author} {\bibfnamefont {Juzar}\ \bibnamefont
  {Thingna}}, \ and\ \bibinfo {author} {\bibfnamefont {Massimiliano}\
  \bibnamefont {Esposito}},\ }\bibfield  {title} {\enquote {\bibinfo {title}
  {Collective power: minimal model for thermodynamics of nonequilibrium phase
  transitions},}\ }\href@noop {} {\bibfield  {journal} {\bibinfo  {journal}
  {Phys. Rev. X}\ }\textbf {\bibinfo {volume} {8}},\ \bibinfo {pages} {031056}
  (\bibinfo {year} {2018})}\BibitemShut {NoStop}%
\bibitem [{\citenamefont {Meibohm}\ and\ \citenamefont
  {Esposito}(2024)}]{meibohm_esposito_2024_minimum-dissipation_principle_stochastic_oscillator}%
  \BibitemOpen
  \bibfield  {author} {\bibinfo {author} {\bibfnamefont {Jan}\ \bibnamefont
  {Meibohm}}\ and\ \bibinfo {author} {\bibfnamefont {Massimiliano}\
  \bibnamefont {Esposito}},\ }\bibfield  {title} {\enquote {\bibinfo {title}
  {Minimum-dissipation principle for synchronized stochastic oscillators far
  from equilibrium},}\ }\href@noop {} {\bibfield  {journal} {\bibinfo
  {journal} {Phys. Rev. E}\ }\textbf {\bibinfo {volume} {110}},\ \bibinfo
  {pages} {L042102} (\bibinfo {year} {2024})}\BibitemShut {NoStop}%
\bibitem [{\citenamefont {Seara}\ \emph {et~al.}(2021)\citenamefont {Seara},
  \citenamefont {Machta},\ and\ \citenamefont
  {Murrell}}]{seara_murrell_2021_irreversibility_dynamical_phases_transitions}%
  \BibitemOpen
  \bibfield  {author} {\bibinfo {author} {\bibfnamefont {Daniel~S}\
  \bibnamefont {Seara}}, \bibinfo {author} {\bibfnamefont {Benjamin~B}\
  \bibnamefont {Machta}}, \ and\ \bibinfo {author} {\bibfnamefont {Michael~P}\
  \bibnamefont {Murrell}},\ }\bibfield  {title} {\enquote {\bibinfo {title}
  {Irreversibility in dynamical phases and transitions},}\ }\href@noop {}
  {\bibfield  {journal} {\bibinfo  {journal} {Nature communications}\ }\textbf
  {\bibinfo {volume} {12}},\ \bibinfo {pages} {392} (\bibinfo {year}
  {2021})}\BibitemShut {NoStop}%
\bibitem [{\citenamefont {Dinelli}\ \emph {et~al.}(2023)\citenamefont
  {Dinelli}, \citenamefont {O’Byrne}, \citenamefont {Curatolo}, \citenamefont
  {Zhao}, \citenamefont {Sollich},\ and\ \citenamefont
  {Tailleur}}]{dinelli_tailleur_2023_non-reciprocity_across_scales}%
  \BibitemOpen
  \bibfield  {author} {\bibinfo {author} {\bibfnamefont {Alberto}\ \bibnamefont
  {Dinelli}}, \bibinfo {author} {\bibfnamefont {J{\'e}r{\'e}my}\ \bibnamefont
  {O’Byrne}}, \bibinfo {author} {\bibfnamefont {Agnese}\ \bibnamefont
  {Curatolo}}, \bibinfo {author} {\bibfnamefont {Yongfeng}\ \bibnamefont
  {Zhao}}, \bibinfo {author} {\bibfnamefont {Peter}\ \bibnamefont {Sollich}}, \
  and\ \bibinfo {author} {\bibfnamefont {Julien}\ \bibnamefont {Tailleur}},\
  }\bibfield  {title} {\enquote {\bibinfo {title} {Non-reciprocity across
  scales in active mixtures},}\ }\href@noop {} {\bibfield  {journal} {\bibinfo
  {journal} {Nat. Commun.}\ }\textbf {\bibinfo {volume} {14}},\ \bibinfo
  {pages} {7035} (\bibinfo {year} {2023})}\BibitemShut {NoStop}%
\bibitem [{\citenamefont {Alston}\ \emph {et~al.}(2023)\citenamefont {Alston},
  \citenamefont {Cocconi},\ and\ \citenamefont
  {Bertrand}}]{alston_bertrand_2023_irreversibility_non-reciprocal_PT-breaking}%
  \BibitemOpen
  \bibfield  {author} {\bibinfo {author} {\bibfnamefont {Henry}\ \bibnamefont
  {Alston}}, \bibinfo {author} {\bibfnamefont {Luca}\ \bibnamefont {Cocconi}},
  \ and\ \bibinfo {author} {\bibfnamefont {Thibault}\ \bibnamefont
  {Bertrand}},\ }\bibfield  {title} {\enquote {\bibinfo {title}
  {Irreversibility across a nonreciprocal pt-symmetry-breaking phase
  transition},}\ }\href@noop {} {\bibfield  {journal} {\bibinfo  {journal}
  {Phys. Rev. Lett.}\ }\textbf {\bibinfo {volume} {131}},\ \bibinfo {pages}
  {258301} (\bibinfo {year} {2023})}\BibitemShut {NoStop}%
\bibitem [{\citenamefont {Suchanek}\ \emph
  {et~al.}(2023{\natexlab{a}})\citenamefont {Suchanek}, \citenamefont {Kroy},\
  and\ \citenamefont
  {Loos}}]{suchanek_loos_2023_irreversible_fluctuations_emergence_dynamical_phases}%
  \BibitemOpen
  \bibfield  {author} {\bibinfo {author} {\bibfnamefont {Thomas}\ \bibnamefont
  {Suchanek}}, \bibinfo {author} {\bibfnamefont {Klaus}\ \bibnamefont {Kroy}},
  \ and\ \bibinfo {author} {\bibfnamefont {Sarah A~M}\ \bibnamefont {Loos}},\
  }\bibfield  {title} {\enquote {\bibinfo {title} {Irreversible mesoscale
  fluctuations herald the emergence of dynamical phases},}\ }\href@noop {}
  {\bibfield  {journal} {\bibinfo  {journal} {Phys. Rev. Lett.}\ }\textbf
  {\bibinfo {volume} {131}},\ \bibinfo {pages} {258302} (\bibinfo {year}
  {2023}{\natexlab{a}})}\BibitemShut {NoStop}%
\bibitem [{\citenamefont {Suchanek}\ \emph
  {et~al.}(2023{\natexlab{b}})\citenamefont {Suchanek}, \citenamefont {Kroy},\
  and\ \citenamefont
  {Loos}}]{suchanek_loos_2023_entropy_production_nonreciprocal_Cahn-Hilliard}%
  \BibitemOpen
  \bibfield  {author} {\bibinfo {author} {\bibfnamefont {Thomas}\ \bibnamefont
  {Suchanek}}, \bibinfo {author} {\bibfnamefont {Klaus}\ \bibnamefont {Kroy}},
  \ and\ \bibinfo {author} {\bibfnamefont {Sarah A~M}\ \bibnamefont {Loos}},\
  }\bibfield  {title} {\enquote {\bibinfo {title} {Entropy production in the
  nonreciprocal {C}ahn-{H}illiard model},}\ }\href@noop {} {\bibfield
  {journal} {\bibinfo  {journal} {Phys. Rev. E}\ }\textbf {\bibinfo {volume}
  {108}},\ \bibinfo {pages} {064610} (\bibinfo {year}
  {2023}{\natexlab{b}})}\BibitemShut {NoStop}%
\bibitem [{\citenamefont {Loos}\ \emph {et~al.}(2023)\citenamefont {Loos},
  \citenamefont {Klapp},\ and\ \citenamefont
  {Martynec}}]{loos_martynec_2023_long-range_order_non-reciprocal_XY_model}%
  \BibitemOpen
  \bibfield  {author} {\bibinfo {author} {\bibfnamefont {Sarah A~M}\
  \bibnamefont {Loos}}, \bibinfo {author} {\bibfnamefont {Sabine H~L}\
  \bibnamefont {Klapp}}, \ and\ \bibinfo {author} {\bibfnamefont {Thomas}\
  \bibnamefont {Martynec}},\ }\bibfield  {title} {\enquote {\bibinfo {title}
  {Long-range order and directional defect propagation in the nonreciprocal
  {XY} model with vision cone interactions},}\ }\href@noop {} {\bibfield
  {journal} {\bibinfo  {journal} {Phys. Rev. Lett.}\ }\textbf {\bibinfo
  {volume} {130}},\ \bibinfo {pages} {198301} (\bibinfo {year}
  {2023})}\BibitemShut {NoStop}%
\bibitem [{\citenamefont {You}\ \emph {et~al.}(2020)\citenamefont {You},
  \citenamefont {Baskaran},\ and\ \citenamefont
  {Marchetti}}]{You_Baskaran_Marchetti_2020_pnas}%
  \BibitemOpen
  \bibfield  {author} {\bibinfo {author} {\bibfnamefont {Zhihong}\ \bibnamefont
  {You}}, \bibinfo {author} {\bibfnamefont {Aparna}\ \bibnamefont {Baskaran}},
  \ and\ \bibinfo {author} {\bibfnamefont {M.~Cristina}\ \bibnamefont
  {Marchetti}},\ }\bibfield  {title} {\enquote {\bibinfo {title}
  {Nonreciprocity as a generic route to traveling states},}\ }\href@noop {}
  {\bibfield  {journal} {\bibinfo  {journal} {Proc. Natl. Acad. Sci. U.S.A.}\
  }\textbf {\bibinfo {volume} {117}},\ \bibinfo {pages} {19767} (\bibinfo
  {year} {2020})}\BibitemShut {NoStop}%
\bibitem [{\citenamefont {Fruchart}\ \emph {et~al.}(2021)\citenamefont
  {Fruchart}, \citenamefont {Hanai}, \citenamefont {Littlewood},\ and\
  \citenamefont {Vitelli}}]{fruchart_2021_non-reciprocal_phase_transitions}%
  \BibitemOpen
  \bibfield  {author} {\bibinfo {author} {\bibfnamefont {Michel}\ \bibnamefont
  {Fruchart}}, \bibinfo {author} {\bibfnamefont {Ryo}\ \bibnamefont {Hanai}},
  \bibinfo {author} {\bibfnamefont {Peter~B}\ \bibnamefont {Littlewood}}, \
  and\ \bibinfo {author} {\bibfnamefont {Vincenzo}\ \bibnamefont {Vitelli}},\
  }\bibfield  {title} {\enquote {\bibinfo {title} {Non-reciprocal phase
  transitions},}\ }\href@noop {} {\bibfield  {journal} {\bibinfo  {journal}
  {Nature}\ }\textbf {\bibinfo {volume} {592}},\ \bibinfo {pages} {363}
  (\bibinfo {year} {2021})}\BibitemShut {NoStop}%
\bibitem [{\citenamefont {Kreienkamp}\ and\ \citenamefont
  {Klapp}(2025)}]{kreienkamp_klapp_2024_synchronization_exceptional_points}%
  \BibitemOpen
  \bibfield  {author} {\bibinfo {author} {\bibfnamefont {Kim~L}\ \bibnamefont
  {Kreienkamp}}\ and\ \bibinfo {author} {\bibfnamefont {Sabine H~L}\
  \bibnamefont {Klapp}},\ }\bibfield  {title} {\enquote {\bibinfo {title}
  {Synchronization and exceptional points in nonreciprocal active polar
  mixtures},}\ }\href@noop {} {\bibfield  {journal} {\bibinfo  {journal}
  {Commun. Phys.}\ }\textbf {\bibinfo {volume} {8}},\ \bibinfo {pages} {307}
  (\bibinfo {year} {2025})}\BibitemShut {NoStop}%
\bibitem [{\citenamefont {Kreienkamp}\ and\ \citenamefont
  {Klapp}(2024{\natexlab{a}})}]{kreienkamp_klapp_2024_non-reciprocal_alignment_induces_asymmetric_clustering}%
  \BibitemOpen
  \bibfield  {author} {\bibinfo {author} {\bibfnamefont {Kim~L}\ \bibnamefont
  {Kreienkamp}}\ and\ \bibinfo {author} {\bibfnamefont {Sabine H~L}\
  \bibnamefont {Klapp}},\ }\bibfield  {title} {\enquote {\bibinfo {title}
  {Nonreciprocal alignment induces asymmetric clustering in active mixtures},}\
  }\href@noop {} {\bibfield  {journal} {\bibinfo  {journal} {Phys. Rev. Lett.}\
  }\textbf {\bibinfo {volume} {133}},\ \bibinfo {pages} {258303} (\bibinfo
  {year} {2024}{\natexlab{a}})}\BibitemShut {NoStop}%
\bibitem [{\citenamefont {Kreienkamp}\ and\ \citenamefont
  {Klapp}(2024{\natexlab{b}})}]{kreienkamp_klapp_2024_dynamical_structures_phase_separating_nonreciprocal_polar_active_mixtures}%
  \BibitemOpen
  \bibfield  {author} {\bibinfo {author} {\bibfnamefont {Kim~L}\ \bibnamefont
  {Kreienkamp}}\ and\ \bibinfo {author} {\bibfnamefont {Sabine H~L}\
  \bibnamefont {Klapp}},\ }\bibfield  {title} {\enquote {\bibinfo {title}
  {Dynamical structures in phase-separating nonreciprocal polar active
  mixtures},}\ }\href@noop {} {\bibfield  {journal} {\bibinfo  {journal} {Phys.
  Rev. E}\ }\textbf {\bibinfo {volume} {110}},\ \bibinfo {pages} {064135}
  (\bibinfo {year} {2024}{\natexlab{b}})}\BibitemShut {NoStop}%
\bibitem [{\citenamefont {Kreienkamp}\ and\ \citenamefont
  {Klapp}(2022)}]{kreienkamp_klapp_2022_clustering_flocking_chiral_active_particles_non-reciprocal_couplings}%
  \BibitemOpen
  \bibfield  {author} {\bibinfo {author} {\bibfnamefont {Kim~L}\ \bibnamefont
  {Kreienkamp}}\ and\ \bibinfo {author} {\bibfnamefont {Sabine H~L}\
  \bibnamefont {Klapp}},\ }\bibfield  {title} {\enquote {\bibinfo {title}
  {Clustering and flocking of repulsive chiral active particles with
  non-reciprocal couplings},}\ }\href@noop {} {\bibfield  {journal} {\bibinfo
  {journal} {New J. Phys.}\ }\textbf {\bibinfo {volume} {24}},\ \bibinfo
  {pages} {123009} (\bibinfo {year} {2022})}\BibitemShut {NoStop}%
\bibitem [{\citenamefont {Cates}\ and\ \citenamefont
  {Tailleur}(2015)}]{cates_tailleur_2015_mips}%
  \BibitemOpen
  \bibfield  {author} {\bibinfo {author} {\bibfnamefont {Michael~E}\
  \bibnamefont {Cates}}\ and\ \bibinfo {author} {\bibfnamefont {Julien}\
  \bibnamefont {Tailleur}},\ }\bibfield  {title} {\enquote {\bibinfo {title}
  {Motility-induced phase separation},}\ }\href@noop {} {\bibfield  {journal}
  {\bibinfo  {journal} {Annu. Rev. Condens. Matter Phys.}\ }\textbf {\bibinfo
  {volume} {6}},\ \bibinfo {pages} {219--244} (\bibinfo {year}
  {2015})}\BibitemShut {NoStop}%
\bibitem [{\citenamefont {Bialk{\'{e}}}\ \emph {et~al.}(2013)\citenamefont
  {Bialk{\'{e}}}, \citenamefont {Löwen},\ and\ \citenamefont
  {Speck}}]{Bialke_2013_microscopic_theory_phase_seperation}%
  \BibitemOpen
  \bibfield  {author} {\bibinfo {author} {\bibfnamefont {Julian}\ \bibnamefont
  {Bialk{\'{e}}}}, \bibinfo {author} {\bibfnamefont {Hartmut}\ \bibnamefont
  {Löwen}}, \ and\ \bibinfo {author} {\bibfnamefont {Thomas}\ \bibnamefont
  {Speck}},\ }\bibfield  {title} {\enquote {\bibinfo {title} {Microscopic
  theory for the phase separation of self-propelled repulsive disks},}\
  }\href@noop {} {\bibfield  {journal} {\bibinfo  {journal} {{EPL}}\ }\textbf
  {\bibinfo {volume} {103}},\ \bibinfo {pages} {30008} (\bibinfo {year}
  {2013})}\BibitemShut {NoStop}%
\bibitem [{\citenamefont {Marchetti}\ \emph {et~al.}(2013)\citenamefont
  {Marchetti}, \citenamefont {Joanny}, \citenamefont {Ramaswamy}, \citenamefont
  {Liverpool}, \citenamefont {Prost}, \citenamefont {Rao},\ and\ \citenamefont
  {Simha}}]{marchetti_simha_2013_hydrodynamics_soft_active_matter}%
  \BibitemOpen
  \bibfield  {author} {\bibinfo {author} {\bibfnamefont {M~Cristina}\
  \bibnamefont {Marchetti}}, \bibinfo {author} {\bibfnamefont
  {Jean-Fran{\c{c}}ois}\ \bibnamefont {Joanny}}, \bibinfo {author}
  {\bibfnamefont {Sriram}\ \bibnamefont {Ramaswamy}}, \bibinfo {author}
  {\bibfnamefont {Tanniemola~B}\ \bibnamefont {Liverpool}}, \bibinfo {author}
  {\bibfnamefont {Jacques}\ \bibnamefont {Prost}}, \bibinfo {author}
  {\bibfnamefont {Madan}\ \bibnamefont {Rao}}, \ and\ \bibinfo {author}
  {\bibfnamefont {R~Aditi}\ \bibnamefont {Simha}},\ }\bibfield  {title}
  {\enquote {\bibinfo {title} {Hydrodynamics of soft active matter},}\
  }\href@noop {} {\bibfield  {journal} {\bibinfo  {journal} {Rev. Mod. Phys.}\
  }\textbf {\bibinfo {volume} {85}},\ \bibinfo {pages} {1143} (\bibinfo {year}
  {2013})}\BibitemShut {NoStop}%
\bibitem [{\citenamefont {Vicsek}\ \emph {et~al.}(1995)\citenamefont {Vicsek},
  \citenamefont {Czir{\'o}k}, \citenamefont {Ben-Jacob}, \citenamefont
  {Cohen},\ and\ \citenamefont
  {Shochet}}]{vicsek_1995_novel_type_phase_transition}%
  \BibitemOpen
  \bibfield  {author} {\bibinfo {author} {\bibfnamefont {Tam{\'a}s}\
  \bibnamefont {Vicsek}}, \bibinfo {author} {\bibfnamefont {Andr{\'a}s}\
  \bibnamefont {Czir{\'o}k}}, \bibinfo {author} {\bibfnamefont {Eshel}\
  \bibnamefont {Ben-Jacob}}, \bibinfo {author} {\bibfnamefont {Inon}\
  \bibnamefont {Cohen}}, \ and\ \bibinfo {author} {\bibfnamefont {Ofer}\
  \bibnamefont {Shochet}},\ }\bibfield  {title} {\enquote {\bibinfo {title}
  {Novel type of phase transition in a system of self-driven particles},}\
  }\href@noop {} {\bibfield  {journal} {\bibinfo  {journal} {Phys. Rev. Lett.}\
  }\textbf {\bibinfo {volume} {75}},\ \bibinfo {pages} {1226} (\bibinfo {year}
  {1995})}\BibitemShut {NoStop}%
\bibitem [{\citenamefont {Gr\'egoire}\ and\ \citenamefont
  {Chat\'e}(2004)}]{gregoire_chate_2004_collective_motion}%
  \BibitemOpen
  \bibfield  {author} {\bibinfo {author} {\bibfnamefont {Guillaume}\
  \bibnamefont {Gr\'egoire}}\ and\ \bibinfo {author} {\bibfnamefont {Hugues}\
  \bibnamefont {Chat\'e}},\ }\bibfield  {title} {\enquote {\bibinfo {title}
  {Onset of collective and cohesive motion},}\ }\href@noop {} {\bibfield
  {journal} {\bibinfo  {journal} {Phys. Rev. Lett.}\ }\textbf {\bibinfo
  {volume} {92}},\ \bibinfo {pages} {025702} (\bibinfo {year}
  {2004})}\BibitemShut {NoStop}%
\bibitem [{\citenamefont {Weeks}\ \emph {et~al.}(1971)\citenamefont {Weeks},
  \citenamefont {Chandler},\ and\ \citenamefont
  {Andersen}}]{weeks_1971_Weeks-Chandler-Andersen_potential}%
  \BibitemOpen
  \bibfield  {author} {\bibinfo {author} {\bibfnamefont {John~D}\ \bibnamefont
  {Weeks}}, \bibinfo {author} {\bibfnamefont {David}\ \bibnamefont {Chandler}},
  \ and\ \bibinfo {author} {\bibfnamefont {Hans~C}\ \bibnamefont {Andersen}},\
  }\bibfield  {title} {\enquote {\bibinfo {title} {Role of repulsive forces in
  determining the equilibrium structure of simple liquids},}\ }\href@noop {}
  {\bibfield  {journal} {\bibinfo  {journal} {J. Chem. Phys.}\ }\textbf
  {\bibinfo {volume} {54}},\ \bibinfo {pages} {5237} (\bibinfo {year}
  {1971})}\BibitemShut {NoStop}%
\bibitem [{\citenamefont {Chat\'e}\ \emph {et~al.}(2008)\citenamefont
  {Chat\'e}, \citenamefont {Ginelli}, \citenamefont {Gr\'egoire},\ and\
  \citenamefont {Raynaud}}]{chate_raynaud_2008_collective_motion}%
  \BibitemOpen
  \bibfield  {author} {\bibinfo {author} {\bibfnamefont {Hugues}\ \bibnamefont
  {Chat\'e}}, \bibinfo {author} {\bibfnamefont {Francesco}\ \bibnamefont
  {Ginelli}}, \bibinfo {author} {\bibfnamefont {Guillaume}\ \bibnamefont
  {Gr\'egoire}}, \ and\ \bibinfo {author} {\bibfnamefont {Franck}\ \bibnamefont
  {Raynaud}},\ }\bibfield  {title} {\enquote {\bibinfo {title} {Collective
  motion of self-propelled particles interacting without cohesion},}\
  }\href@noop {} {\bibfield  {journal} {\bibinfo  {journal} {Phys. Rev. E}\
  }\textbf {\bibinfo {volume} {77}},\ \bibinfo {pages} {046113} (\bibinfo
  {year} {2008})}\BibitemShut {NoStop}%
\bibitem [{\citenamefont {El-Ganainy}\ \emph {et~al.}(2018)\citenamefont
  {El-Ganainy}, \citenamefont {Makris}, \citenamefont {Khajavikhan},
  \citenamefont {Musslimani}, \citenamefont {Rotter},\ and\ \citenamefont
  {Christodoulides}}]{el-ganainy_demetrios_2018_non-Hermitian_physics_PT-symmetry_breaking}%
  \BibitemOpen
  \bibfield  {author} {\bibinfo {author} {\bibfnamefont {Ramy}\ \bibnamefont
  {El-Ganainy}}, \bibinfo {author} {\bibfnamefont {Konstantinos~G}\
  \bibnamefont {Makris}}, \bibinfo {author} {\bibfnamefont {Mercedeh}\
  \bibnamefont {Khajavikhan}}, \bibinfo {author} {\bibfnamefont {Ziad~H}\
  \bibnamefont {Musslimani}}, \bibinfo {author} {\bibfnamefont {Stefan}\
  \bibnamefont {Rotter}}, \ and\ \bibinfo {author} {\bibfnamefont
  {Demetrios~N}\ \bibnamefont {Christodoulides}},\ }\bibfield  {title}
  {\enquote {\bibinfo {title} {Non-{H}ermitian physics and {PT} symmetry},}\
  }\href@noop {} {\bibfield  {journal} {\bibinfo  {journal} {Nat. Phys.}\
  }\textbf {\bibinfo {volume} {14}},\ \bibinfo {pages} {11--19} (\bibinfo
  {year} {2018})}\BibitemShut {NoStop}%
\bibitem [{\citenamefont {Suchanek}\ \emph
  {et~al.}(2023{\natexlab{c}})\citenamefont {Suchanek}, \citenamefont {Kroy},\
  and\ \citenamefont
  {Loos}}]{suchanek_loos_2023_time-reversal_PT_symmetry_breaking_non-Hermitian_field_theories}%
  \BibitemOpen
  \bibfield  {author} {\bibinfo {author} {\bibfnamefont {Thomas}\ \bibnamefont
  {Suchanek}}, \bibinfo {author} {\bibfnamefont {Klaus}\ \bibnamefont {Kroy}},
  \ and\ \bibinfo {author} {\bibfnamefont {Sarah A~M}\ \bibnamefont {Loos}},\
  }\bibfield  {title} {\enquote {\bibinfo {title} {Time-reversal and
  parity-time symmetry breaking in non-{H}ermitian field theories},}\
  }\href@noop {} {\bibfield  {journal} {\bibinfo  {journal} {Phys. Rev. E}\
  }\textbf {\bibinfo {volume} {108}},\ \bibinfo {pages} {064123} (\bibinfo
  {year} {2023}{\natexlab{c}})}\BibitemShut {NoStop}%
\bibitem [{\citenamefont
  {Seifert}(2005)}]{seifert_2005_entropy_production_along_stochastic_trajectory}%
  \BibitemOpen
  \bibfield  {author} {\bibinfo {author} {\bibfnamefont {Udo}\ \bibnamefont
  {Seifert}},\ }\bibfield  {title} {\enquote {\bibinfo {title} {Entropy
  production along a stochastic trajectory and an integral fluctuation
  theorem},}\ }\href@noop {} {\bibfield  {journal} {\bibinfo  {journal} {Phys.
  Rev. Lett.}\ }\textbf {\bibinfo {volume} {95}},\ \bibinfo {pages} {040602}
  (\bibinfo {year} {2005})}\BibitemShut {NoStop}%
\bibitem [{\citenamefont
  {Seifert}(2008)}]{seifert_2008_stochastic_thermodynamics}%
  \BibitemOpen
  \bibfield  {author} {\bibinfo {author} {\bibfnamefont {Udo}\ \bibnamefont
  {Seifert}},\ }\bibfield  {title} {\enquote {\bibinfo {title} {Stochastic
  thermodynamics: principles and perspectives},}\ }\href@noop {} {\bibfield
  {journal} {\bibinfo  {journal} {Eur. Phys. J. B}\ }\textbf {\bibinfo {volume}
  {64}},\ \bibinfo {pages} {423--431} (\bibinfo {year} {2008})}\BibitemShut
  {NoStop}%
\bibitem [{\citenamefont {Spinney}\ and\ \citenamefont
  {Ford}(2012)}]{spinney_ford_2012_entropy_production_full_phase_space}%
  \BibitemOpen
  \bibfield  {author} {\bibinfo {author} {\bibfnamefont {Richard~E}\
  \bibnamefont {Spinney}}\ and\ \bibinfo {author} {\bibfnamefont {Ian~J}\
  \bibnamefont {Ford}},\ }\bibfield  {title} {\enquote {\bibinfo {title}
  {Entropy production in full phase space for continuous stochastic
  dynamics},}\ }\href@noop {} {\bibfield  {journal} {\bibinfo  {journal} {Phys.
  Rev. E}\ }\textbf {\bibinfo {volume} {85}},\ \bibinfo {pages} {051113}
  (\bibinfo {year} {2012})}\BibitemShut {NoStop}%
\bibitem [{\citenamefont {Crosato}\ \emph {et~al.}(2019)\citenamefont
  {Crosato}, \citenamefont {Prokopenko},\ and\ \citenamefont
  {Spinney}}]{crosato_spinney_2019_irreversibility_active_matter}%
  \BibitemOpen
  \bibfield  {author} {\bibinfo {author} {\bibfnamefont {Emanuele}\
  \bibnamefont {Crosato}}, \bibinfo {author} {\bibfnamefont {Mikhail}\
  \bibnamefont {Prokopenko}}, \ and\ \bibinfo {author} {\bibfnamefont
  {Richard~E}\ \bibnamefont {Spinney}},\ }\bibfield  {title} {\enquote
  {\bibinfo {title} {Irreversibility and emergent structure in active
  matter},}\ }\href@noop {} {\bibfield  {journal} {\bibinfo  {journal} {Phys.
  Rev. E}\ }\textbf {\bibinfo {volume} {100}},\ \bibinfo {pages} {042613}
  (\bibinfo {year} {2019})}\BibitemShut {NoStop}%
\bibitem [{\citenamefont {Shankar}\ and\ \citenamefont
  {Marchetti}(2018)}]{shankar_marchetti_2018_hidden_entropy_production}%
  \BibitemOpen
  \bibfield  {author} {\bibinfo {author} {\bibfnamefont {Suraj}\ \bibnamefont
  {Shankar}}\ and\ \bibinfo {author} {\bibfnamefont {M~Cristina}\ \bibnamefont
  {Marchetti}},\ }\bibfield  {title} {\enquote {\bibinfo {title} {Hidden
  entropy production and work fluctuations in an ideal active gas},}\
  }\href@noop {} {\bibfield  {journal} {\bibinfo  {journal} {Phys. Rev. E}\
  }\textbf {\bibinfo {volume} {98}},\ \bibinfo {pages} {020604} (\bibinfo
  {year} {2018})}\BibitemShut {NoStop}%
\bibitem [{Note1()}]{Note1}%
  \BibitemOpen
  \bibinfo {note} {Note that the sum of the translational contributions to the
  entropy production for odd and even interpretations is a constant \cite
  {pietzonka_seifert_2017_entropy_production_active_particles,dadhichi_ramaswamy_2018_origins_nonequilibrium_character_active_systems}.
  The alignment contribution is independent of the choice of time-reversal
  convention for self-propulsion, such that peaks in the alignment contribution
  to the entropy production rate with the odd interpretation also occur for the
  even interpretation. Nevertheless, the dependence on the choice of
  self-propulsion time reversal symmetry -- in combination with the fact that
  we calculate an informatic entropy production rate -- makes a direct
  interpretation of the absolute value of the entropy production rate
  challenging.}\BibitemShut {Stop}%
\bibitem [{Note2()}]{Note2}%
  \BibitemOpen
  \bibinfo {note} {To evaluate a Stratonovich product of form $f(x(t),t) \circ
  {\protect \rm d}x(t)$ with stochastic variable $x(t)$, one computes
  $f(x_{\protect \rm mp}, t_{\protect \rm mp}) \cdot (x(t+{\protect \rm d}t) -
  x(t))$, where $f(x)$ is evaluated at the mid points $x_{\protect \rm mp} =
  (x(t)+x(t+{\protect \rm d}t))/2$, $t_{\protect \rm mp} = (t+t+{\protect \rm
  d}t)/2$ and ${\protect \rm d}t \rightarrow 0$ \cite
  {risken_1996_fokker-planck_equation}.}\BibitemShut {Stop}%
\bibitem [{\citenamefont {Cates}\ \emph {et~al.}(2022)\citenamefont {Cates},
  \citenamefont {Fodor}, \citenamefont {Markovich}, \citenamefont {Nardini},\
  and\ \citenamefont
  {Tjhung}}]{cates_tjhung_2022_stochastic_hydrodynamics_complex_fluids}%
  \BibitemOpen
  \bibfield  {author} {\bibinfo {author} {\bibfnamefont {Michael~E}\
  \bibnamefont {Cates}}, \bibinfo {author} {\bibfnamefont {{\'E}tienne}\
  \bibnamefont {Fodor}}, \bibinfo {author} {\bibfnamefont {Tomer}\ \bibnamefont
  {Markovich}}, \bibinfo {author} {\bibfnamefont {Cesare}\ \bibnamefont
  {Nardini}}, \ and\ \bibinfo {author} {\bibfnamefont {Elsen}\ \bibnamefont
  {Tjhung}},\ }\bibfield  {title} {\enquote {\bibinfo {title} {Stochastic
  hydrodynamics of complex fluids: Discretisation and entropy production},}\
  }\href@noop {} {\bibfield  {journal} {\bibinfo  {journal} {Entropy}\ }\textbf
  {\bibinfo {volume} {24}},\ \bibinfo {pages} {254} (\bibinfo {year}
  {2022})}\BibitemShut {NoStop}%
\bibitem [{\citenamefont {Mohite}\ and\ \citenamefont
  {Rieger}(2025)}]{mohite_rieger_2025_stochastic_thermodynamics_non-reciprocity}%
  \BibitemOpen
  \bibfield  {author} {\bibinfo {author} {\bibfnamefont {Atul~Tanaji}\
  \bibnamefont {Mohite}}\ and\ \bibinfo {author} {\bibfnamefont {Heiko}\
  \bibnamefont {Rieger}},\ }\bibfield  {title} {\enquote {\bibinfo {title}
  {Stochastic thermodynamics of non-reciprocally interacting particles and
  fields},}\ }\href@noop {} {\bibfield  {journal} {\bibinfo  {journal} {arXiv
  preprint arXiv:2504.10515}\ } (\bibinfo {year} {2025})}\BibitemShut {NoStop}%
\bibitem [{\citenamefont {Yu}\ and\ \citenamefont
  {Tu}(2022)}]{yu_tu_2022_energy_cost_flocking_active_spins}%
  \BibitemOpen
  \bibfield  {author} {\bibinfo {author} {\bibfnamefont {Qiwei}\ \bibnamefont
  {Yu}}\ and\ \bibinfo {author} {\bibfnamefont {Yuhai}\ \bibnamefont {Tu}},\
  }\bibfield  {title} {\enquote {\bibinfo {title} {Energy cost for flocking of
  active spins: the cusped dissipation maximum at the flocking transition},}\
  }\href@noop {} {\bibfield  {journal} {\bibinfo  {journal} {Phys. Rev. Lett.}\
  }\textbf {\bibinfo {volume} {129}},\ \bibinfo {pages} {278001} (\bibinfo
  {year} {2022})}\BibitemShut {NoStop}%
\bibitem [{Note3()}]{Note3}%
  \BibitemOpen
  \bibinfo {note} {Note that, in one-species systems, the entropy production
  rate associated with the alignment couplings exhibits a peak at the
  transition from disordered to flocking motion \cite
  {ferretti_walczak_2022_signatures_irreversibility_flocking,yu_tu_2022_energy_cost_flocking_active_spins}.
  Here, we do not study the conventional ordered-disorder transition. Instead,
  we consider a binary mixture transitioning from antiflocking to flocking as
  $\kappa $ is varied.}\BibitemShut {Stop}%
\bibitem [{\citenamefont {Toner}\ \emph {et~al.}(2005)\citenamefont {Toner},
  \citenamefont {Tu},\ and\ \citenamefont
  {Ramaswamy}}]{toner_tu_2005_hydrodynamics_phases_of_flocks}%
  \BibitemOpen
  \bibfield  {author} {\bibinfo {author} {\bibfnamefont {John}\ \bibnamefont
  {Toner}}, \bibinfo {author} {\bibfnamefont {Yuhai}\ \bibnamefont {Tu}}, \
  and\ \bibinfo {author} {\bibfnamefont {Sriram}\ \bibnamefont {Ramaswamy}},\
  }\bibfield  {title} {\enquote {\bibinfo {title} {Hydrodynamics and phases of
  flocks},}\ }\href@noop {} {\bibfield  {journal} {\bibinfo  {journal} {Ann.
  Phys.}\ }\textbf {\bibinfo {volume} {318}},\ \bibinfo {pages} {170} (\bibinfo
  {year} {2005})}\BibitemShut {NoStop}%
\bibitem [{\citenamefont {Dadhichi}\ \emph {et~al.}(2018)\citenamefont
  {Dadhichi}, \citenamefont {Maitra},\ and\ \citenamefont
  {Ramaswamy}}]{dadhichi_ramaswamy_2018_origins_nonequilibrium_character_active_systems}%
  \BibitemOpen
  \bibfield  {author} {\bibinfo {author} {\bibfnamefont {Lokrshi~Prawar}\
  \bibnamefont {Dadhichi}}, \bibinfo {author} {\bibfnamefont {Ananyo}\
  \bibnamefont {Maitra}}, \ and\ \bibinfo {author} {\bibfnamefont {Sriram}\
  \bibnamefont {Ramaswamy}},\ }\bibfield  {title} {\enquote {\bibinfo {title}
  {Origins and diagnostics of the nonequilibrium character of active
  systems},}\ }\href@noop {} {\bibfield  {journal} {\bibinfo  {journal} {J.
  Stat. Mech.: Theory Exp.}\ }\textbf {\bibinfo {volume} {2018}},\ \bibinfo
  {pages} {123201} (\bibinfo {year} {2018})}\BibitemShut {NoStop}%
\bibitem [{\citenamefont {Onsager}\ and\ \citenamefont
  {Machlup}(1953)}]{onsager_machlup_1953_fluctuations_irreversible_processes}%
  \BibitemOpen
  \bibfield  {author} {\bibinfo {author} {\bibfnamefont {Lars}\ \bibnamefont
  {Onsager}}\ and\ \bibinfo {author} {\bibfnamefont {Stefan}\ \bibnamefont
  {Machlup}},\ }\bibfield  {title} {\enquote {\bibinfo {title} {Fluctuations
  and irreversible processes},}\ }\href@noop {} {\bibfield  {journal} {\bibinfo
   {journal} {Phys. Rev.}\ }\textbf {\bibinfo {volume} {91}},\ \bibinfo {pages}
  {1505} (\bibinfo {year} {1953})}\BibitemShut {NoStop}%
\bibitem [{\citenamefont {Cugliandolo}\ and\ \citenamefont
  {Lecomte}(2017)}]{cugliandolo_lecomte_2017_rules_calculus_path_intergral_representation_langevin_equations}%
  \BibitemOpen
  \bibfield  {author} {\bibinfo {author} {\bibfnamefont {Leticia~F}\
  \bibnamefont {Cugliandolo}}\ and\ \bibinfo {author} {\bibfnamefont {Vivien}\
  \bibnamefont {Lecomte}},\ }\bibfield  {title} {\enquote {\bibinfo {title}
  {Rules of calculus in the path integral representation of white noise
  {L}angevin equations: the {O}nsager--{M}achlup approach},}\ }\href@noop {}
  {\bibfield  {journal} {\bibinfo  {journal} {J. Phys. A-Math. Theor.}\
  }\textbf {\bibinfo {volume} {50}},\ \bibinfo {pages} {345001} (\bibinfo
  {year} {2017})}\BibitemShut {NoStop}%
\bibitem [{Note4()}]{Note4}%
  \BibitemOpen
  \bibinfo {note} {Note that $\protect \mathcal {A}_{\protect \bm
  {w}^a}^{\protect \rm conv}$ and $\protect \overleftarrow {\protect \mathcal
  {A}}_{\protect \bm {w}^a}^{\protect \rm conv}$ cancel when the functional
  $\protect \mathcal {F}^a_{k=0}$ is fully anti-symmetric in time (as in this
  case). This is not generally the case.}\BibitemShut {Stop}%
\bibitem [{\citenamefont
  {Gardiner}(1985)}]{gardiner_1985_handbook_stochastic_methods}%
  \BibitemOpen
  \bibfield  {author} {\bibinfo {author} {\bibfnamefont {Crispin~W}\
  \bibnamefont {Gardiner}},\ }\href@noop {} {\emph {\bibinfo {title} {Handbook
  of stochastic methods}}},\ Vol.~\bibinfo {volume} {3}\ (\bibinfo  {publisher}
  {Springer Berlin},\ \bibinfo {year} {1985})\BibitemShut {NoStop}%
\bibitem [{Note5()}]{Note5}%
  \BibitemOpen
  \bibinfo {note} {In It\^o calculus, the delta-correlated noise $\protect
  \tilde {\protect \bm {\eta }}^a$ only affects $\protect \bm {w}^c$ at later
  time steps, making $\protect \tilde {\protect \bm {\eta }}^a$ and $\protect
  \bm {w}^c$ uncorrelated at equal times.}\BibitemShut {Stop}%
\bibitem [{Note6()}]{Note6}%
  \BibitemOpen
  \bibinfo {note} {This conversion is commonly presented for particles and
  unit-variance noise, e.g., \cite
  {gardiner_1985_handbook_stochastic_methods,cates_tjhung_2022_stochastic_hydrodynamics_complex_fluids}:
  for a Langevin equation \begin {equation} \protect \frac {{\protect \rm
  d}x_i}{{\protect \rm d}t} = f_i(\{x_i\}) + g_{ij}(\{x_i\})\protect \,\eta
  _j(t) \end {equation} for $\{x_i(t)\}$, where $i=1,2,...,d$, and $\langle
  \eta _i(t)\rangle = 0$, $\langle \eta _i(t)\protect \,\eta _j(t') \rangle =
  \delta _{ij}\protect \,\delta (t-t')$, one has \cite
  {cates_tjhung_2022_stochastic_hydrodynamics_complex_fluids} \begin {equation}
  \begin {split} \DOTSI \intop \ilimits@ _{t_0}^{t_N} h_{i}(\{x_i\}) \circ \eta
  _j(t) \protect \, {\protect \rm d}t = & \DOTSI \intop \ilimits@ _{t_0}^{t_N}
  h_{i}(\{x_i\}) \cdot \eta _j(t) \protect \, {\protect \rm d}t + \protect
  \frac {1}{2}\protect \, \DOTSI \intop \ilimits@ _{t_0}^{t_N} \protect \frac
  {\partial h_{i}(\{x_i\})}{\partial x_k} \protect \, g_{kj} \protect \,
  {\protect \rm d}t, \end {split} \end {equation} where $h_i$ is a function of
  $\{x_i(t)\}$. In our field-theoretical case, the noise $\protect \tilde
  {\protect \bm {\eta }}^b$ has no unit-variance but `includes' the $g_{jk}$.
  Therefore, the additional prefactor $2\protect \,D_{\protect \rm h}\protect
  \,\delta (\protect \bm {r}-\protect \bm {r}')$ needs to be taken into
  account.}\BibitemShut {Stop}%
\bibitem [{\citenamefont {Soriani}\ \emph {et~al.}(2025)\citenamefont
  {Soriani}, \citenamefont {Tjhung}, \citenamefont {Fodor},\ and\ \citenamefont
  {Markovich}}]{soriani_markovich_2025_control_active_field_theories}%
  \BibitemOpen
  \bibfield  {author} {\bibinfo {author} {\bibfnamefont {Artur}\ \bibnamefont
  {Soriani}}, \bibinfo {author} {\bibfnamefont {Elsen}\ \bibnamefont {Tjhung}},
  \bibinfo {author} {\bibfnamefont {{\'E}tienne}\ \bibnamefont {Fodor}}, \ and\
  \bibinfo {author} {\bibfnamefont {Tomer}\ \bibnamefont {Markovich}},\
  }\bibfield  {title} {\enquote {\bibinfo {title} {Control of active field
  theories at minimal dissipation},}\ }\href@noop {} {\bibfield  {journal}
  {\bibinfo  {journal} {arXiv preprint arXiv:2504.19285}\ } (\bibinfo {year}
  {2025})}\BibitemShut {NoStop}%
\bibitem [{\citenamefont {Garcia-Millan}\ \emph {et~al.}(2024)\citenamefont
  {Garcia-Millan}, \citenamefont {Sch{\"u}ttler}, \citenamefont {Cates},\ and\
  \citenamefont
  {Loos}}]{garcia-millan_loos_2024_optimal_closed-loop_control_active_particles}%
  \BibitemOpen
  \bibfield  {author} {\bibinfo {author} {\bibfnamefont {Rosalba}\ \bibnamefont
  {Garcia-Millan}}, \bibinfo {author} {\bibfnamefont {Janik}\ \bibnamefont
  {Sch{\"u}ttler}}, \bibinfo {author} {\bibfnamefont {Michael~E}\ \bibnamefont
  {Cates}}, \ and\ \bibinfo {author} {\bibfnamefont {Sarah A~M}\ \bibnamefont
  {Loos}},\ }\bibfield  {title} {\enquote {\bibinfo {title} {Optimal
  closed-loop control of active particles and a minimal information engine},}\
  }\href@noop {} {\bibfield  {journal} {\bibinfo  {journal} {arXiv preprint
  arXiv:2407.18542}\ } (\bibinfo {year} {2024})}\BibitemShut {NoStop}%
\bibitem [{\citenamefont {Loos}\ \emph {et~al.}(2024)\citenamefont {Loos},
  \citenamefont {Monter}, \citenamefont {Ginot},\ and\ \citenamefont
  {Bechinger}}]{loos_bechinger_2024_universal_symmetry_optimal_control_microscale}%
  \BibitemOpen
  \bibfield  {author} {\bibinfo {author} {\bibfnamefont {Sarah~AM}\
  \bibnamefont {Loos}}, \bibinfo {author} {\bibfnamefont {Samuel}\ \bibnamefont
  {Monter}}, \bibinfo {author} {\bibfnamefont {Felix}\ \bibnamefont {Ginot}}, \
  and\ \bibinfo {author} {\bibfnamefont {Clemens}\ \bibnamefont {Bechinger}},\
  }\bibfield  {title} {\enquote {\bibinfo {title} {Universal symmetry of
  optimal control at the microscale},}\ }\href@noop {} {\bibfield  {journal}
  {\bibinfo  {journal} {Phys. Rev. X}\ }\textbf {\bibinfo {volume} {14}},\
  \bibinfo {pages} {021032} (\bibinfo {year} {2024})}\BibitemShut {NoStop}%
\bibitem [{\citenamefont {Li}\ \emph {et~al.}(2024)\citenamefont {Li},
  \citenamefont {Liu}, \citenamefont {Szurek},\ and\ \citenamefont
  {Fakhri}}]{li_fakhri_2024_measuring_irreversibility_learned_representations_biological_patterns}%
  \BibitemOpen
  \bibfield  {author} {\bibinfo {author} {\bibfnamefont {Junang}\ \bibnamefont
  {Li}}, \bibinfo {author} {\bibfnamefont {Chih-Wei~Joshua}\ \bibnamefont
  {Liu}}, \bibinfo {author} {\bibfnamefont {Michal}\ \bibnamefont {Szurek}}, \
  and\ \bibinfo {author} {\bibfnamefont {Nikta}\ \bibnamefont {Fakhri}},\
  }\bibfield  {title} {\enquote {\bibinfo {title} {Measuring irreversibility
  from learned representations of biological patterns},}\ }\href@noop {}
  {\bibfield  {journal} {\bibinfo  {journal} {Phys. Rev. X Life}\ }\textbf
  {\bibinfo {volume} {2}},\ \bibinfo {pages} {033013} (\bibinfo {year}
  {2024})}\BibitemShut {NoStop}%
\bibitem [{\citenamefont {Tan}\ \emph {et~al.}(2021)\citenamefont {Tan},
  \citenamefont {Watson}, \citenamefont {Chao}, \citenamefont {Li},
  \citenamefont {Gingrich}, \citenamefont {Horowitz},\ and\ \citenamefont
  {Fakhri}}]{tan_fakhri_2021_scale-dependent_irreversibility_living_matter}%
  \BibitemOpen
  \bibfield  {author} {\bibinfo {author} {\bibfnamefont {Tzer~Han}\
  \bibnamefont {Tan}}, \bibinfo {author} {\bibfnamefont {Garrett~A}\
  \bibnamefont {Watson}}, \bibinfo {author} {\bibfnamefont {Yu-Chen}\
  \bibnamefont {Chao}}, \bibinfo {author} {\bibfnamefont {Junang}\ \bibnamefont
  {Li}}, \bibinfo {author} {\bibfnamefont {Todd~R}\ \bibnamefont {Gingrich}},
  \bibinfo {author} {\bibfnamefont {Jordan~M}\ \bibnamefont {Horowitz}}, \ and\
  \bibinfo {author} {\bibfnamefont {Nikta}\ \bibnamefont {Fakhri}},\ }\bibfield
   {title} {\enquote {\bibinfo {title} {Scale-dependent irreversibility in
  living matter},}\ }\href@noop {} {\bibfield  {journal} {\bibinfo  {journal}
  {arXiv preprint arXiv:2107.05701}\ } (\bibinfo {year} {2021})}\BibitemShut
  {NoStop}%
\bibitem [{\citenamefont
  {Seifert}(2019)}]{seifert_2019_thermodynamic_inference}%
  \BibitemOpen
  \bibfield  {author} {\bibinfo {author} {\bibfnamefont {Udo}\ \bibnamefont
  {Seifert}},\ }\bibfield  {title} {\enquote {\bibinfo {title} {From stochastic
  thermodynamics to thermodynamic inference},}\ }\href@noop {} {\bibfield
  {journal} {\bibinfo  {journal} {Annu. Rev. Condens. Matter Phys.}\ }\textbf
  {\bibinfo {volume} {10}},\ \bibinfo {pages} {171--192} (\bibinfo {year}
  {2019})}\BibitemShut {NoStop}%
\bibitem [{\citenamefont {Liu}\ \emph {et~al.}(2025)\citenamefont {Liu},
  \citenamefont {Hanai},\ and\ \citenamefont
  {Littlewood}}]{liu_littlewood_2025_universal_scaling_non-reciprocity}%
  \BibitemOpen
  \bibfield  {author} {\bibinfo {author} {\bibfnamefont {Shuoguang}\
  \bibnamefont {Liu}}, \bibinfo {author} {\bibfnamefont {Ryo}\ \bibnamefont
  {Hanai}}, \ and\ \bibinfo {author} {\bibfnamefont {Peter~B}\ \bibnamefont
  {Littlewood}},\ }\bibfield  {title} {\enquote {\bibinfo {title} {Universal
  scaling in one-dimensional non-reciprocal matter},}\ }\href@noop {}
  {\bibfield  {journal} {\bibinfo  {journal} {arXiv preprint arXiv:2503.14384}\
  } (\bibinfo {year} {2025})}\BibitemShut {NoStop}%
\bibitem [{\citenamefont {Risken}\ and\ \citenamefont
  {Risken}(1996)}]{risken_1996_fokker-planck_equation}%
  \BibitemOpen
  \bibfield  {author} {\bibinfo {author} {\bibfnamefont {Hannes}\ \bibnamefont
  {Risken}}\ and\ \bibinfo {author} {\bibfnamefont {Hannes}\ \bibnamefont
  {Risken}},\ }\href@noop {} {\emph {\bibinfo {title} {Fokker-{P}lanck
  equation}}}\ (\bibinfo  {publisher} {Springer},\ \bibinfo {year}
  {1996})\BibitemShut {NoStop}%
\bibitem [{\citenamefont {Canuto}\ \emph {et~al.}(2007)\citenamefont {Canuto},
  \citenamefont {Hussaini}, \citenamefont {Quarteroni},\ and\ \citenamefont
  {Zang}}]{canuto_zang_2007_spectral_methods}%
  \BibitemOpen
  \bibfield  {author} {\bibinfo {author} {\bibfnamefont {Claudio}\ \bibnamefont
  {Canuto}}, \bibinfo {author} {\bibfnamefont {M~Yousuff}\ \bibnamefont
  {Hussaini}}, \bibinfo {author} {\bibfnamefont {Alfio}\ \bibnamefont
  {Quarteroni}}, \ and\ \bibinfo {author} {\bibfnamefont {Thomas~A}\
  \bibnamefont {Zang}},\ }\href@noop {} {\emph {\bibinfo {title} {Spectral
  methods: fundamentals in single domains}}}\ (\bibinfo  {publisher} {Springer
  Science \& Business Media},\ \bibinfo {year} {2007})\BibitemShut {NoStop}%
\bibitem [{\citenamefont {Kloeden}\ and\ \citenamefont
  {Platen}(2011)}]{kloeden_platen_2011_numerical_solution_SDE}%
  \BibitemOpen
  \bibfield  {author} {\bibinfo {author} {\bibfnamefont {P.E.}\ \bibnamefont
  {Kloeden}}\ and\ \bibinfo {author} {\bibfnamefont {E.}~\bibnamefont
  {Platen}},\ }\href@noop {} {\emph {\bibinfo {title} {Numerical Solution of
  Stochastic Differential Equations}}},\ Stochastic Modelling and Applied
  Probability\ (\bibinfo  {publisher} {Springer Berlin Heidelberg},\ \bibinfo
  {year} {2011})\BibitemShut {NoStop}%
\bibitem [{\citenamefont {Mart{\'\i}n-G{\'o}mez}\ \emph
  {et~al.}(2018)\citenamefont {Mart{\'\i}n-G{\'o}mez}, \citenamefont {Levis},
  \citenamefont {D{\'\i}az-Guilera},\ and\ \citenamefont
  {Pagonabarraga}}]{martin_gomez_pagonabarraga_2018_collective_motion_ABP_alignment_volume_exclusion}%
  \BibitemOpen
  \bibfield  {author} {\bibinfo {author} {\bibfnamefont {Aitor}\ \bibnamefont
  {Mart{\'\i}n-G{\'o}mez}}, \bibinfo {author} {\bibfnamefont {Demian}\
  \bibnamefont {Levis}}, \bibinfo {author} {\bibfnamefont {Albert}\
  \bibnamefont {D{\'\i}az-Guilera}}, \ and\ \bibinfo {author} {\bibfnamefont
  {Ignacio}\ \bibnamefont {Pagonabarraga}},\ }\bibfield  {title} {\enquote
  {\bibinfo {title} {Collective motion of active {B}rownian particles with
  polar alignment},}\ }\href@noop {} {\bibfield  {journal} {\bibinfo  {journal}
  {Soft Matter}\ }\textbf {\bibinfo {volume} {14}},\ \bibinfo {pages}
  {2610--2618} (\bibinfo {year} {2018})}\BibitemShut {NoStop}%
\bibitem [{\citenamefont {Baglietto}\ and\ \citenamefont
  {Albano}(2008)}]{Baglietto_Albano_2008_Finite-size_scaling_analysis_self-driven_indiviudals}%
  \BibitemOpen
  \bibfield  {author} {\bibinfo {author} {\bibfnamefont {Gabriel}\ \bibnamefont
  {Baglietto}}\ and\ \bibinfo {author} {\bibfnamefont {Ezequiel~V}\
  \bibnamefont {Albano}},\ }\bibfield  {title} {\enquote {\bibinfo {title}
  {Finite-size scaling analysis and dynamic study of the critical behavior of a
  model for the collective displacement of self-driven individuals},}\
  }\href@noop {} {\bibfield  {journal} {\bibinfo  {journal} {Phys. Rev. E}\
  }\textbf {\bibinfo {volume} {78}},\ \bibinfo {pages} {021125} (\bibinfo
  {year} {2008})}\BibitemShut {NoStop}%
\bibitem [{\citenamefont {Adhikary}\ and\ \citenamefont
  {Santra}(2022)}]{Adhikary_Santra_2022_pattern_formation_phase_transition_binary_mixture}%
  \BibitemOpen
  \bibfield  {author} {\bibinfo {author} {\bibfnamefont {Sagarika}\
  \bibnamefont {Adhikary}}\ and\ \bibinfo {author} {\bibfnamefont {S~B}\
  \bibnamefont {Santra}},\ }\bibfield  {title} {\enquote {\bibinfo {title}
  {Pattern formation and phase transition in the collective dynamics of a
  binary mixture of polar self-propelled particles},}\ }\href@noop {}
  {\bibfield  {journal} {\bibinfo  {journal} {Phys. Rev. E}\ }\textbf {\bibinfo
  {volume} {105}},\ \bibinfo {pages} {064612} (\bibinfo {year}
  {2022})}\BibitemShut {NoStop}%
\bibitem [{\citenamefont {Hansen}\ and\ \citenamefont
  {McDonald}(2013)}]{hansen_mcDonald_2013_theory_simple_liquids}%
  \BibitemOpen
  \bibfield  {author} {\bibinfo {author} {\bibfnamefont {Jean-Pierre}\
  \bibnamefont {Hansen}}\ and\ \bibinfo {author} {\bibfnamefont {Ian~Ranald}\
  \bibnamefont {McDonald}},\ }\href@noop {} {\emph {\bibinfo {title} {Theory of
  simple liquids: with applications to soft matter}}}\ (\bibinfo  {publisher}
  {Academic press},\ \bibinfo {year} {2013})\BibitemShut {NoStop}%
\bibitem [{\citenamefont {Cavagna}\ \emph {et~al.}(2018)\citenamefont
  {Cavagna}, \citenamefont {Giardina},\ and\ \citenamefont
  {Grigera}}]{cavagna_grigera_2018_physics_of_flocking_correlations}%
  \BibitemOpen
  \bibfield  {author} {\bibinfo {author} {\bibfnamefont {Andrea}\ \bibnamefont
  {Cavagna}}, \bibinfo {author} {\bibfnamefont {Irene}\ \bibnamefont
  {Giardina}}, \ and\ \bibinfo {author} {\bibfnamefont {Tom{\'a}s~S}\
  \bibnamefont {Grigera}},\ }\bibfield  {title} {\enquote {\bibinfo {title}
  {The physics of flocking: Correlation as a compass from experiments to
  theory},}\ }\href@noop {} {\bibfield  {journal} {\bibinfo  {journal} {Physics
  Reports}\ }\textbf {\bibinfo {volume} {728}},\ \bibinfo {pages} {1--62}
  (\bibinfo {year} {2018})}\BibitemShut {NoStop}%
\bibitem [{\citenamefont
  {Kardar}(2007)}]{kardar_2007_statistical_physics_fields}%
  \BibitemOpen
  \bibfield  {author} {\bibinfo {author} {\bibfnamefont {Mehran}\ \bibnamefont
  {Kardar}},\ }\href@noop {} {\emph {\bibinfo {title} {Statistical physics of
  fields}}}\ (\bibinfo  {publisher} {Cambridge University Press},\ \bibinfo
  {year} {2007})\BibitemShut {NoStop}%
\bibitem [{Note7()}]{Note7}%
  \BibitemOpen
  \bibinfo {note} {In two dimensions: $\nabla ^{-2} f(\protect \bm {r}) =
  \DOTSI \intop \ilimits@ _{\protect \mathbb {R}^2} \protect \frac
  {1}{2\protect \,\pi } {\protect \rm ln}(\vert \protect \bm {r} - \protect \bm
  {r}' \vert ) \protect \, f(\protect \bm {r}') \protect \, {\protect \rm
  d}^2\protect \bm {r}'$.}\BibitemShut {Stop}%
\bibitem [{\citenamefont {Pietzonka}\ and\ \citenamefont
  {Seifert}(2017)}]{pietzonka_seifert_2017_entropy_production_active_particles}%
  \BibitemOpen
  \bibfield  {author} {\bibinfo {author} {\bibfnamefont {Patrick}\ \bibnamefont
  {Pietzonka}}\ and\ \bibinfo {author} {\bibfnamefont {Udo}\ \bibnamefont
  {Seifert}},\ }\bibfield  {title} {\enquote {\bibinfo {title} {Entropy
  production of active particles and for particles in active baths},}\
  }\href@noop {} {\bibfield  {journal} {\bibinfo  {journal} {J. Phys. A-Math.
  Theor.}\ }\textbf {\bibinfo {volume} {51}},\ \bibinfo {pages} {01LT01}
  (\bibinfo {year} {2017})}\BibitemShut {NoStop}%
\end{thebibliography}

\end{document}